\documentclass[aps,preprint,amsmath,amssymb,showpacs,floatfix,pre,nofootinbib]{
revtex4}
\usepackage{amsmath,amssymb,graphicx,subfigure,color,psfrag}

\newcommand{\dd}{\mbox{d}}
\begin{document}

\title{The quasilinear theory in the approach of long-range systems to
quasi-stationary states}
\author{Alessandro Campa$^1$ and Pierre-Henri Chavanis$^2$}
\affiliation{\footnotesize $^1$Complex Systems and Theoretical
Physics Unit, Health and
Technology
Department, Istituto Superiore di Sanit\`{a}, and INFN Roma1, Gruppo
Collegato Sanit\`{a}, 00161 Roma, Italy \\
$^2$ Laboratoire de Physique Th\'eorique, Universit\'e Paul Sabatier, 118 route
de Narbonne 31062 Toulouse, France}

\begin{abstract}

{\footnotesize We develop a quasilinear theory of the Vlasov
equation in order to describe the
approach of systems with long-range interactions to
quasi-stationary states. We derive a diffusion equation
governing the evolution of the velocity distribution of the
system towards a steady state. This steady state is expected
to correspond to the angle-averaged quasi-stationary distribution function
reached by the Vlasov equation as a result of a violent
relaxation. We compare the prediction of the
quasilinear theory to direct numerical simulations of the Hamiltonian Mean Field
model, starting from an unstable  spatially homogeneous distribution, either
Gaussian or semi-elliptical. In the Gaussian case, we find that the quasilinear
theory works reasonably well for weakly unstable
initial conditions (i.e. close to the critical energy $\epsilon_{\rm
c}=3/4=0.75$)
and that it is able to predict the energy $\epsilon_{\rm t}\simeq
0.735$ marking the out-of-equilibrium phase transition between unmagnetized and
magnetized quasi-stationary states. Similarly, the quasilinear theory
works well for energies close to the instability threshold of the
semi-elliptical case
$\epsilon^*_c =5/8=0.625$, and it predicts the out-of-equilibrium transition at
$\epsilon_{t}\simeq 0.619$. In both situations,
at energies lower
than the out-of-equilibrium transition the quasilinear theory works less
well, the disagreement
with the numerical simulations increasing by decreasing the energy.
 In that case, we
observe, in agreement with our previous numerical  study [A. Campa and P.-H.
Chavanis, Eur. Phys. J. B {\bf 86}, 170 (2013)], that the quasi-stationary
states are remarkably
well fitted by polytropic distributions (Tsallis distributions) with index
$n=2$ (Gaussian case) or $n=1$ (semi-elliptical case). In particular, these
polytropic distributions are able to account for the
region of negative specific heats in the out-of-equilibrium caloric curve,
unlike the Boltzmann and Lynden-Bell distributions.}

\end{abstract}
%\date{\today}
\maketitle

\section{Introduction}

Systems with long-range interactions exhibit peculiar properties both at equilibrium and
out-of-equilibrium. Among the equilibrium features that are not present in short-range systems,
probably the most striking are the ensemble inequivalence and the negative specific heat in the
microcanonical ensemble. Even richer is the phenomenology one observes out-of-equilibrium:
long-lived quasi-stationary states (QSSs), non-mixing behavior, phase
transitions between
non-Boltzmannian distributions, etc
\cite{houches,assisebook,campabook,physrep}. Many
results have been established, or illustrated, in the context of a toy model
called the Hamiltonian Mean Field (HMF) model \cite{hmf}. Although progress has
constantly been made in the last couple
of decades, there are still points very poorly understood. One of these issues concerns the
approach to QSSs.

In the statistical dynamics study of long-range systems, one of the main tools
is the Vlasov equation
that represents the interaction through a mean-field. For this reason the Vlasov equation describes
a ``collisionless'' dynamics. It can be shown that this is a very good
approximation that improves
more and more when the number of components increases (the Vlasov equation
becomes essentially exact when the number of particles $N\rightarrow +\infty$).
More precisely, the larger the system the larger
the time range for which the true dynamics can be studied with the Vlasov
equation. Eventually,
``collisional'' effects, i.e. finite $N$ effects, will take over, driving the
system towards Boltzmann-Gibbs
equilibrium. However, in some cases, especially in the astrophysical context, the time scale of these
``collisional'' effects can be orders of magnitudes larger that the age of the
Universe. It is therefore
important to characterize the various stages of the Vlasov dynamics. This equation admits both
an infinite number of conserved quantitites (the Casimirs) and an infinite number of stationary states.
The QSS reached by the system after an initial transient is just one of these stationary states,
obviously one which is stable with respect to perturbations. The infinite number of stationary states
of the Vlasov equation makes the prediction of which one is selected by the
system very difficult \cite{ybbdr,incomplete}. Actually, it is fair to say that,
theoretically, this is
still a completely unsolved problem.

The initial state is in general a non-stationary, or unstable stationary, state
of the Vlasov
equation. In that case, the system 
evolves very rapidly, with the dynamics governed by the Vlasov equation, towards
a QSS. Simulations show that,
as one could argue on the basis of the existence of an infinite number of stationary states, the selected QSS
strongly depends on the initial state of the system. A predictive theory was
built by Lynden-Bell \cite{lb} in an astrophysical context. According
to this theory, the QSS reached by the system is the one that maximizes an
entropy functional while preserving the energy and
all the Casimirs. Although perfectly defined, this problem is amenable to a computable solution only when
the distribution of the initial state has a small number of phase levels, i.e., it is piecewise constant. However,
even when the computation is possible, the comparison with the simulations shows
that the prediction of the Lynden-Bell
theory works only in some cases.
The explanation of its failure\footnote{Although we shall
focus in this paper on situations where the Lynden-Bell theory does not
provide a good description of the QSSs, we would like to emphasize
that there exist situations where the Lynden-Bell theory works remarkably well
and is able to predict the existence of out-of-equilibrium phase transitions
and re-entrant phases that would not have been detected without the help of
this theory; see in particular Refs. \cite{precommun,stan} and Fig. 36 of Ref.
\cite{epjb2013} for successful predictions of the Lynden-Bell theory.} has to be
sought in the
absence
of a complete mixing of the dynamics (under given constraints),
a property that is implicitly assumed in the theory \cite{incomplete}. Even
worse, in
self-gravitating systems, e.g. elliptical galaxies,
the Lynden-Bell distribution has infinite mass, thus it cannot represent the actual QSS of the system.

In the attempt to find alternatives, we have proposed \cite{epjb2013} to
compare the QSS found in numerical simulations
of the HMF model with polytropic
distributions
(sometimes called Tsallis distributions \cite{tsallisbook}). These distributions
are critical
points of functionals of the one-particle
distribution function that generalize
the usual Boltzmann entropy. This by itself does not give to polytropes
a privileged status with respect to other possible stationary states of the Vlasov equation. The reasons to employ them
can be summarized as follows. They have been introduced long ago in astrophysics, where they are called stellar
polytropes, in the attempt to describe the non-Boltzmannian distributions
observed \cite{btnew}. Being the critical points (at fixed particle number and
energy) of
functionals of the form $S[f]= -\int \dd \mathbf{x} \dd \mathbf{p} \, C(f)$,
where $C(f)$ is a convex function, they
determine distributions of the form $f(e)$, where $e$ is the individual energy,
with $f'(e)<0$ \cite{thlb}. The stability
of these Vlasov stationary states has to be determined in each case (see, e.g.,
the comments in this respect in Ref.
\cite{jsm2010,epjb2010}). 
The function $C(f)$ corresponding to the
Boltzmann entropy is $C(f) =-f \ln f$. Under suitable assumptions the function
$C(f) = f^q/(1-q)$ that gives rise to polytropic distributions can be obtained
from a modification of the Lynden-Bell theory that takes into account the
absence of complete
mixing \cite{assise,epjb2010}. In
a
numerical study \cite{epjb2013}, we have
shown that in some cases in which the prediction of the Lynden-Bell theory
fails to obtain the distribution of the QSSs reached by the system after the
violent relaxation, polytropic
distributions are a good approximation of the QSSs. However, we have found that
the quality of the approximation
worsens when the initial distribution is only weakly unstable.

In the case of weak instability of initially homogeneous distributions, another
type of approximation could be more suitable,
i.e. the quasilinear (QL) approximation.\footnote{The QL
theory is well-known in plasma physics for the Coulombian interaction
\cite{nicholson}. We apply it here to a new situation, namely the HMF model, in
which the interaction between particles is attractive.} This approach is
based on the
assumption that, although the initial
distribution is not Vlasov stable, nevertheless its evolution towards a Vlasov stable stationary state is such that
it is all the time only slightly inhomogeneous. The assumption will be made more precise below. In this paper we have
tested the validity of the QL theory in the HMF model, motivated
by the fact that the simulations performed
in \cite{epjb2013} to study the polytropic approximation have shown that, when
the initial distribution is only weakly unstable and the
polytropic approximation is not good, the QSS is only weakly
inhomogeneous. Therefore, we make the hypothesis that the distribution
is only slightly inhomogeneous all the time and compare the prediction of the
QL theory to the results of numerical simulations.

In Section \ref{sec_ql} we give the derivation of the QL
approximation for a generic mean field system composed of particles
moving on a circle. In Section \ref{sec_app}, we apply the theory to the HMF
model, where the particular form of the
interaction allows a simplification of the expressions. In Sections
\ref{sec_results}, \ref{sec_comp} and \ref{sec_comp_bis} we show
the results for a Gaussian and semi-elliptical initial condition and determine
the domain of validity of the QL theory and the polytropic fit.

\section{The QL approximation}
\label{sec_ql}

We describe here the derivation of the QL approximation for the
evolution of the
one-particle distribution function as determined from the Vlasov equation. In view of the
application to the HMF model, we consider a generic one dimensional system on
the circle,
i.e., a system with Hamiltonian
\begin{equation}\label{hamgen}
H = \sum_{i=1}^N \frac{p_i}{2} + \frac{1}{N}\underset{1\le i<j\le N}{\sum\sum}
V\left( \theta_i -\theta_j\right) \,\, ,
\end{equation}
where $\theta \in [0,2\pi]$, and the potential $V(\theta)$, which has $2\pi$
periodicity, is supposed to be continuous (the mass of the particles has been
put equal to $1$ without loss of generality). The Vlasov equation for
the one-particle distribution function
$f(\theta,p,t)$ is
\begin{equation}\label{vlasovequat}
\frac{\partial f(\theta,p,t)}{\partial t} + p\frac{\partial f(\theta,p,t)}{\partial \theta}
-\frac{\partial \Phi(\theta,t;f)}{\partial \theta}\frac{\partial f(\theta,p,t)}
{\partial p} = 0 \,\, ,
\end{equation}
where the mean field potential is given by
\begin{equation}\label{meanfieldpot}
\Phi(\theta,t;f) = \int \dd p' \, \int_0^{2\pi} \dd \theta' \, V(\theta - \theta')
f(\theta',p',t).
\end{equation}

For large long-range systems with $N \gg 1$ the Vlasov equation is a very good approximation
for times of order $N$
\cite{houches,assisebook,campabook,physrep}. Functions $f_0(p)$
that depend only on $p$ are particular stationary
solution of the Vlasov equation.\footnote{It is easy to see
that the mean field potential $\Phi$ is constant in $\theta$ for a uniform
distribution, i.e., a function $f$ that does not depend on $\theta$.} The
stability of these stationary solutions is
studied by putting $f(\theta,p,t) = f_0(p) + f_1(\theta,p,t)$ and linearizing the Vlasov
equation around $f_0(p)$, i.e., keeping only the terms at most linear in $f_1$. We then obtain
the following linear equation for $f_1(\theta,p,t)$:
\begin{equation}\label{linearvlasov}
\frac{\partial f_1(\theta,p,t)}{\partial t} + p\frac{\partial f_1(\theta,p,t)}{\partial \theta}
-\frac{\partial \Phi(\theta,t;f_1)}{\partial \theta}\frac{\partial f_0(p)}
{\partial p} = 0 \,\, ,
\end{equation}
which is valid as long as $f_1(\theta,p,t) \ll f_0(p)$. This is the
linearized Vlasov equation. The proper frequencies
$\omega$ of the dynamics
determined by this equation can be studied by inserting solutions where the time dependence is of the
form ${\rm e}^{-i\omega t}$. The function $f_0(p)$ is stable if all the solution
for $\omega$ have a non
positive imaginary part. In that case, the distribution
function  $f_0(p)$  can change only because of the effects of ``collisions''
(finite $N$ effects) that are not considered in the Vlasov
equation \cite{physrep,epjp}. We recall
that in the case of one
dimensional systems 
the time range of validity of the Vlasov equation grows like $N$
(for inhomogeneous systems) or with a higher power of
$N$ (for homogeneous systems) \cite{physrep,epjp}. If, on the contrary, some of
the
frequencies have a positive imaginary part, $f_0(p)$ is
Vlasov unstable, and the distribution $f(\theta,p,t)$ will depart from $f_0(p)$ rapidly.

There is an approximation that can work if the positive imaginary parts
of the proper frequencies are small enough (the scale with which to measure this
smallness will be precised later): the so-called QL approximation. We
first derive the corresponding
equations, then we describe their solutions, and finally we explain in which case one can expect
that the approximation is good. We begin by defining the angle average
of the distribution $f(\theta,p,t)$:
\begin{equation}\label{angleaverage}
f_0(p,t) = \frac{1}{2\pi}\int_0^{2\pi} \dd \theta \, f(\theta,p,t) \,\, .
\end{equation}
The angle averaged distribution is then used to define $f_1(\theta,p,t)$ by
\begin{equation}\label{f1define}
f(\theta,p,t)= f_0(p,t) + f_1(\theta,p,t) \,\, .
\end{equation}
Clearly the angle average of $f_1$ is zero.
Substituting Eq. (\ref{f1define}) in the Vlasov equation (\ref{vlasovequat}) we
have
\begin{equation}\label{vlasovequatf0f1}
\frac{\partial f_0(p,t)}{\partial t} + \frac{\partial f_1(\theta,p,t)}{\partial t}
+ p\frac{\partial f_1(\theta,p,t)}{\partial \theta}
-\frac{\partial \Phi(\theta,t;f_1)}{\partial \theta}
\left( \frac{\partial f_0(p,t)}{\partial p} + \frac{\partial f_1(\theta,p,t)}{\partial p} \right)
= 0 \,\, .
\end{equation}
We can derive an equation for the time variation of $f_0$ by taking the angle
average of the
last equation. We obtain
\begin{equation}\label{vlasovaverage}
\frac{\partial f_0(p,t)}{\partial t} - \frac{1}{2\pi}\frac{\partial}{\partial p}
\int_0^{2\pi} \dd \theta \, \frac{\partial \Phi(\theta,t;f_1)}{\partial \theta}f_1(\theta,p,t) = 0
\,\, .
\end{equation}
Using this equation to substitute the time derivative of $f_0$ in Eq.
(\ref{vlasovequatf0f1}),
we have
\begin{eqnarray}\label{vlasovequatf0f1_b}
\frac{\partial f_1(\theta,p,t)}{\partial t} &+& p\frac{\partial f_1(\theta,p,t)}{\partial \theta}
-\frac{\partial \Phi(\theta,t;f_1)}{\partial \theta}
\left( \frac{\partial f_0(p,t)}{\partial p} + \frac{\partial f_1(\theta,p,t)}{\partial p} \right)
\nonumber \\ &+& \frac{1}{2\pi}\frac{\partial}{\partial p}
\int_0^{2\pi} \dd \theta \, \frac{\partial \Phi(\theta,t;f_1)}{\partial \theta}f_1(\theta,p,t)
= 0 \,\, .
\end{eqnarray}

Let us now suppose that the following disequality is satisfied during the time evolution
\begin{equation}\label{smallapprox}
f_1(\theta,p,t) \ll f_0(p,t) \,\, .
\end{equation}
Then Eq. (\ref{vlasovequatf0f1_b}) can be approximated by keeping only the first order terms
in $f_1$, i.e.
\begin{equation}\label{vlasovequatf1}
\frac{\partial f_1(\theta,p,t)}{\partial t} + p\frac{\partial f_1(\theta,p,t)}{\partial \theta}
-\frac{\partial \Phi(\theta,t;f_1)}{\partial \theta} \frac{\partial f_0(p,t)}{\partial p}
= 0 \,\, ,
\end{equation}
which is a linear equation in $f_1$ similar to Eq. (\ref{linearvlasov}), but in which $f_0$
is not constant in time. The procedure now is to write the solution of this linear equation
in $f_1$, which depends on $f_0$, and to insert it in Eq.
(\ref{vlasovaverage}), to
obtain a closed equation for $f_0$.\footnote{We note the
formal similarity between Eqs. (\ref{vlasovaverage}) and
(\ref{vlasovequatf1})  and those arising in the kinetic theory of systems with
long-range interactions leading to the Lenard-Balescu equation (see Eqs. (12)
and (13) in \cite{epjp}). Both are based on a ``quasilinear approximation''.
However, there are crucial physical
differences. We are considering here the collisionless evolution
($N\rightarrow +\infty$) of a Vlasov unstable distribution function while the
Lenard-Balescu kinetic theory is concerned with the collisional evolution
(induced by finite $N$ effects) of a Vlasov stable distribution. Thus, our
kinetic theory is based on the Vlasov equation (\ref{vlasovequat}) instead of
the Klimontovich equation (see Eq. (3) in \cite{epjp}). Our kinetic
theory is also different from the kinetic theory of the Vlasov equation
(based on a maximum entropy production principle (MEPP) \cite{csr} or on another
quasilinear approximation \cite{cg}) leading to the equilibrium Lynden-Bell
distribution. The kinetic theory in \cite{csr,cg} implicitly assumes an
efficient mixing while our theory, as we shall see, assumes a weak mixing.} We
remark the conceptual difference between
Eq. (\ref{linearvlasov}) on the one hand, and the couple of equations given by
Eqs.
(\ref{vlasovaverage}) and (\ref{vlasovequatf1}) on the other hand. In the former case $f_0(p)$
is a stationary solution of the Vlasov equation (\ref{vlasovequat}) and, as
such, it is by
definition constant in time even if from the proper frequencies of Eq.
(\ref{linearvlasov}) we
could find that it is unstable. In the latter case, even if solving Eq. (\ref{vlasovequatf1})
we find that $f_1(\theta,p,t)$ does not have exponential growth, in general $f_0(p,t)$
defined by Eq. (\ref{vlasovaverage}) depends on time.

To analyze the solution of Eq. (\ref{vlasovequatf1}) we first consider Eq.
(\ref{linearvlasov}). Then, we generalize our results to Eq.
(\ref{vlasovequatf1}). The solution of Eq. (\ref{linearvlasov}) is best
obtained by decomposing $f_1(\theta,p,t)$ in Fourier components. To study its $k$-th Fourier
component,  we substitute in Eq. (\ref{linearvlasov}) the plane wave
expressions 
\begin{eqnarray}\label{fouriercomp_1}
f_1(\theta,p,t) &=& a_k(p) {\rm e}^{i(k\theta - \omega t)},
\\
\label{fouriercomp_2}
\Phi(\theta,t) &=& b_k {\rm e}^{i(k\theta - \omega t)}.
\end{eqnarray}
As usual, the physically meaningful solutions are, separately, the real
and the imaginary parts
of these expressions. The wavenumber $k$ takes all integer values from
$-\infty$ to $+\infty$.
Since $f_1(\theta,p,t)$ and $\Phi(\theta,t)$ are real, in their
Fourier expansion (as in Eq. (\ref{initialf1}) below)
there will be both components with $k$ and $-k$. Substituting Eq.
(\ref{fouriercomp_1}) and
Eq. (\ref{fouriercomp_2}) in Eq. (\ref{linearvlasov}) we have
\begin{equation}\label{linearvlasovfourier}
-i\omega a_k(p) +ipk a_k(p) -ikb_k \frac{\partial f_0}{\partial p} = 0.
\end{equation}
Obviously $b_k$ and $a_k(p)$ are related. Using the Fourier decomposition of the potential
$V(\theta)$
\begin{equation}\label{fourierpot}
V(\theta) = \frac{1}{2\pi} \sum_{k=-\infty}^\infty V_k {\rm e}^{ik\theta} \,\, ,
\end{equation}
where
\begin{equation}\label{fourierpotcomp}
V_k = \int_0^{2\pi} \dd \theta \, V(\theta) {\rm e}^{-ik\theta} \,\, ,
\end{equation}
we get
\begin{equation}\label{relat_a_b}
b_k = V_k \int \dd p \, a_k(p) \,\, .
\end{equation}
With the natural assumption $V(-\theta) = V(\theta)$ we have that $V_k$ is real and that
$V_{-k}=V_k$.  Substituting
Eq. (\ref{relat_a_b}) in
Eq. (\ref{linearvlasovfourier}) this
equation becomes
\begin{equation}\label{linearvlasovfourier_b}
(pk - \omega) a_k(p) -kV_k \frac{\partial f_0}{\partial p} \int \dd p' \, a_k(p') = 0
\,\, .
\end{equation}
We have to look for which values of $\omega$ this equation admits solutions $a_k(p)$ which are not
identically zero. Since this equation is linear, we can freely impose the normalization condition
\begin{equation}\label{normcondak}
\int \dd p' \, a_k(p') = 1 \,\, ,
\end{equation}
so that the equation becomes
\begin{equation}\label{linearvlasovfourier_c}
(pk - \omega) a_k(p) -kV_k \frac{\partial f_0}{\partial p} = 0
\,\, .
\end{equation}
As a matter of fact this equation is satified (for $k\neq 0$) for any real value of
$\omega$, that we denote by $\omega_{{\rm R}}$, by putting
\begin{equation}\label{solreal}
a_k(p;\omega_{{\rm R}}) = kV_k \, {\rm P} \frac{\frac{\partial f_0}{\partial p}}{pk-
\omega_{{\rm R}}} + |k|c(k,\omega_{{\rm R}}) \delta (pk-\omega_{{\rm R}} ) \,\, ,
\end{equation}
where ${\rm P}$ denotes that the principal value is taken when integrating
over $p$.  The
normalization condition for $a_k(p;\omega_{{\rm R}})$ provided by Eq.
(\ref{normcondak}) implies that
\begin{equation}\label{factorc}
c(k,\omega_{{\rm R}}) = 1 - kV_k \, {\rm P} \int \dd p \, \frac{\frac{\partial f_0}{\partial p}}
{pk-\omega_{{\rm R}}} \,\, .
\end{equation}
Therefore the real values of $\omega$ are not solutions of a dispersion relation, and thus
they are not proper frequencies of the linear equation (\ref{linearvlasovfourier_c}) in the usual
sense. These solutions are present also when the Fourier component $V_k$ is zero; in that
case $a_k(p;\omega_{{\rm R}})$ is a simple delta function and $c(k,\omega_{{\rm R}})=1$.
Let us note that $a_{-k}(p,-\omega_{{\rm R}}) = a_k(p,\omega_{{\rm R}})$. For $k=0$ the only
solution of Eq. (\ref{linearvlasovfourier_c}) is $\omega = 0$.

Another type of solutions of Eq. (\ref{linearvlasovfourier_c}) can be obtained
for
particular complex values of $\omega$. They are given by
\begin{equation}\label{solforak}
a_k(p;\omega) = kV_k \frac{\frac{\partial f_0}{\partial p}}{pk-\omega}
\,\, ,
\end{equation}
provided the normalization condition (\ref{normcondak}) is satified. Integrating both sides
with respect to $p$ we see that the complex value of $\omega$ must satisfy
\begin{equation}\label{dielectric}
D(k,\omega) \equiv 1 - kV_k \int \dd p \, \frac{\frac{\partial f_0}{\partial p}}{pk-\omega}
= 0 \,\, .
\end{equation}
The function $D(k,\omega)$ is called the response dielectric function. There are
isolated
complex values of $\omega$ such that $D(k,\omega)=0$ (dispersion relation).
These are the proper frequencies of
Eq. (\ref{linearvlasovfourier_c}). There are two important relations to
note. Since $V_k = V_{-k}$ is real, then if $D(k,\omega) = 0$ we have that also
$D(k,\omega^*)=0$ and $D(-k,-\omega)=0$. Therefore, if for a given wavenumber $k$ there is a
complex proper frequency $\omega$, there is also the complex conjugate proper frequency
$\omega^*$, while for the wavenumber $-k$ there will be the proper frequencies $-\omega$
and $-\omega^*$. Correspondingly, we have $a_k(p,\omega^*) = a_k^*(p,\omega)$ and
$a_{-k}(p,-\omega) = a_k(p,\omega)$.

From the Plemelj formula
\begin{equation}\label{plemelj}
\lim_{\eta \to 0^+} \frac{1}{x - x_0 \mp i \eta}
= {\rm P} \frac{1}{x - x_0} \pm i\pi \delta (x - x_0)
\end{equation}
we can obtain the value of $D(k,\omega)$ for real values of $\omega$. If
$\omega_{{\rm R}}$ and
$\omega_{{\rm I}}$ are the real and imaginary parts of $\omega$, we have
\begin{equation}\label{dielectriclimit}
\lim_{\omega_{{\rm I}} \to 0^{\pm}} D(k,\omega) =
1 - kV_k {\rm P} \int \dd p \, \frac{\frac{\partial f_0}{\partial p}}{pk-\omega_{{\rm R}}}
%+ \frac{i\pi}{k} \frac{\partial f_0}{\partial p}\left(\frac{\omega_{{\rm R}}}{k}\right)
%+ \frac{i\pi}{k} \frac{\partial f_0}{\partial p}\left(\omega_{{\rm R}}/k\right)
\pm \frac{i\pi}{k} \left. \frac{\partial f_0}{\partial p}\right|_{p=\omega_{{\rm R}}/k}
= c(k,\omega_{{\rm R}})
\pm \frac{i\pi}{k} \left. \frac{\partial f_0}{\partial p}\right|_{p=\omega_{{\rm R}}/k}
\,\, .
\end{equation}
The different limit from above and below the real line shows that Eq. (\ref{dielectric})
defines two different analytic functions of $\omega$, one in the upper plane and one in
the lower plane. The two limits coincide only for the particular values of $\omega_{{\rm R}}$
for which the last term in the right-hand side of Eq. (\ref{dielectriclimit}) is zero, i.e.,
for $\omega_{{\rm R}}/k$ equal to the values of $p$ where $f_0(p)$ is minimum or maximum.
If it happens that for these particular values $\omega_{{\rm R}}$ we also have
$c(k,\omega_{{\rm R}})=0$, then these real frequencies belong to the proper frequencies,
since they satisfy $D(k,\omega)=0$. 
Therefore among the isolated solutions there can also be real $\omega_{{\rm R}}$ such that
$c(k,\omega_{{\rm R}})=0$ and $f_0'(\omega_{{\rm R}}/k)=0$. In this case the normalized
$a_k(p;\omega_{{\rm R}})$ is given by Eq. (\ref{solforak}) with $\omega = \omega_{{\rm R}}$
and without the necessity of the principal value.

It is possible to show \cite{case} that, for any given $k \ne 0$, the functions
$a_k(p;\omega_{{\rm R}})$
in Eq. (\ref{solreal}) and $a_k(p;\omega)$ in Eq. (\ref{solforak}) constitute a complete orthogonal
system (with
the usual scalar product), when
$\omega_{{\rm R}}$ in the former runs over the real axis and $\omega$ in the
latter
runs over the discrete solutions of the dispersion relation
$D(k,\omega)=0$.\footnote{Equation
(\ref{linearvlasov}) is not self-adjoint, and for each solution
$a_k(p;\omega_{{\rm R}})$
or $a_k(p;\omega)$ there is a corresponding solution
$\tilde{a}_k(p;\omega_{{\rm R}})$ or $\tilde{a}_k(p;\omega)$ of the adjoint
equation.
The orthogonality of the system is expressed by the fact that each given
solution $a_k$ of Eq. (\ref{linearvlasov})
is orthogonal to all the solutions $\tilde{a}_k$ of the adjoint equation,
except the one corresponding to
it \cite{case}.}
Therefore,
any function $h(p)$ can be expanded as
\begin{equation}\label{hexpansion}
h(p) = \int \dd \omega_{{\rm R}} \, \alpha(\omega_{{\rm R}};k)a_k(p;\omega_{{\rm R}}) +
\sum_j \beta_j(k) a_k(p;\omega_j(k)) 
\end{equation}
with proper coefficients $\alpha(\omega_{{\rm R}};k)$ and $\beta_j(k)$. As just
mentioned,
the sum runs on the discrete values (generally both complex and real) satifying $D(k,\omega)=0$.
Therefore, if the initial value of $f_1(\theta,p,t)$ of Eq. (\ref{linearvlasov}) is
$h(p) \exp \left( ik\theta \right)$, the following evolution is given by
\begin{equation}\label{f1later}
f_1(\theta,p,t) = {\rm e}^{ik\theta}\left[ \int \dd \omega_{{\rm R}} \,
\alpha(\omega_{{\rm R}};k)a_k(p;\omega_{{\rm R}})
{\rm e}^{-i\omega_{{\rm R}} t} + \sum_j \beta_j(k) a_k(p;\omega_j(k)){\rm e}^{-i\omega_j(k) t} \right]
\,\, .
\end{equation}
The corresponding evolution of ${\partial \Phi(\theta,t)}/{\partial \theta}$
is
\begin{equation}\label{philater}
\frac{\partial \Phi(\theta,t)}{\partial \theta} = ikV_k{\rm e}^{ik\theta}
\left[ \int \dd \omega_{{\rm R}} \, \alpha(\omega_{{\rm R}};k) {\rm e}^{-i\omega_{{\rm R}} t}
+ \sum_j \beta_j(k) {\rm e}^{-i\omega_j(k) t} \right] \,\, .
\end{equation}
A general initial value of $f_1$ can be expanded in Fourier series as
\begin{equation}\label{initialf1}
f_1(\theta,p,0) = \frac{1}{2\pi} \sum_{k=-\infty}^\infty f_k(p) {\rm e}^{ik\theta}
\,\, ,
\end{equation}
where the term with $k=0$ can be assumed to be absent (it simply corresponds
to a constant
in time), and where $f_{-k} = f_k^*$. Expanding any $f_k(p)$ in the form given by Eq. (\ref{hexpansion}),
the evolution of $f_1$ is given by
\begin{equation}\label{f1totlater}
f_1(\theta,p,t) = \frac{1}{2\pi} \sum_{k=-\infty}^\infty {\rm e}^{ik\theta}
\left[ \int \dd \omega_{{\rm R}} \,
\alpha(\omega_{{\rm R}};k)a_k(p;\omega_{{\rm R}})
{\rm e}^{-i\omega_{{\rm R}} t} + \sum_j \beta_j(k) a_k(p;\omega_j(k)){\rm e}^{-i\omega_j(k) t} \right]
\,\, .
\end{equation}
The reality of this expression follows from $\alpha(-\omega_{{\rm R}};-k) = \alpha^*(\omega_{{\rm R}};k)$
and $\beta_j(-k) = \beta_j^*(k)$ when the proper frequencies for $k$ and $-k$ are numbered so that
for the same $j$ we have $\omega$ for $k$ and $-\omega^*$ for $-k$.
Finally, the evolution of ${\partial \Phi(\theta,t)}/{\partial \theta}$ is
\begin{equation}\label{phitotlater}
\frac{\partial \Phi(\theta,t)}{\partial \theta} = \frac{1}{2\pi} \sum_{k=-\infty}^\infty ikV_k
{\rm e}^{ik\theta} \left[ \int \dd \omega_{{\rm R}} \, \alpha(\omega_{{\rm R}};k)
{\rm e}^{-i\omega_{{\rm R}} t} + \sum_j \beta_j(k) {\rm e}^{-i\omega_j(k) t} \right]
\,\, ,
\end{equation}
which is also real.

Let us now compute the integral in $\theta$ of the product of
${\partial \Phi(\theta,t)}/{\partial \theta}$ and $f_1(\theta,p,t)$, the
expression that
appears in Eq. (\ref{vlasovaverage}). Multiplying Eqs. (\ref{f1totlater}) and (\ref{phitotlater})
and integrating, we obtain
\begin{eqnarray}\label{phif1int}
&&\int_0^{2\pi} \dd \theta \, \frac{\partial \Phi(\theta,t)}{\partial \theta}f_1(\theta,p,t) \nonumber \\
&=& -i \frac{1}{2\pi} \sum_{k=-\infty}^\infty kV_k
\left[ \int \dd \omega_{{\rm R}}' \, \alpha(\omega_{{\rm R}}';-k)
{\rm e}^{-i\omega_{{\rm R}}' t} + \sum_n \beta_n(-k) {\rm e}^{-i\omega_n(-k) t} \right] \nonumber \\
&\times&\left[ \int \dd \omega_{{\rm R}} \,
\alpha(\omega_{{\rm R}};k)a_k(p;\omega_{{\rm R}})
{\rm e}^{-i\omega_{{\rm R}} t} + \sum_j \beta_j(k) a_k(p;\omega_j(k)){\rm e}^{-i\omega_j(k) t} \right]
\,\, .
\end{eqnarray}
When $f_0(p)$ is stable, there are no complex solutions of Eq. (\ref{dielectric}). This follows from the fact
that its complex solutions come in complex conjugate pairs, and therefore if there are complex solutions there
are necessarily unstable modes.\footnote{If
$\omega=\omega_R+i\omega_I$ is a solution of the dispersion relation
(\ref{dielectric}), then
$\omega=\omega_R-i\omega_I$ is also a solution. If the first one is stable the
second
one is unstable, and {\it vice versa}.} Thus, for a stable $f_0(p)$, the term
with the sum in Eqs. (\ref{f1totlater}) and (\ref{phitotlater}) is either absent or with only
real proper frequencies. Then, $f_1(\theta,p,t)$, ${\partial
\Phi(\theta,t)}/{\partial \theta}$
and the integral of their product, Eq. (\ref{phif1int}), will not have terms with exponential
growth. Actually, the interferences between the real frequencies will give rise
to an exponential decay (the Landau damping).\footnote{We remark
that, here, we are treating the linearized Vlasov equation {\it \`a la} Van
Kampen \cite{vankampen}, i.e.,
with the Fourier transform in time. This treatment is the most natural one to
find the proper frequencies of a linear problem. However, it is also possible to
treat the problem {\it \`a la} Landau \cite{landau}, i.e., with the Laplace
transform in time. The latter procedure is more suitable to study an initial
value problem and to obtain the collective damped modes that result from the
interference effects (in particular the Landau damping). In the treatment with
the Laplace transform, the dielectric function is defined by Eq.
(\ref{dielectric}) only in the upper $\omega$ plane; in the rest of the complex
plane, it is defined by the analytic continuation. In the treatment {\it
\`a la} Landau
one sees how the Landau damping stems from the real frequencies of the treatment
{\it \`a la} Van Kampen (see, e.g., \cite{nicholson} for mathematical
details.)}  On the other hand, when there
are complex proper frequencies, and then with positive imaginary parts (since they come in complex
conjugate pairs), there will be terms with exponential growths. In that case,
one could approximate
the right-hand side of Eq. (\ref{phif1int}) by keeping only those terms. With
the further assumption
that for each $k$ there is only one complex proper
frequency\footnote{According
to the Nyquist theorem \cite{nyquisthmf,nyquistaa}, for an unstable single
humped distribution function $f_0(p)$, there is only one unstable mode.
In more general
cases, if there are several complex proper frequencies, one selects the
largest one, i.e., the one that corresponds to the maximum growth rate.}
 $\omega(k)=\omega_{{\rm R}}(k) + i
\omega_{{\rm I}}(k)$ with $\omega_{{\rm I}}(k)>0$ (and its complex conjugate) we
obtain
\begin{equation}\label{phif1intapprox}
\int_0^{2\pi} \dd \theta \, \frac{\partial \Phi(\theta,t)}{\partial \theta}f_1(\theta,p,t)
\approx -i \frac{1}{2\pi} \sum_{k=-\infty}^\infty kV_k
|\beta(k)|^2 a_k(p;\omega(k)) {\rm e}^{-i\left(\omega(k)-\omega^*(k)\right) t}
\,\, .
\end{equation}
Using the expression of $a_k(p;\omega)$ from Eq. (\ref{solforak}) the previous
expression can be transformed in
\begin{equation}\label{phif1intapprox_b}
\int_0^{2\pi} \dd \theta \, \frac{\partial \Phi(\theta,t)}{\partial \theta}f_1(\theta,p,t)
\approx \frac{1}{2\pi} \sum_{k>0} k^2 V_k |\beta(k)|^2 \frac{2\omega_{{\rm I}}(k)}
{\left(pk - \omega_{{\rm R}}(k)\right)^2 + \omega_{{\rm I}}^2(k)}
\frac{\partial f_0(p)}{\partial p} {\rm e}^{2 \omega_{{\rm I}}(k) t}
\,\, .
\end{equation}

We now consider Eq. (\ref{vlasovequatf1}). For a linear equation with time
dependent
coefficients there are not proper frequencies in the usual sense, and the solutions are not of
the same form as in Eqs. (\ref{fouriercomp_1}) and (\ref{fouriercomp_2}).
However, we can try,
for a given Fourier component, similar expressions, i.e.
\begin{eqnarray}\label{fouriercompgena}
f_1(\theta,p,t) &=& a_k(p,t) {\rm e}^{ik\theta}{\rm e}^{-i\int_0^t {\rm d} t' \,
\omega(t')},
\end{eqnarray}
\begin{eqnarray}\label{fouriercompgenb}
\Phi(\theta,t) &=& b_k(t) {\rm e}^{ik\theta}{\rm e}^{-i\int_0^t {\rm d} t' \,
\omega(t')}.
\end{eqnarray}
We substitute Eqs. (\ref{fouriercompgena}) and (\ref{fouriercompgenb}) in Eq.
(\ref{vlasovequatf1}) and we neglect the time derivatives of $a_k(p,t)$
and $b_k(t)$, so that we obtain an equation similar to
(\ref{linearvlasovfourier}):
\begin{equation}\label{linearvlasovfouriertime}
-i\omega a_k(p,t) +ipk a_k(p,t) -ikb_k(t) \frac{\partial f_0(p,t)}{\partial p} =
0.
\end{equation}
With this adiabatic approximation, in which the time
derivatives of $a_k(p,t)$ and $b_k(t)$ are
neglected, the last equation is solved as before, with the only difference that $f_0'(p,t)$
and thus $\omega(t)$ depend on time, and the time dependence of $a_k(p,t)$ is only through
$\omega(t)$. Without repeating the whole procedure, we write
directly the
expression for the
integral of the product of ${\partial \Phi(\theta,t)}/{\partial \theta}$ and
$f_1(\theta,p,t)$ in the approximation that for each $k$ there is only one proper frequency
with positive imaginary part (see footnote 8):
\begin{equation}\label{phif1intapprox_time}
\int_0^{2\pi} \dd \theta \, \frac{\partial \Phi(\theta,t)}{\partial \theta}f_1(\theta,p,t)
\approx \frac{1}{2\pi} \sum_{k>0} k^2 V_k |\beta(k)|^2 \frac{2\omega_{{\rm I}}(k,t)}
{\left(pk - \omega_{{\rm R}}(k,t)\right)^2 + \omega_{{\rm I}}^2(k,t)}
\frac{\partial f_0(p,t)}{\partial p}
{\rm e}^{2 \int_0^t {\rm d} t' \, \omega_{{\rm I}}(k,t')}
\,\, .
\end{equation}
This equation can be written in a different form by defining a diffusion
coefficient
\begin{equation}\label{defdiff}
D(p,t) = \sum_{k>0} \frac{2\chi_k(t)\omega_{{\rm I}}(k,t)}
{\left(pk - \omega_{{\rm R}}(k,t)\right)^2 + \omega_{{\rm I}}^2(k,t)}
\,\, 
\end{equation}
with
\begin{equation}\label{derchi}
\frac{\partial \chi_k(t)}{\partial t} = 2\omega_{{\rm I}}(k,t)\chi_k(t) \,\, ,
\end{equation}
where some constants have been incorporated in the definition of $\chi_k(0)$.
Therefore,
Eq. (\ref{phif1intapprox_time}) becomes
\begin{equation}\label{phif1intapprox_time_b}
\int_0^{2\pi} \dd \theta \, \frac{\partial \Phi(\theta,t)}{\partial \theta}f_1(\theta,p,t)
\approx D(p,t) \frac{\partial f_0(p,t)}{\partial p}
\,\, .
\end{equation}
Substituting the right-hand side of this expression in Eq. (\ref{vlasovaverage})
we obtain
\begin{equation}\label{vlasovaveragediff}
\frac{\partial f_0(p,t)}{\partial t} = \frac{\partial}{\partial p} \left(
D(p,t) \frac{\partial f_0(p,t)}{\partial p} \right)
\,\, ,
\end{equation}
which is of the form of a diffusion equation. This is the QL
approximation for
the evolution of the angle averaged distribution function $f_0(p,t)$.

Let us finally discuss when we expect that this approximation is good, i.e.,
when
Eq. (\ref{smallapprox}) is satisfied throughout the time evolution. If we assume that
$f_0(p,t)$ is of order $1$, we can argue that the approximation is good as long as
$\chi_k^{{1}/{2}}(t)\ll 1$ for each $k$. This defines the smallness of
$\omega_{{\rm I}}(k)$ also in relation to the initial values $\chi_k(0)$. We
note that,
in general, we expect that $\omega_{{\rm I}}(k)$ decreases in time since
the
dynamics  drives the function $f_0(p,t)$ towards a Vlasov (marginally) stable
distribution.

\section{Application to the HMF model}
\label{sec_app}

The expressions simplify for the HMF model where the potential is $V(\theta)
= 1 - \cos \theta$. Then, we have
\begin{equation}\label{vkhmf}
V_k = 2\pi \delta_{k,0} - \pi \left( \delta_{k,1} + \delta_{k,-1}\right).
\end{equation}
As a consequence, the dielectric function $D(k,\omega)$ has a structure only for
$k=\pm 1$, and there is a collective dynamics only for these wavenumbers:
\begin{equation}\label{dielectrichmf}
D(k,\omega) = 1 + \pi k \left( \delta_{k,1} + \delta_{k,-1}\right)
\int \dd p \, \frac{\frac{\partial f_0}{\partial p}}{pk-\omega}
\,\, .
\end{equation}
The diffusion coefficient of the QL approximation is then
\begin{equation}\label{defdiffhmf}
D(p,t) = \frac{2 \chi(t) \omega_{{\rm I}}(t)}
{(p-\omega_{{\rm R}}(t))^2 + \omega_{{\rm I}}^2(t)} \,\, ,
\end{equation}
where the evolution of $\chi(t)$ is given by Eq. (\ref{derchi}). It is
sufficient to study the dispersion
relation only for $k=1$. It is given by
\begin{equation}\label{disperhmf}
1 +\pi \int \dd p \, \frac{\frac{\partial f_0(p,t)}{\partial p}}{p -\omega}
= 0.
\end{equation}
Things simplify further if we consider a function $f_0(p)$ that is initially even in $p$
and with only a single maximum at $p=0$. These properties are conserved by the diffusion
equation for $f_0(p,t)$ (we will prove this below). It is then easy to show that if there is
a proper frequency with a positive imaginary part (unstable case), its real part
must be zero. Indeed, we
can write the dispersion relation as
\begin{equation}\label{disperhmfeven}
1 +\pi \int \dd p \, \frac{f_0'(p,t)(p-\omega_{{\rm R}}(t))}
{(p-\omega_{{\rm R}}(t))^2 + \omega_{{\rm I}}^2(t)} +i\pi \omega_{{\rm I}}
\int \dd p \, \frac{f_0'(p,t)}
{(p-\omega_{{\rm R}}(t))^2 + \omega_{{\rm I}}^2(t)} = 0 \,\, .
\end{equation}
The real and imaginary parts must be separately equal to zero. However, the last
integral is negative (positive) definite if $\omega_{{\rm R}}(t)>0 \,\, (<0)$, and it
is zero when $\omega_{{\rm R}}(t) = 0$. This can be seen as
follows: (i) for $\omega_{{\rm R}}(t)=0$ it vanishes since $f_0'(p,t)$ is odd;
(ii) for $\omega_{{\rm R}}(t)>0$, exploiting that
$f_0'(p,t)$ is odd and that it is negative (positive) for $p>0$ ($<0$), the sum of the values
of the integrand for $p$ and $-p$ is negative; (iii) analogously for
$\omega_{{\rm R}}(t)<0$ this sum is positive. To be more explicit, consider for
example the case
$\omega_{{\rm R}}(t)>0$. Then, for any
$p>0$, we have $|p-\omega_{{\rm R}}(t)|<|-p-\omega_{{\rm R}}(t)|$, i.e.,
$(p-\omega_{{\rm R}}(t))^2 + \omega_{{\rm I}}^2(t) <(-p-\omega_{{\rm
R}}(t))^2+\omega_{{\rm I}}^2(t)$.
Thus, since for $p>0$ one has $f_0'(p,t)=-f_0'(-p,t)<0$, the sum of the
contributions of $p$ and $-p$
to the last integral in Eq. (\ref{disperhmfeven}) is negative. Therefore, since
we must have $\omega_{{\rm R}}(t)=0$, the dispersion relation
becomes\footnote{Another proof of this result is given in Sec.
2.8. of
\cite{nyquisthmf}. Considering an unstable single humped symmetric distribution
$f_0(p)$, and using the Nyquist theorem, one can show that the 
dispersion relation (\ref{disperhmfeven}) has a unique solution with
$\omega_I>0$; it is such that $\omega_R=0$.} 
\begin{equation}\label{disperhmfeven_b}
1 +\pi \int \dd p \, \frac{pf_0'(p,t)}
{p^2 + \omega_{{\rm I}}^2(t)} = 0 \,\, ,
\end{equation}
and the diffusion coefficient
\begin{equation}\label{defdiffhmf_b}
D(p,t) = \frac{2 \chi(t) \omega_{{\rm I}}(t)}
{p^2 + \omega_{{\rm I}}^2(t)} \,\, .
\end{equation}
Summarizing, for the HMF model, the integration of the diffusion equation
(\ref{vlasovaveragediff}) is performed with $D(p,t)$ given by Eq. (\ref{defdiffhmf_b}),
where $\omega_{{\rm I}}(t)$ is the unique positive solution of Eq. (\ref{disperhmfeven_b}),
with $f_0(p,t)$ single humped and even, and where the evolution equation for $\chi(t)$ is
\begin{equation}\label{derchi_hmf}
\frac{\dd \chi(t)}{\dd t} = 2\omega_{{\rm I}}(t)\chi(t) \,\, .
\end{equation}
We remark that Eq. (\ref{vlasovaveragediff}) is a diffusion equation with a time
(and velocity) dependent diffusion coefficient given, for the HMF model, by Eq.
(\ref{defdiffhmf_b}), which is different
from $0$ as long as $\omega_{{\rm I}}(t)$ does not vanish. If it happens that 
$\omega_{{\rm I}}$ vanishes at a certain time, then the solution of the
diffusion equation reaches a stationary state.
Actually, as already noted, we expect that the diffusion equation will drive the function
$f_0(p,t)$ towards a stationary state by monotonically decreasing $\omega_{{\rm
I}}(t)$ up to $0$.  We do not have a general proof of this, but we can show
that, if at time $t=0$ the imaginary pulsation $\omega_{{\rm I}}(t)$ is
positive, it has to tend to $0$ for increasing time [see Eq. (\ref{prooftend0})
below]. We cannot prove that $\omega_{{\rm I}}(t)$ decreases monotonically;
however, we can give an approximate argument, described in
Appendix
\ref{omegadecrease}, to show that this is likely to be the case. The argument
is based
on the result
presented below
in Eq. (\ref{derkinener_b}), and on the fact, found in the numerical integration of the diffusion
equation and shown in the next Section, that the function $f_0(p,t)$ keeps approximately the same
functional form while it varies as driven by the diffusion equation.

It is useful to make the following evaluation. If we suppose that, at a given time $t$, $f_0(p,t)$
is a Gaussian
\begin{equation}\label{fzerogauss}
f_0(p,t) = \frac{1}{2\pi} \sqrt{\frac{\beta(t)}{2\pi}} {\rm
e}^{-\frac{1}{2}\beta(t) p^2} \,\, ,
\end{equation}
which is normalized such that $\int_{-\infty}^{+\infty} f_0(p,t) \, dp=1/2\pi$,
then the dispersion relation (\ref{disperhmfeven_b}) becomes
\begin{equation}\label{disperhmfgauss}
1 - \sqrt{\frac{\beta^3}{8\pi}} \int \dd p \, \frac{p^2 {\rm
e}^{-\frac{1}{2}\beta p^2}}
{p^2 + \omega_{{\rm I}}^2} = 0 \,\, ,
\end{equation}
where we have not indicated the time dependence of $\beta$ and $\omega_{{\rm
I}}$. The integral
in the last expression can be expressed in term of the error function
\begin{equation}\label{erfdef}
{\rm erf}(x) = \frac{2}{\sqrt{\pi}} \int_0^x \dd y \, {\rm e}^{-y^2}
\,\, .
\end{equation}
We have
\begin{equation}\label{disperhmfgauss_b}
1 - \frac{1}{2}\beta + \omega_{{\rm I}}\sqrt{\frac{\pi\beta^3}{8}} {\rm
e}^{\frac{1}{2}\beta \omega_{{\rm I}}^2}
\left[ 1 - {\rm erf}\left (\sqrt{\frac{\beta}{2}} \omega_{{\rm I}}\right )
\right] = 0 \,\, .
\end{equation}
For a general distribution function $f_0(p,t)$, we define the kinetic
temperature $T(t)$ by
\begin{equation}
\label{tkin}
T(t)=\langle p^2\rangle=2\pi\int_{-\infty}^{+\infty} {\rm d}p\, p^2 f_0(p,t).
\end{equation}
This is the variance of the distribution. For the Gaussian
distribution (\ref{fzerogauss}), we have $\beta=1/T$.
Therefore, Eq. (\ref{disperhmfgauss_b}) relates the growth rate $\omega_I$ to
the temperature $T$ for the Gaussian distribution (\ref{fzerogauss}). Let us
suppose that at time $t=0$ the function $f_0$ is a Gaussian. Its evolution, as
determined by Eq. (\ref{vlasovaveragediff}), together with Eqs.
(\ref{defdiffhmf_b})
and (\ref{derchi_hmf}), will not maintain this structure, i.e., we
cannot hope that
there will simply be a change with time of the parameter $\beta$. However, in
our
simulations we will see that the true evolution does not depart very much
from
a Gaussian, at least for the cases considered.

We now prove that an initial $f_0(p)$ which is even in $p$ and with only a single
maximum at $p=0$ will keep these properties during the dynamics determined by
Eq. (\ref{vlasovaveragediff}). We have seen that, as long as $f_0(p,t)$ is of this
form, then $\omega_{{\rm R}}(t)=0$, and the diffusion coefficient is given by
Eq. (\ref{defdiffhmf_b}). This is an even function of $p$, with a single maximum
at $p=0$. Therefore, the right-hand side of Eq. (\ref{vlasovaveragediff}) is
even in $p$.
Furthermore, taking the derivative of Eq. (\ref{vlasovaveragediff}) with
respect to $p$ we have
\begin{equation}\label{derdiffeq}
\frac{\partial }{\partial t}\frac{\partial f_0}{\partial p} =
\frac{\partial^2 D}{\partial p^2}\frac{\partial f_0}{\partial p}+
2\frac{\partial D}{\partial p}\frac{\partial^2 f_0}{\partial p^2}+
D\frac{\partial^3 f_0}{\partial p^3} \,\, .
\end{equation}
If $f_0(p,t)$ develops a point $p > 0$ with $\frac{\partial f_0}{\partial p} = 0$,
at that moment, by continuity, we will have $\frac{\partial^2 f_0}{\partial
p^2}=0$
and $\frac{\partial^3 f_0}{\partial p^3} < 0$. Then, we will have
$\frac{\partial }{\partial t}\frac{\partial f_0}{\partial p} < 0$. This proves that
the function $f_0(p,t)$ will remain with a single maximum at $p=0$.

We also prove that the average kinetic energy increases with time. In fact, we
have
\begin{eqnarray}\label{derkinener}
\frac{\partial }{\partial t} \int \dd p \, \frac{p^2}{2} f_0(p,t)
&=& \int \dd p \, \frac{p^2}{2} \frac{\partial f_0(p,t)}{\partial t}
\nonumber \\
= \int \dd p \, \frac{p^2}{2} \frac{\partial}{\partial p} \left( D(p,t)
\frac{\partial f_0(p,t)}{\partial p} \right) &=& - \int \dd p \, pD(p,t)
\frac{\partial f_0(p,t)}{\partial p} \,\, .
\end{eqnarray}
Substituting the expression of the diffusion coefficient (\ref{defdiffhmf_b})
in Eq. (\ref{derkinener}), and
using the dispersion relation (\ref{disperhmfeven_b}), we obtain
\begin{equation}\label{derkinener_b}
\frac{\partial }{\partial t} \int \dd p \, \frac{p^2}{2} f_0(p,t) =
-2 \chi(t) \omega_{{\rm I}}(t) \int \dd p \, \frac{p\frac{\partial f_0(p,t)}{\partial p}}
{p^2 + \omega_{{\rm I}}^2(t)} = \frac{2\chi(t) \omega_{{\rm I}}(t)}{\pi} > 0 \,\, .
\end{equation}
Since
\begin{equation}
\epsilon_{\rm kin}(t)=2\pi \int \dd p \, \frac{p^2}{2} f_0(p,t)=\frac{1}{2}T(t),
\end{equation}
we find that
\begin{equation}
\frac{dT}{dt} =
8\chi(t) \omega_{{\rm I}}(t) > 0 \,\, ,
\end{equation}
and we conclude that the kinetic temperature $T(t)$ increases with time.
Therefore,
the velocity distribution has the tendency to spread. The increase of the
kinetic temperature also qualitatively confirms that the system tends to be less
unstable as time goes on.

As anticipated, we can prove that $\omega_I(t)$ must necessarily tend to
$0$.
In fact, the relation
\begin{eqnarray}
\frac{\partial }{\partial t} \int \dd p \, f_0(p,t)^2
&=& 2 \int \dd p \, f_0(p,t) \frac{\partial f_0(p,t)}{\partial t}
\nonumber \\
= 2\int \dd p \,   f_0(p,t)\frac{\partial}{\partial p} \left( D(p,t)
\frac{\partial f_0(p,t)}{\partial p} \right) &=& - 2\int \dd p \, D(p,t)
\left (\frac{\partial f_0(p,t)}{\partial p}\right )^2\le 0 \,\, ,
\label{prooftend0}
\end{eqnarray}
is a sort of $H$-theorem for the functional $\Gamma_2=\int \dd p \,
f_0(p,t)^2$ similar to the enstrophy in
two-dimensional turbulence.\footnote{This inequality is actually
true for any $H$-function  \cite{thlb} of the form $H=-\int C(f)\, dp$ where
$C(f)$ is convex ($C''\ge 0$) since $\dot H=\int \dd p \, D(p,t) C''(f_0(p,t))
\left (\frac{\partial f_0(p,t)}{\partial p}\right )^2\ge
0$.} Since $\dot \Gamma_2\le 0$ and since $\Gamma_2$ is bounded
from below by $0$, we
conclude that $\dot \Gamma_2\rightarrow 0$ for $t\rightarrow +\infty$. According
to the last equality in Eq. (\ref{prooftend0}), this implies that
$D(p,t)\rightarrow 0$ for $t\rightarrow +\infty$ at least for some $p$'s (note
that $f_0(p,t)$ cannot become independent on $p$ otherwise it would not be
normalizable). Considering the expression (\ref{defdiffhmf_b}) of $D(p,t)$ this
in turn implies that $\omega_I(t)$ must necessarily tend to $0$.

Finally, we derive the approximate expression of the dispersion relation that
holds
when $\omega_{{\rm I}}(t)$ is very small. This can be useful in the numerical
computations. In fact, as emphasized previously, we expect that during the
dynamics
$\omega_{{\rm I}}(t)$ decreases towards zero, so that $f_0(p,t)$ becomes
(marginally) Vlasov stable. It is not difficult to see that at first order
in $\omega_{{\rm I}}(t)$ the dispersion relation (\ref{disperhmfeven_b})
(we remind that this expression is valid for an even $f_0(p,t)$ with a
single maximum at $p=0$) becomes for $\omega_{{\rm I}}(t)\ge
0$:\footnote{For
$\omega_{{\rm I}}(t)<0$, in the last term of the left hand side of Eq.
(\ref{disperhmfsmall})
$\omega_{{\rm I}}(t)$ must be substituted by $|\omega_{{\rm I}}(t)|$. This is
consistent with the fact that
the expression in Eq. (\ref{disperhmfeven_b}) is even in $\omega_{{\rm I}}(t)$
(but not differentiable
at $\omega_{{\rm I}}(t)=0$). Equation (\ref{disperhmfsmall}) is obtained from a
straightforward adaptation to the HMF model of the method developed in Sec. 3.7
of \cite{nyquistaa} for self-gravitating systems. Another, more
direct, proof of Eq.
(\ref{disperhmfsmall})
is given in Appendix \ref{proofomegasmall}.}
\begin{equation}\label{disperhmfsmall}
1 + \pi \int \dd p \, \frac{f_0'(p,t)}{p} - \pi^2 \omega_{{\rm I}}(t)
f_0''(0,t) = 0 \,\, .
\end{equation}
Specializing to the case of the Gaussian (\ref{fzerogauss}), Eq.
(\ref{disperhmfsmall}) becomes
\begin{equation}\label{disperhmfgausssmall}
1 - \frac{1}{2}\beta + \sqrt{\frac{\pi\beta^3}{8}} \omega_{{\rm I}}(t) = 0
\,\, .
\end{equation}
This relation can also be obtained by a power expansion of Eq.
(\ref{disperhmfgauss_b}). From Eq. (\ref{disperhmfsmall}) we obtain
\begin{equation}\label{disperhmfsmallsol}
\omega_{{\rm I}}(t) = \frac{1}{\pi^2 f_0''(0,t)} \left[ 1 + \pi
\int \dd p \, \frac{f_0'(p,t)}{p} \right] \,\, .
\end{equation}
We remind that the term in square brackets is exactly the expression that one
obtains from the Nyquist criterion for the Vlasov stability of a normalized
distribution function $f_0(p)$ which is even in $p$ and with a single maximum at
$p=0$ \cite{nyquisthmf}. Such a function is stable if, and only if, that
expression
is positive:
\begin{equation}\label{nyc}
f_0(p) \qquad {\rm stable} \qquad \Leftrightarrow \qquad 1 + \pi
\int \dd p \, \frac{f_0'(p,t)}{p}>0.
\end{equation}
Then, we expect to have a positive $\omega_{{\rm I}}(t)$ if that expression
is negative. Since $f_0''(0,t) <0$ we therefore have a consistent expression.

{\it Remark: } We note that the stability criterion (\ref{nyc})
can be rewritten as
\begin{equation}\label{stabeff}
T_{\rm eff}\equiv \frac{-1}{2\pi
\int \dd p \, \frac{f_0'(p,t)}{p}}>\frac{1}{2},
\end{equation}
where $T_{\rm eff}$ is an effective temperature (it coincides with the ordinary
temperature when $f_0(p)$ is the Gaussian distribution). This formula shows that
stability is achieved when the width of the distribution function is
sufficiently large, except that the width is not measured by the kinetic
temperature (\ref{tkin}) but by the effective temperature (\ref{stabeff}).

\section{Results}
\label{sec_results}

The HMF model exhibits a second order phase transition between
a homogeneous unmagnetized phase and a magnetized phase at the critical
temperature $T_c = 1/2$,
corresponding to the
critical energy $\epsilon_c = 3/4= 0.75$ (with $\epsilon$ we denote the energy
per particle) \cite{physrep}. The homogeneous phase is thermodynamically stable
if, and only if, $T\ge T_c$. The critical temperature $T_c = 1/2$ also coincides
with the dynamical
stability threshold
of the homogeneous Gaussian distribution
\begin{equation}\label{gausshmf}
f_0(p) = \frac{1}{2\pi} \sqrt{\frac{\beta}{2\pi}} {\rm e}^{-\frac{\beta}{2} p^2}
\,\, .
\end{equation}
Indeed, one finds that in this case
the expression in the right hand side of Eq. (\ref{nyc})
vanishes for $\beta=\beta_c=2$, while it is positive (resp. negative) for $\beta
<2$ (resp. $\beta >2$). Therefore, a Gaussian distribution at temperature
smaller than $1/2$, i.e., at energy smaller than $0.75$,
is Vlasov unstable.

Other forms of homogeneous distribution functions have other
(dynamical) stability thresholds. We will consider
a semi-elliptical distribution, given by
\begin{equation}\label{semiellf}
f_0(p)= \frac{1}{2\pi^2 b} \sqrt{2b - p^2}\, \Theta (2b-p^2) \,\, ,
\end{equation}
where $\Theta(\cdot)$ is the Heaviside step function, and where the parameter $b$ is related to the energy
$\epsilon$ by $b = 4\epsilon -2$. A semi-elliptical distribution is a polytrope
of index $n=1$ \cite{epjb2010,epjb2013}. The stability threshold for this
function is
$\epsilon^*_c = \epsilon_{n=1}= 5/8=0.625$, therefore
we will study the dynamics when $\epsilon$ is somewhat smaller than this value.

We have performed simulations of both the diffusion equation and of the full $N$-body
dynamics of the HMF model at several energies smaller than the critical energy
for both the cases in which the initial homogeneous distribution is a Gaussian and a semi-ellipse.
The analysis and comparison of the results has been as follows.
Starting from a Vlasov unstable state, the $N$-body system has a violent relaxation that ends
at a QSS, a Vlasov stationary and stable state. Afterwards,
the system evolves slowly,
driven by finite-size effects, a phase of the dynamics in which we are not interested in this work. The
QSS is characterized by a single particle distribution
function that in principle will
not be homogeneous, depending on both $\theta$ and $p$. On the other hand, the diffusion equation concerns
the evolution of the angle-averaged distribution function $f_0(p,t)$, and this
evolution stops when $f_0(p,t)$
reaches a Vlasov stable state for which $\omega_{{\rm I}} =0$. Therefore, we
have compared the angle-averaged
distribution function of the $N$-body QSS with the function
$f_0(p)$ obtained when the
evolution of the diffusion equation reaches an end. To characterize this comparison with a single parameter,
we have considered the temperature associated to these distributions, i.e., the expected value of $p^2/2$
(since we consider only distributions that are even in $p$, the expected value of $p$ is zero). We note that
for the HMF model the kinetic temperature $T$ and the magnetization $m$ of the
system,
for a given energy, are uniquely
related. In fact we have
\begin{equation}\label{ener_m_temp_rel}
\epsilon = \frac{1}{2}T +\frac{1}{2}\left( 1-m^2\right) \,\, .
\end{equation}
Then, the analysis of the temperature is equivalent to that of the
magnetization, i.e., the order parameter. From Eq. (\ref{ener_m_temp_rel}) we
note that the smallest possible energy for a homogeneous initial condition
($m=0$) is $\epsilon=1/2$. 

\subsection{Integration of the diffusion
equation}

For convenience we rewrite here the diffusion equation
\begin{equation}\label{vlasovaveragediff_bis}
\frac{\partial f_0(p,t)}{\partial t} = \frac{\partial}{\partial p} \left(
D(p,t) \frac{\partial f_0(p,t)}{\partial p} \right)
\,\, ,
\end{equation}
and the diffusion coefficient for the case of the HMF model with an even and single humped distribution function
\begin{equation}\label{defdiffhmf_b_bis}
D(p,t) = \frac{2 \chi(t) \omega_{{\rm I}}(t)}
{p^2 + \omega_{{\rm I}}^2(t)} \,\, .
\end{equation}
Eq. (\ref{disperhmfeven_b}) determines the value of $\omega_{{\rm I}}(t)$, while $\chi(t)$ evolves according
to Eq. (\ref{derchi_hmf}). It follows that the time scale of the dynamics of the
diffusion equation depends on the
chosen initial value $\chi(0)$. We want to be sure that, independently from this value, provided that it is sufficiently
small for the QL approximation to be valid, the final distribution is
practically the same. We provide
details about this by showing the diffusion equation results for the initial Gaussian distribution.

\begin{figure}[htpb]
\begin{center}
\includegraphics[width=0.45\linewidth]{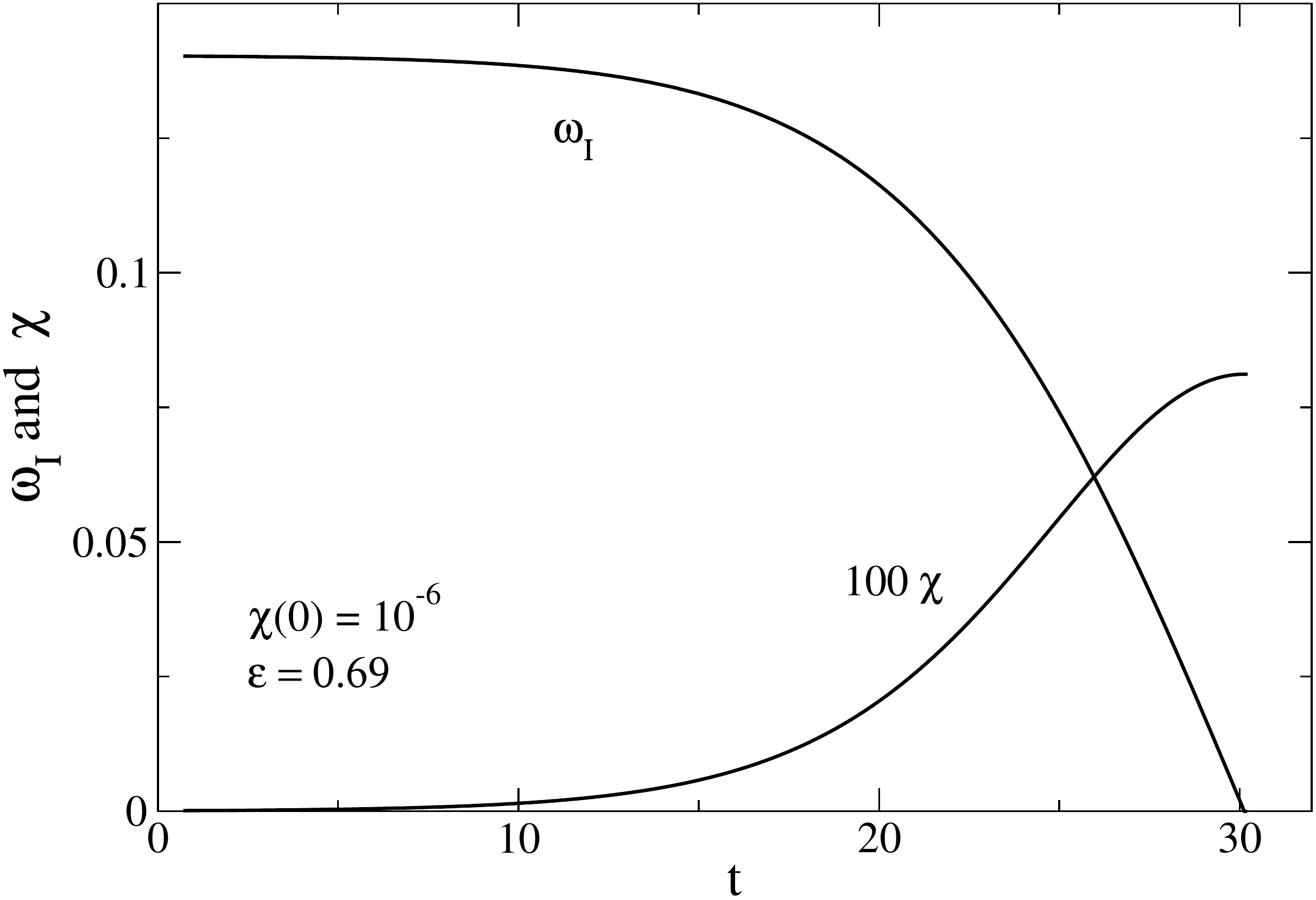}
\includegraphics[width=0.45\linewidth]{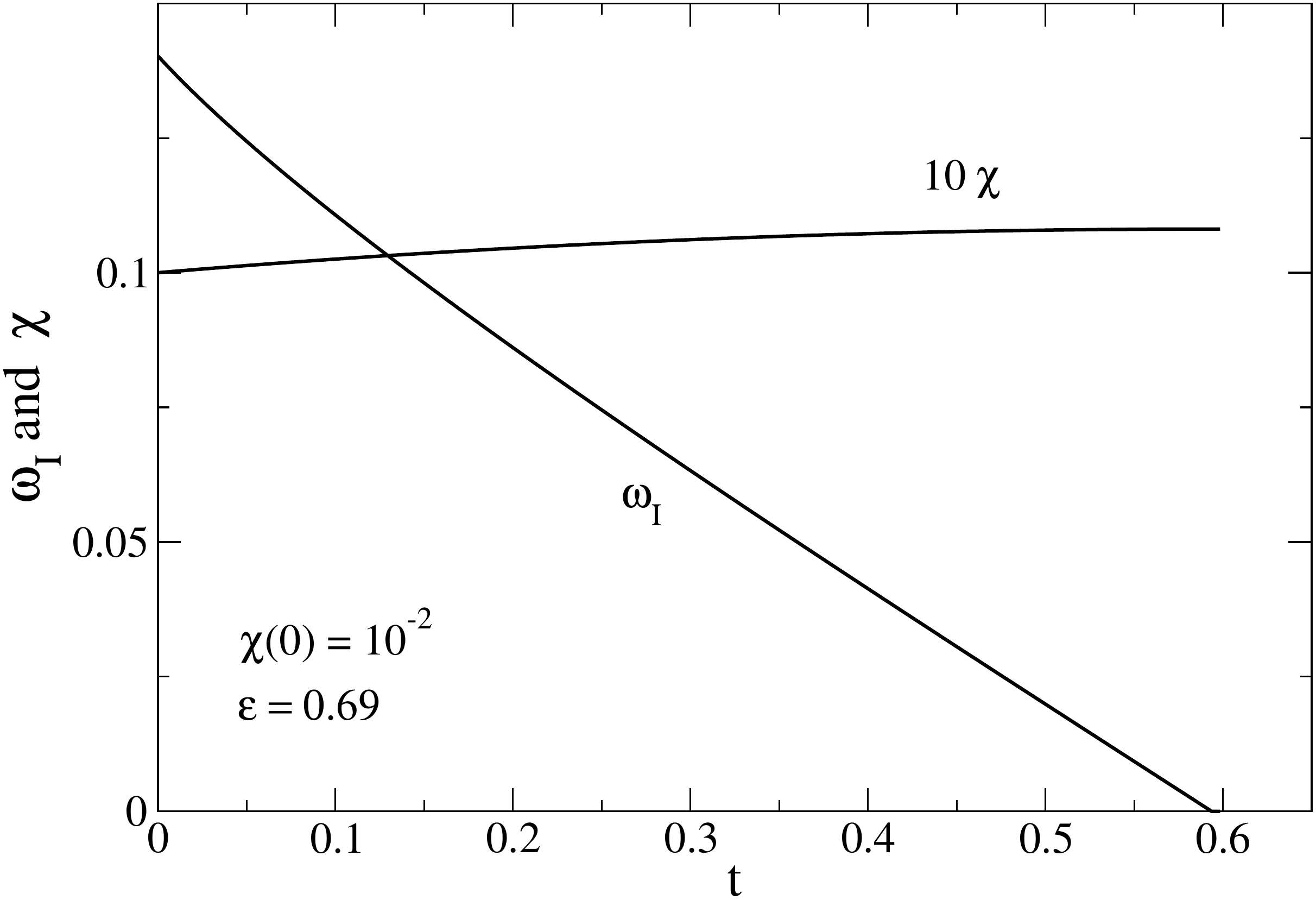}
\end{center}
\caption
{\label{omegachi} Evolution of $\omega_{{\rm I}}(t)$ and of $\chi(t)$ as
determined by the diffusion equation
(\ref{vlasovaveragediff_bis}) together with Eqs. (\ref{defdiffhmf_b_bis}),
(\ref{disperhmfeven_b}) and (\ref{derchi_hmf}),
for two different initial values $\chi(0)$. The initial distribution function is
the Gaussian (\ref{gausshmf}) at $T=1/\beta=0.38$.
Left panel: $\chi(0)=10^{-6}$; right panel: $\chi(0)=10^{-2}$. For graphical
convenience in both cases a multiple of $\chi(t)$ is
plotted.}
\end{figure}

In Fig. \ref{omegachi}
we show the evolution of $\omega_{{\rm I}}(t)$
and of $\chi(t)$, according to the diffusion equation, starting from an initial
$f_0(p,0)$ given by the Gaussian (\ref{gausshmf})
at temperature $T=1/\beta=0.38$, corresponding, for $m=0$, to
$\epsilon=0.69<\epsilon_c$. The left and right panels refer to the integration
with initial value $\chi(0)=10^{-6}$ and $\chi(0)=10^{-2}$, respectively.

\begin{figure}[htpb]
\begin{center}
\includegraphics[width=0.45\linewidth]{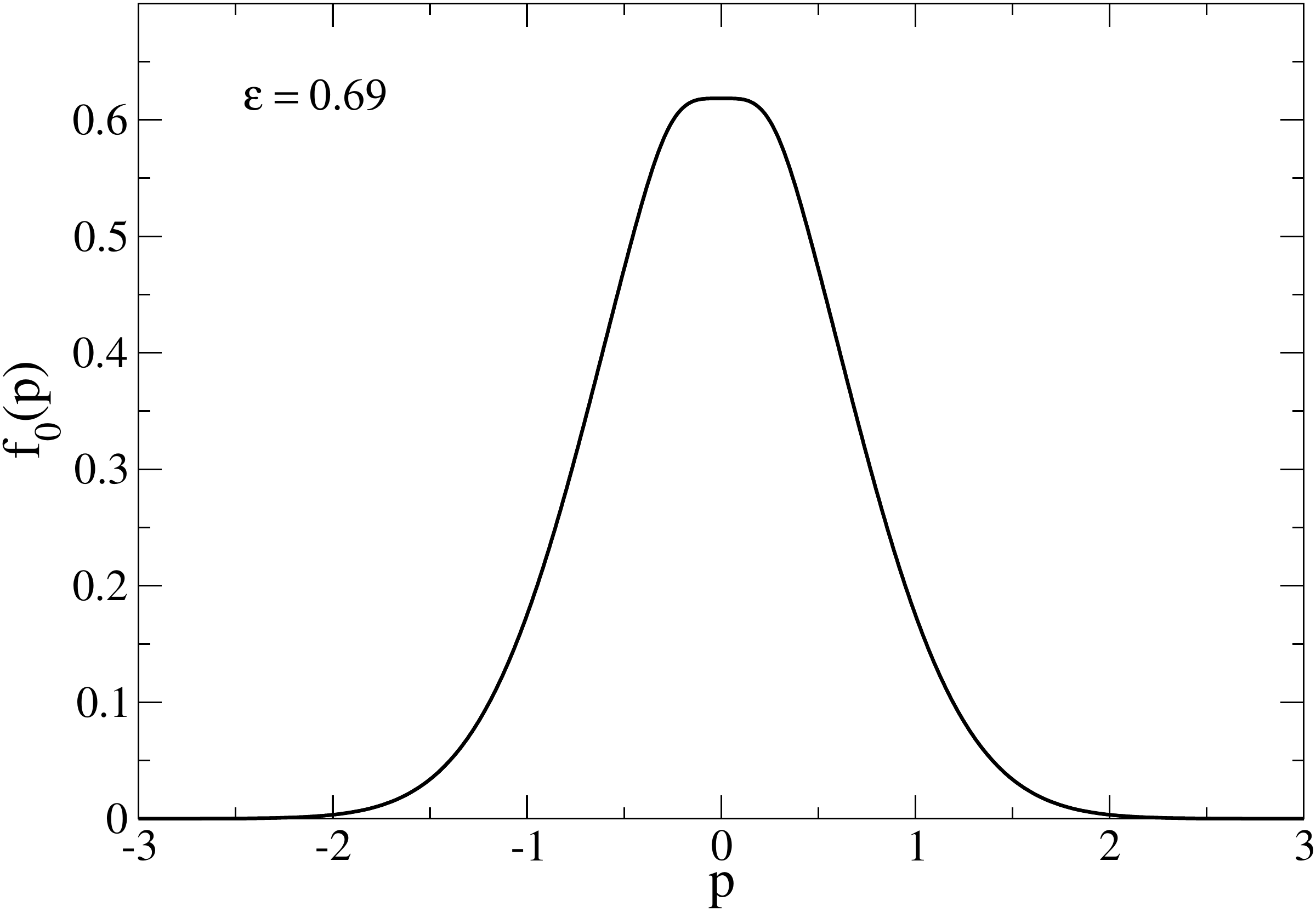}
 \includegraphics[width=0.45\linewidth]{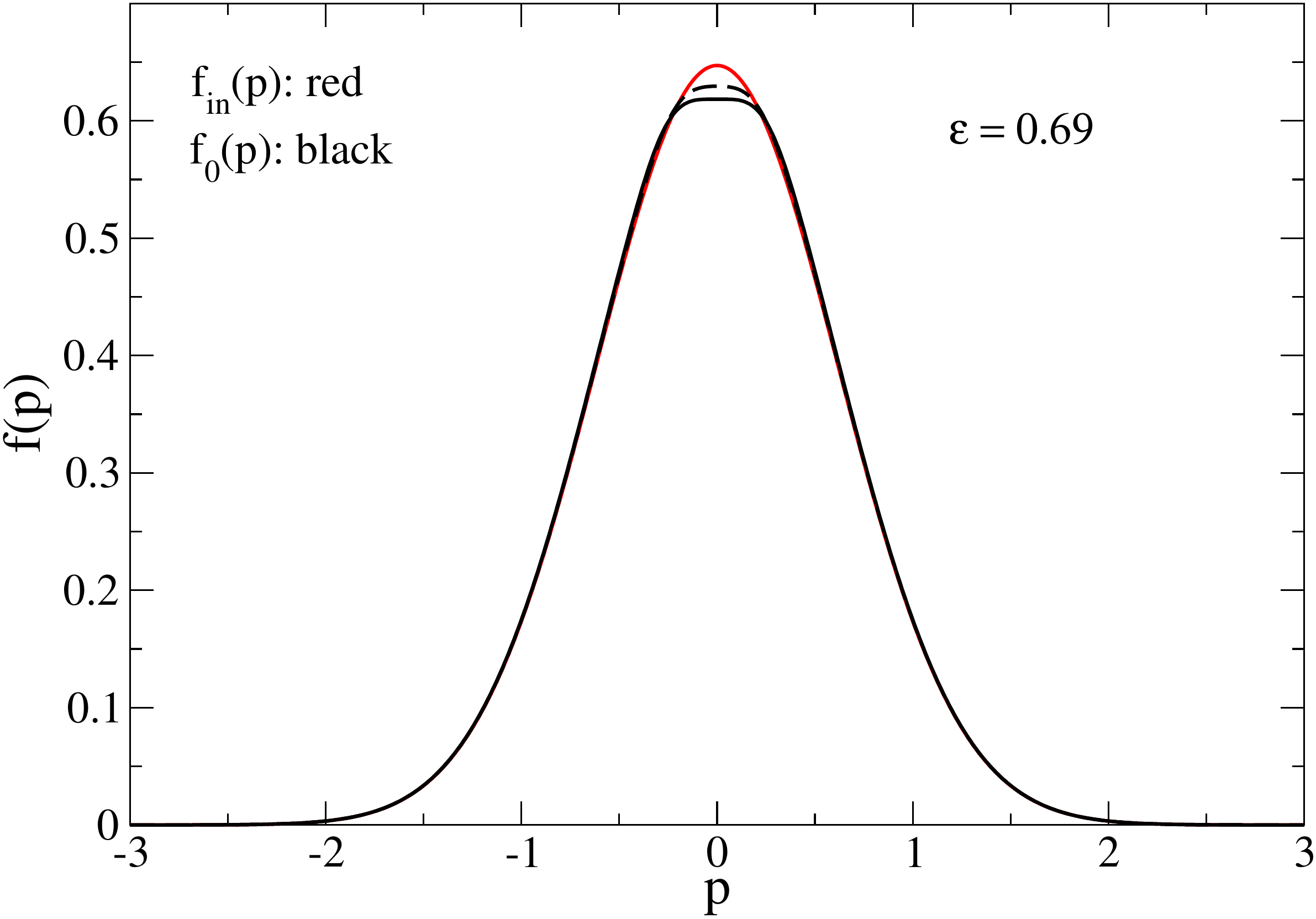}
\end{center}
\caption
{\label{distrfin_ql} Left panel: the final distribution functions $f_0(p)$,
obtained when $\omega_{{\rm I}}(t)$ vanishes, for the
two cases plotted in Fig. \ref{omegachi}. They are indistinguishable. Right
panel: the initial and the final distributions. We have also added an
intermediate distribution to illustrate the progressive formation of the
plateau. The
distribution function is normalized such that $\int f(\theta,p)\, d\theta dp=1$
and $2\pi\int f(p)\, dp=1$.}
\end{figure}

We see that the evolutions are qualitatively different. In particular,
the convergence towards a stationary solution $(\omega_I=0$) is more
rapid when $\chi(0)$ is large. However, the distribution $f_0(p)$ obtained at
the
end of the evolution does not depend on $\chi(0)$. In the left panel of Fig. \ref{distrfin_ql} we plot this distribution
for both cases. It is evident that the two final distributions actually plotted
are virtually indistinguishable.\footnote{We note that the final values of
$\chi$ are different in the two cases (by a factor $10$) even if this
difference has been reduced with respect to the initial condition.} In the
right
panel we plot the initial and the final distributions. This figure suggests
that the form of the final distribution is not very
different from a Gaussian,
except in the central
region, where a sort of flat region seems to have developed. This can be
understood as follows. In the core of the distribution $p\ll 1$, the diffusion
coefficient can be approximated by $D(p,t)\simeq 2\chi(t)/\omega_I(t)$. Since
$\omega_I(t)$ decreases to $0$, the diffusion coefficient in the core becomes
very large and this implies $\partial f_0(p,t)/\partial p\simeq 0$, so that the
distribution function in the core forms a plateau. On the other hand, in the
tail of the  distribution $p\gg 1$,
the diffusion coefficient can be approximated by $D(p,t)\simeq
2\chi(t)\omega_I(t)/p^2$. Since $\omega_I(t)$ decreases to $0$, and since $p\gg
1$, the diffusion
coefficient in the tail is very small and this implies $\partial
f_0(p,t)/\partial t\simeq 0$, so that the distribution function in the tail
does not evolve substantially. As a matter of fact, under the
assumption that
the final distribution
has a Gaussian structure apart from a central region, which is flat, it is possible to obtain exactly its
parameters. Such an assumption is equivalent to assuming that the diffusion
equation gradually develops a flat portion in the
central part of the distribution, while maintaining the Gaussian form outside that portion. The possibility to derive exactly
this distribution stems from the fact that there are as many equations to satisfy as parameters characterizing the distribution.
In fact, the assumed form can be written as
\begin{eqnarray}\label{flatgauss}
f_{\rm mG}(p) &=& C \quad\quad\quad\quad\quad\quad |p| \le p_1, \nonumber \\
&=& A {\rm e}^{-\beta \frac{p^2}{2}} \quad\quad\quad\quad |p| \ge p_1 \, ,
\end{eqnarray}
where the subscript denotes a ``modified Gaussian'', and where $p_1 \ge 0$.
Continuity implies that $C = A {\rm e}^{-\beta \frac{p_1^2}{2}}$. Therefore there
are three parameters to be determined, i.e., $A$, $\beta$ and $p_1$. On the other hand, there are also three equations to
satisfy. They are the normalization, the marginal stability of the distribution
and the relation expressing the average kinetic
energy:
\begin{eqnarray}\label{eqforparameters_1}
2\pi \int \dd p \, f_{\rm mG}(p) &=& 1, \\ 
\label{eqforparameters_2}
1 +\pi \int \dd p \, \frac{f_{\rm mG}'(p)}{p} &=& 0, \\
\label{eqforparameters_3}
2\pi \int \dd p \, \frac{p^2}{2} f_{\rm mG}(p) &=& \epsilon_{\rm kin} \, ,
\end{eqnarray}
where $\epsilon_{\rm kin}$ is the final average kinetic energy, determined
numerically from the distribution at the end of the
integration.  In Appendix \ref{appmodgau} we
show how to solve this system of equations for $A$, $\beta$ and $p_1$.

\begin{figure}[htpb]
\begin{center}
\includegraphics[width=0.45\linewidth]{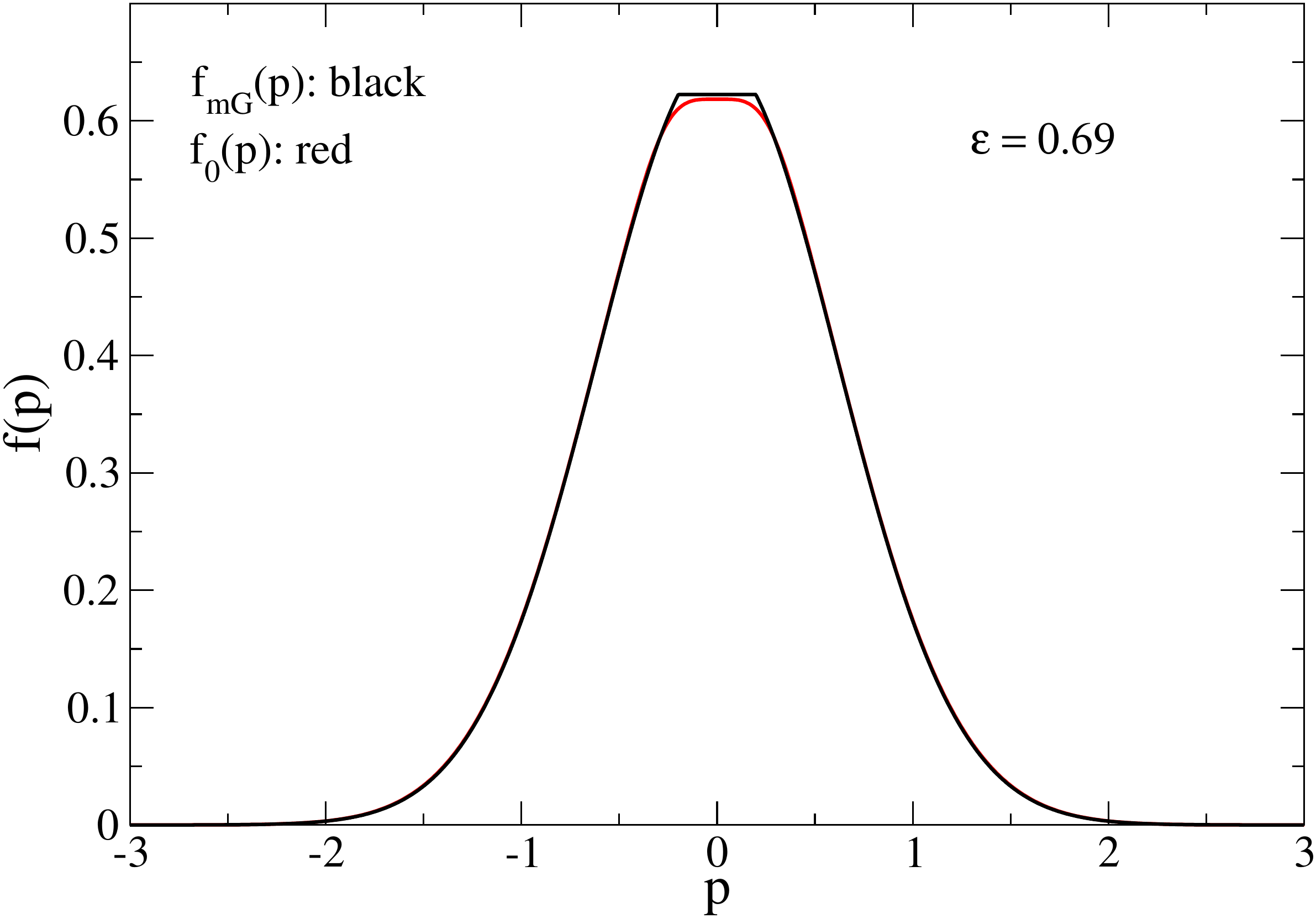}
\includegraphics[width=0.45\linewidth]{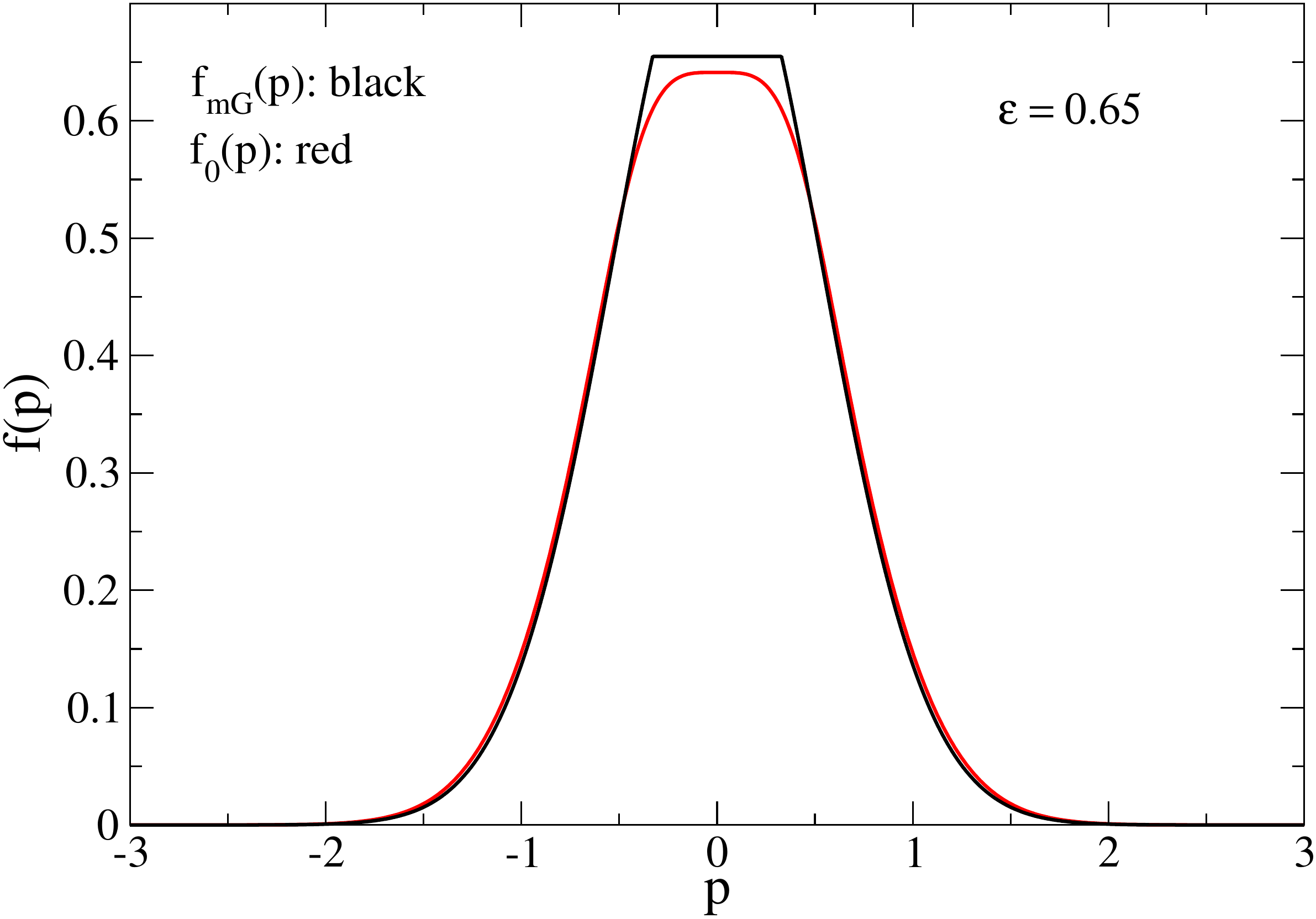}
\end{center}
\caption
{\label{fql_fplat} The final distribution function $f_0(p)$ and the modified
Gaussian $f_{\rm mG}(p)$ for
energies $\epsilon = 0.69$ (left panel) and $\epsilon = 0.65$ (right panel). The
function $f_{\rm mG}(p)$ is computed for
the same kinetic energy $\epsilon_{\rm kin}$ resulting from $f_0(p)$.}
\end{figure}

In Fig. \ref{fql_fplat} we show the comparison, for the energies $\epsilon = 0.69$ and $\epsilon = 0.65$, between the final
distribution of the diffusion equation and the approximation with a modified Gaussian. It is evident that for the larger
energy the approximation with a modified Gaussian is rather good, while for the smaller energy there is a clear difference
in the region of the plateau.

\subsection{The $N$-body simulations and the comparison with the results of the
diffusion equation}
\label{sec_fc}

In this section, we compare the results of the QL theory with the results of
$N$-body simulations. We begin the analysis by showing the evolution of the
magnetization for both the diffusion equation and the
$N$-body simulations. These simulations have been performed with $N=2^{18}$ particles. As remarked above, for the diffusion
equation the magnetization can be computed using
Eq. (\ref{ener_m_temp_rel}), obtaining the kinetic temperature at time $t$ from
the variance of $f_0(p,t)$ using Eq. (\ref{tkin}).

Let us first comment on the comparison between the time scales in the two dynamical evolutions. As
explained above and as
evidenced in Fig. \ref{omegachi}, the time scale of the dynamics of the diffusion equation depends strongly on the
initial value $\chi(0)$. This is of course reflected  in the temporal
evolution of the magnetization.
%in Fig. \ref{mql1} we plot the dynamics
%of the magnetization as obtained from the diffusion equation for energy $\epsilon = 0.69$, for the two cases
%of Fig. \ref{omegachi}, i.e. $\chi(0)=10^{-6}$ and $\chi(0)=10^{-2}$.
%We see that the final magnetization is the same
%in both cases, as expected from the fact that the final distributions are the same, although the time scale
%is different, as shown in the two panels of Fig. \ref{omegachi}.

\begin{figure}[htpb]
\begin{center}
\includegraphics[width=0.45\linewidth]{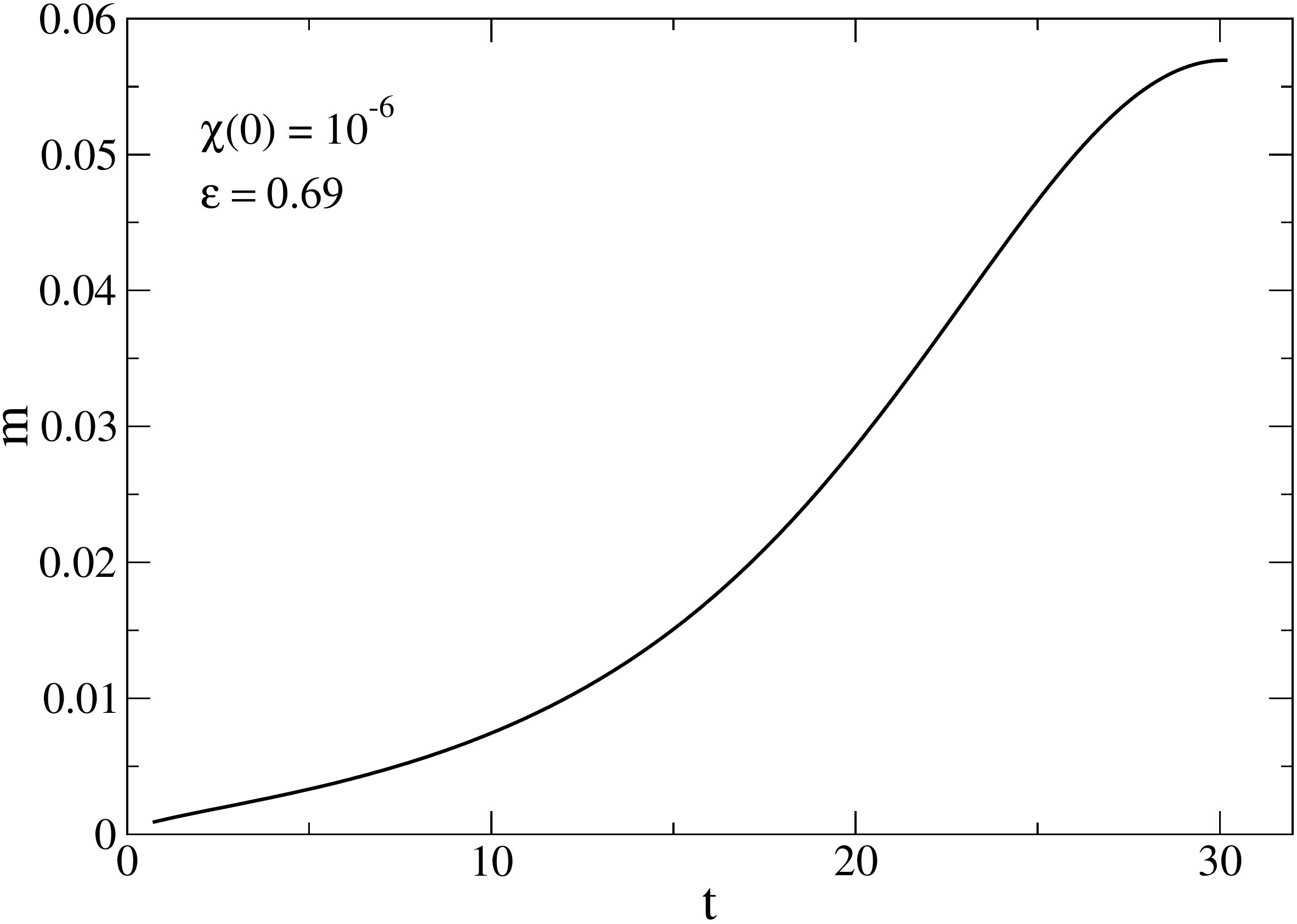}
\includegraphics[width=0.45\linewidth]{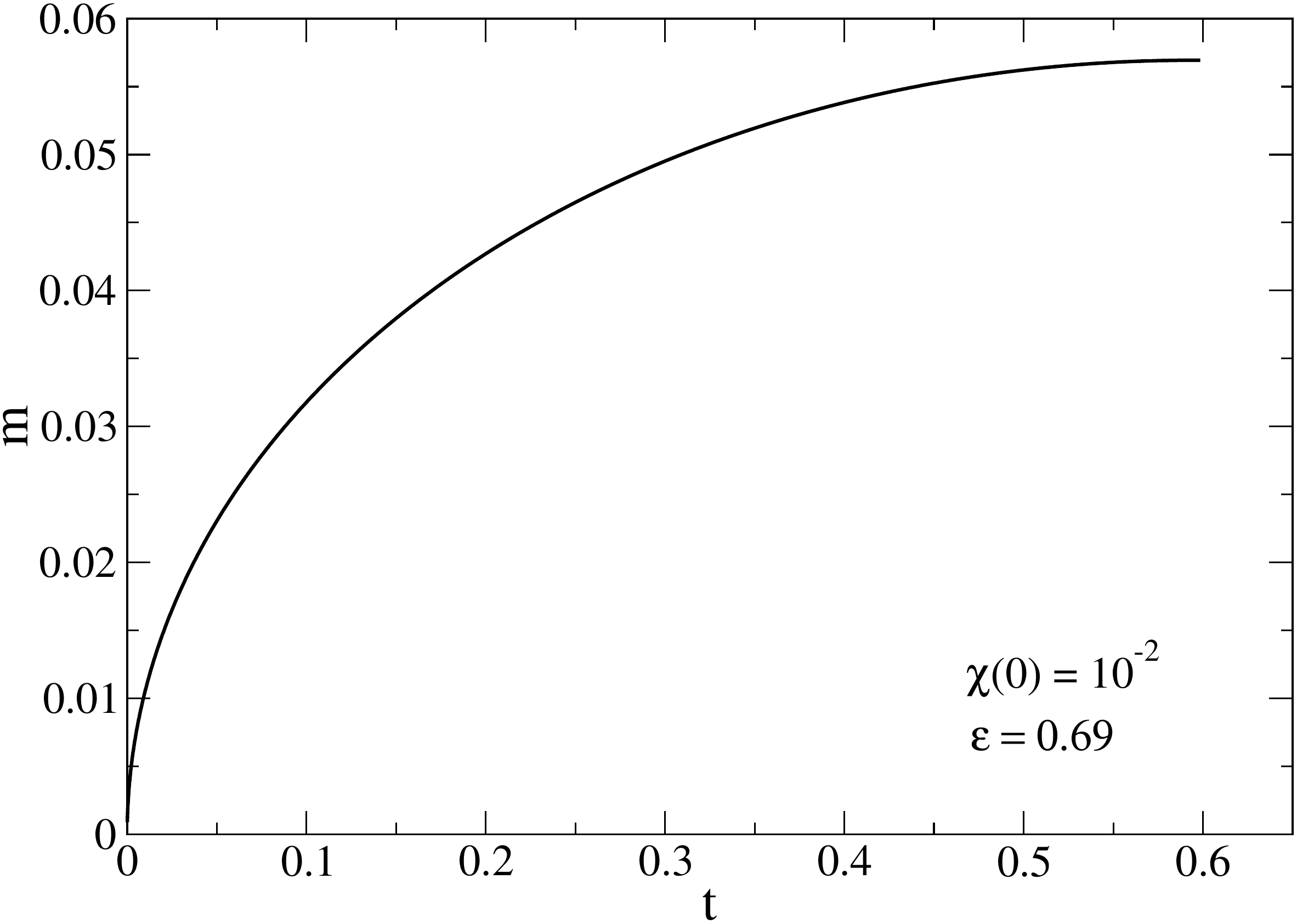}
\end{center}
\caption
{\label{mql1} Magnetization as a function of time from the diffusion equation at
energy $\epsilon=0.69$, obtained from the
distribution function $f_0(p,t)$ using Eqs. (\ref{tkin})
and (\ref{ener_m_temp_rel}). Left panel:
$\chi(0)=10^{-6}$. Right panel:
$\chi(0)=10^{-2}$. }
\end{figure}

\begin{figure}[htpb]
\begin{center}
\includegraphics[width=0.45\linewidth]{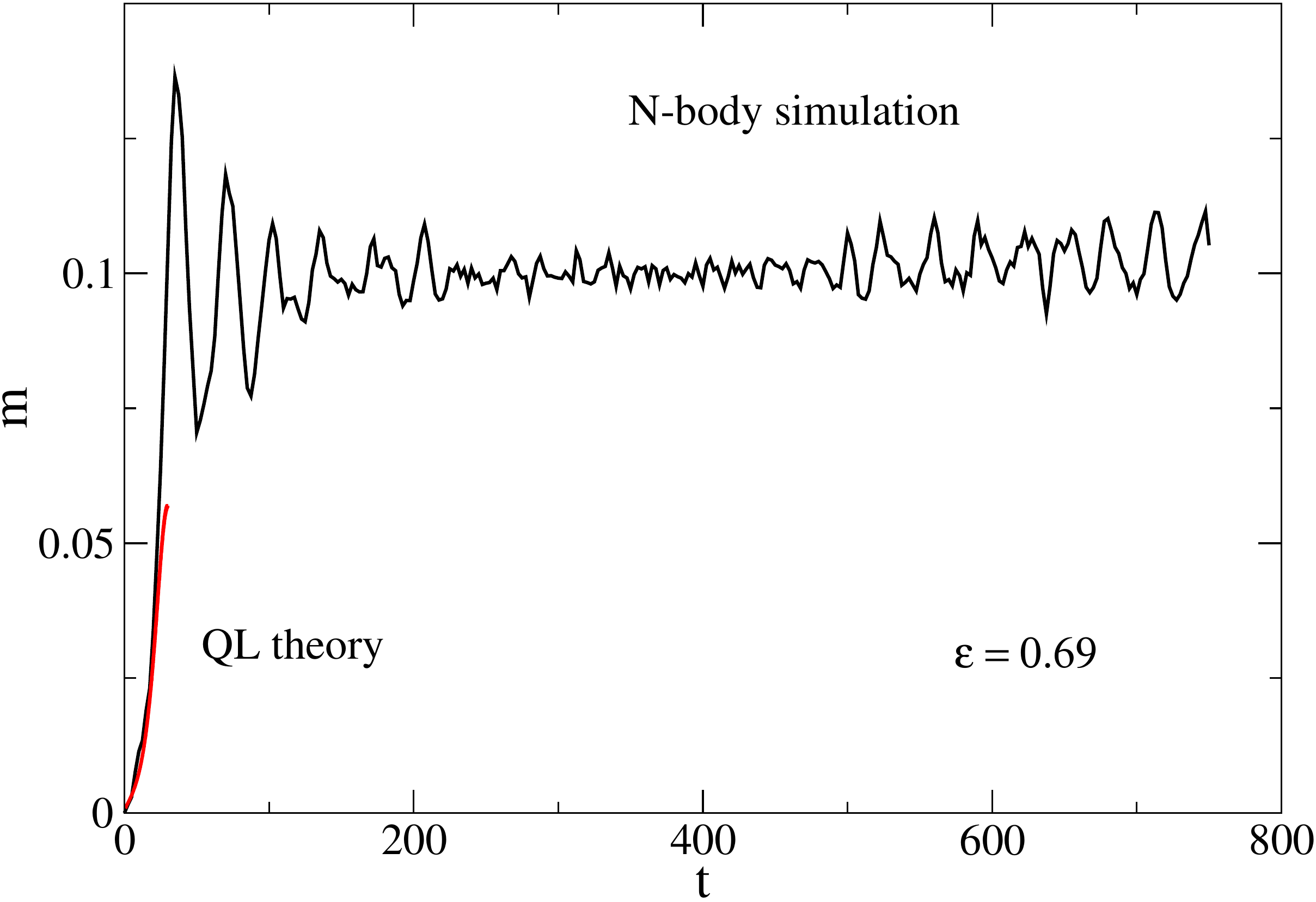}
\includegraphics[width=0.45\linewidth]{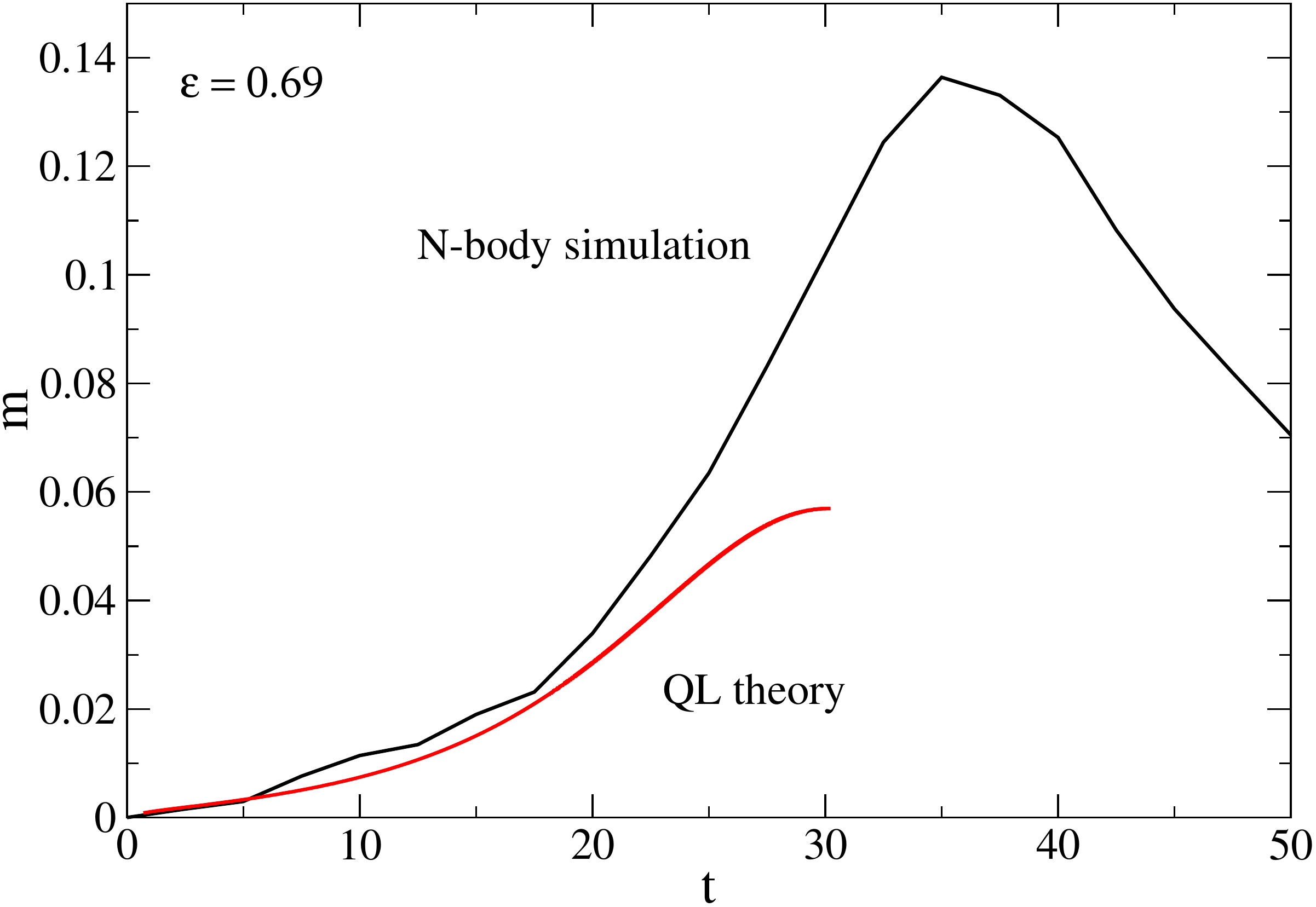}
\end{center}
\caption
{\label{mql1_mnbody} Magnetization as a function of time at energy
$\epsilon=0.69$ for a homogeneous Gaussian
velocity initial distribution. Both the $N$-body simulation with $N=2^{18}$
particles (black)  and the diffusion
equation (red) curves are shown (we have taken $\chi(0)=10^{-6}$). Left panel:
plot of the magnetization during
the whole duration of the simulation.  Right
panel:  plot of the magnetization during the early times of the simulation. In
the diffusion equation, the magnetization remains
constant at times larger than those
plotted since the evolution stops when $\omega_{{\rm I}}(t)$ reaches the value
$0$.}
\end{figure}

In Fig. \ref{mql1}, we plot the dynamics
of the magnetization as obtained from the diffusion equation for the Gaussian initial condition and energy
$\epsilon = 0.69$ for the two cases of Fig. \ref{omegachi}, i.e.,
$\chi(0)=10^{-6}$ and $\chi(0)=10^{-2}$.
We see that the final magnetization is the same
in both cases, as expected from the fact that the final distributions are the same, although the time scale
is different. 
In the diffusion equation, $\chi(0)$ is related to the initial perturbation to
the stationary but Vlasov unstable
distribution $f_0(p,0)$. It therefore depends on the order of magnitude of
$f_1(\theta,p,0)$ in Eq. (\ref{f1define}).
In the $N$-body simulation, this role is played by the finite size effects in the initial conditions determined according
to $f_0(p,0)$. From the definition of $\chi(t)$, given implicitly by the comparison of Eqs. (\ref{phif1intapprox_time}),
(\ref{defdiff}) and (\ref{derchi}), it is clear that the order of magnitude of
$\chi(0)$ is that of $f_1^2(\theta,p,0)$.
In turn, in the $N$-body simulations we expect that the order of magnitude of
$f_1(\theta,p,0)$, due to finite size effects,
is $N^{-\frac{1}{2}}$. Therefore, we expect that the simulations performed with $N=2^{18}=0.262144\times 10^6$ particles
should evolve, as far as the relaxation to the stationary state is concerned, on a time scale similar to that of the diffusion
equation with a value of $\chi(0)$ of the order $10^{-6}$. In Fig. \ref{mql1_mnbody} we show that this is indeed the case.
We stress, however, that the $N$ dependence of this time scale which is due to
the dynamical instability of the initial state is expected to behave like $\log
N$, contrary to the power law dependence occurring for the slow evolution in the
QSS due to collisional effects \cite{physrep,ccgm}.

The analysis of Fig. \ref{mql1_mnbody} puts in evidence several important
things. Firstly, there is no perfect agreement on the final
value of the magnetization, which is about $0.057$ for the diffusion equation
and about $0.10$ for the $N$-body simulation. However, these magnetization
values are comparable, and they are substantially  different from that of
the BG equilibrium state corresponding to $\epsilon =0.69$
which is about $0.31$. On the other hand, the left panel shows that in the
state reached by the $N$-body system at the end of the fast
relaxation the magnetization has relevant oscillations. Therefore the system is
not in a proper QSS, although one
could extend the definition of such a state also to cases that present these
oscillations. The QL theory does not account for these
oscillations because they appear in the regime where the system is
dynamically stable. They may correspond to small perturbations about
a stable steady state of the Vlasov equation. A possibility  is that
these oscillations will decay by Landau damping. Another
possibility considered in \cite{epjb2013} is
that these oscillations will survive during the whole collisionless regime like
in the simulations performed by Morita and Kaneko \cite{mk}.
Apart from the presence of the oscillations, we  conclude that, at the level of
the final magnetization value, there is a reasonable but not a perfect
agreement between the
full $N$-body simulation and the approximation represented by the QL
theory since, in relative terms, the difference is
large. However, we have to consider that for such small values of the magnetization, a very small change in $T$ causes a strong
variation in $m$, since the derivative of $m$ with respect to $T$, that can be computed from Eq. (\ref{ener_m_temp_rel}), diverges
for $T\to 1-2\epsilon$. As a matter of fact, while the temperature of the QSS of
the $N$-body simulation is
about $0.390$, the temperature of the final state of the diffusion equation is
about $0.383$. At this energy, $T$
is about $0.475$ at BG equilibrium. Therefore the relative error in the
temperature value of the QSS is much smaller
than the error in the magnetization.

\begin{figure}[htpb]
\begin{center}
\includegraphics[width=0.45\linewidth]{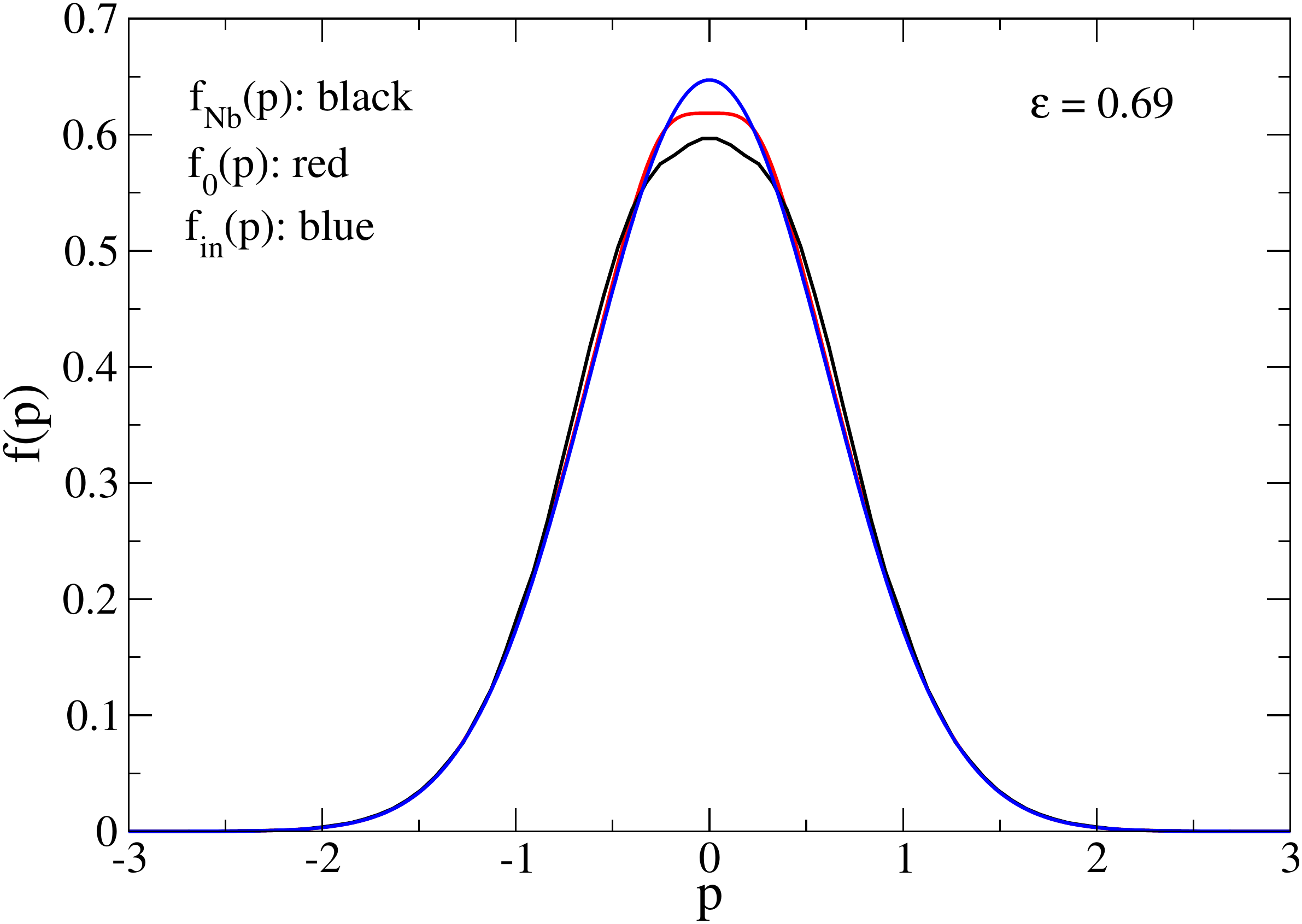}
\includegraphics[width=0.45\linewidth]{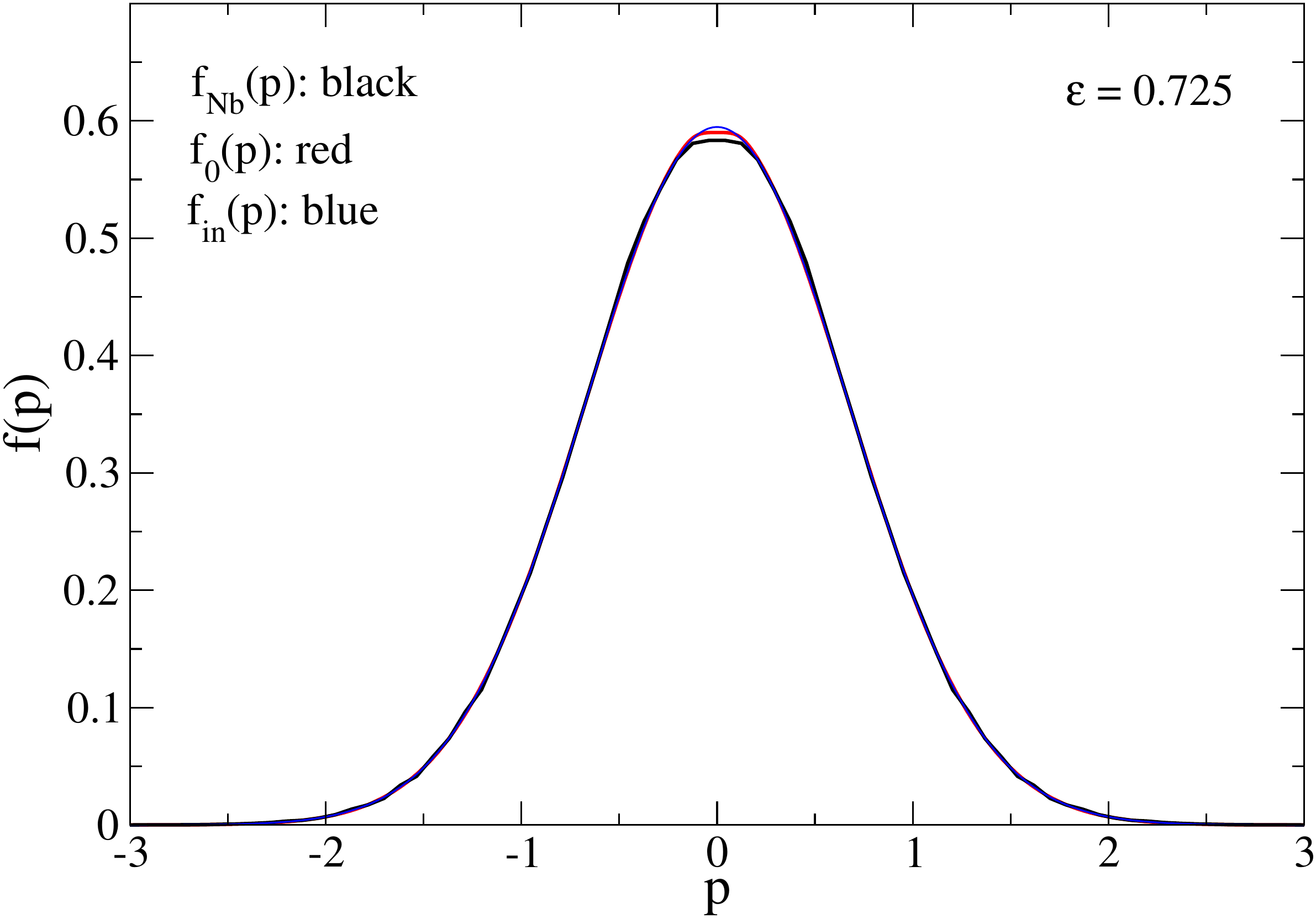}
\end{center}
\caption
{\label{distrfin_ql_nbody} The final distribution function $f_0(p)$ of the
diffusion equation and the velocity distribution
$f_{\rm Nb}(p)$ of the QSS of the $N$-body simulation, starting
from a Gaussian initial distribution. Left panel:
$\epsilon = 0.69$. Right panel: and $\epsilon = 0.725$. We have also
plotted the initial condition $f_{\rm in}(p)$ for comparison. We
emphasize that the
initial
distribution is Vlasov unstable while the final distribution is
(marginally) Vlasov stable
even when the profile does not seem to have changed a lot.}
\end{figure}

Let us now compare the velocity distribution functions of the QSS
reached in the $N$-body simulation
and the final distribution function of the diffusion equation. In Fig. \ref{distrfin_ql_nbody} we show this comparison for the two cases
of $\epsilon =0.69$ and $\epsilon =0.725$ for the initial Gaussian distribution.
We observe a very good agreement  for the energy
$\epsilon=0.725$ which is closer to the instability threshold $\epsilon_c=3/4$.
For the
case $\epsilon =0.69$, there is a very good agreement in the tail of the
distribution  but a disagreement concerns the central region of the
distribution. The
distribution function obtained from the diffusion equation clearly exhibits a
central plateau while there is only the hint of a plateau in the distribution
function obtained from the $N$-body simulation. We also note that the QL theory
predicts that the central distribution function $f(0,t)$ decreases\footnote{This
is
consistent with the notion of ``coarse-graining''. The maximum
value of the coarse-grained
distribution can only decreases by phase mixing \cite{lb}.} with respect
to
its initial value (see Fig. \ref{distrfin_ql}), in qualitative
agreement with
the $N$-body simulation, although the decrease is stronger in the  $N$-body
simulation.

We note that the distribution of the QSS obtained from $N$-body
simulations and the distribution predicted by the QL theory
have a form very similar to the initial distribution. Actually, the distribution
function almost does not change in the tail of the distribution. Nevertheless,
the slight change in the central region (core) is crucial because the initial
distribution is
Vlasov unstable while the QSS and the distribution predicted by the QL theory
are Vlasov stable. Therefore, close to the instability threshold,
stability can be regained by a very slight modification of the distribution
function that affects essentially the core of the distribution (small
velocities) while preserving the tail (large velocities).

\subsection{The nonequilibrium phase transition predicted by the QL
theory}

In Ref. \cite{epjb2013} we studied the evolution of several unstable homogeneous
initial conditions in
situations where the Lynden-Bell theory fails to predict
with good approximation the QSS reached by the system. We found that, in most
of these cases, the QSS could be well approximated by a polytropic
distribution. Furthermore, the index of the polytrope, even if it was
not possible to predict if from first principles, happened to be the
same for a given class of initial conditions (e.g. Gaussian,
semi-elliptical...) for a wide range of energies. However, we remarked that 
the polytropic approximation was not good for energies close to the
instability threshold pertaining to the class of the initial distribution. We
have seen here that, in this case, the QL theory provides a better
approximation. In fact, we have seen that the final distribution computed by
the QL theory is close to the angle-averaged
distribution of the QSS obtained in the $N$-body simulations. Furthermore, the
values of the kinetic temperature and  magnetization predicted by the QL theory
are relatively  close to those
obtained from the direct $N$-body simulations, even if the quality of
the agreement deteriorates if we are too far away from the instability
threshold.

\begin{figure}[htpb]
\begin{center}
\includegraphics[width=0.45\linewidth]{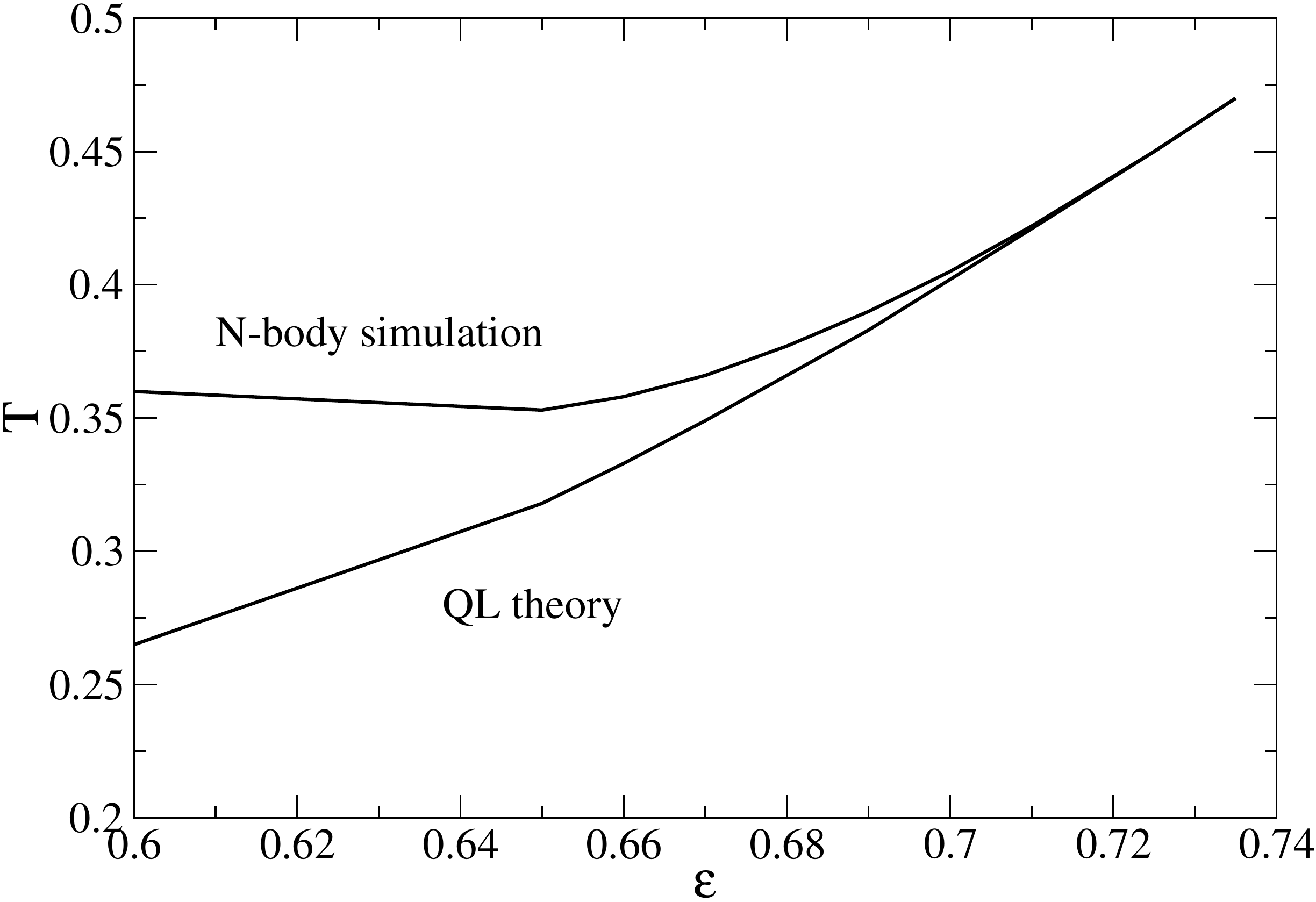}
\includegraphics[width=0.45\linewidth]{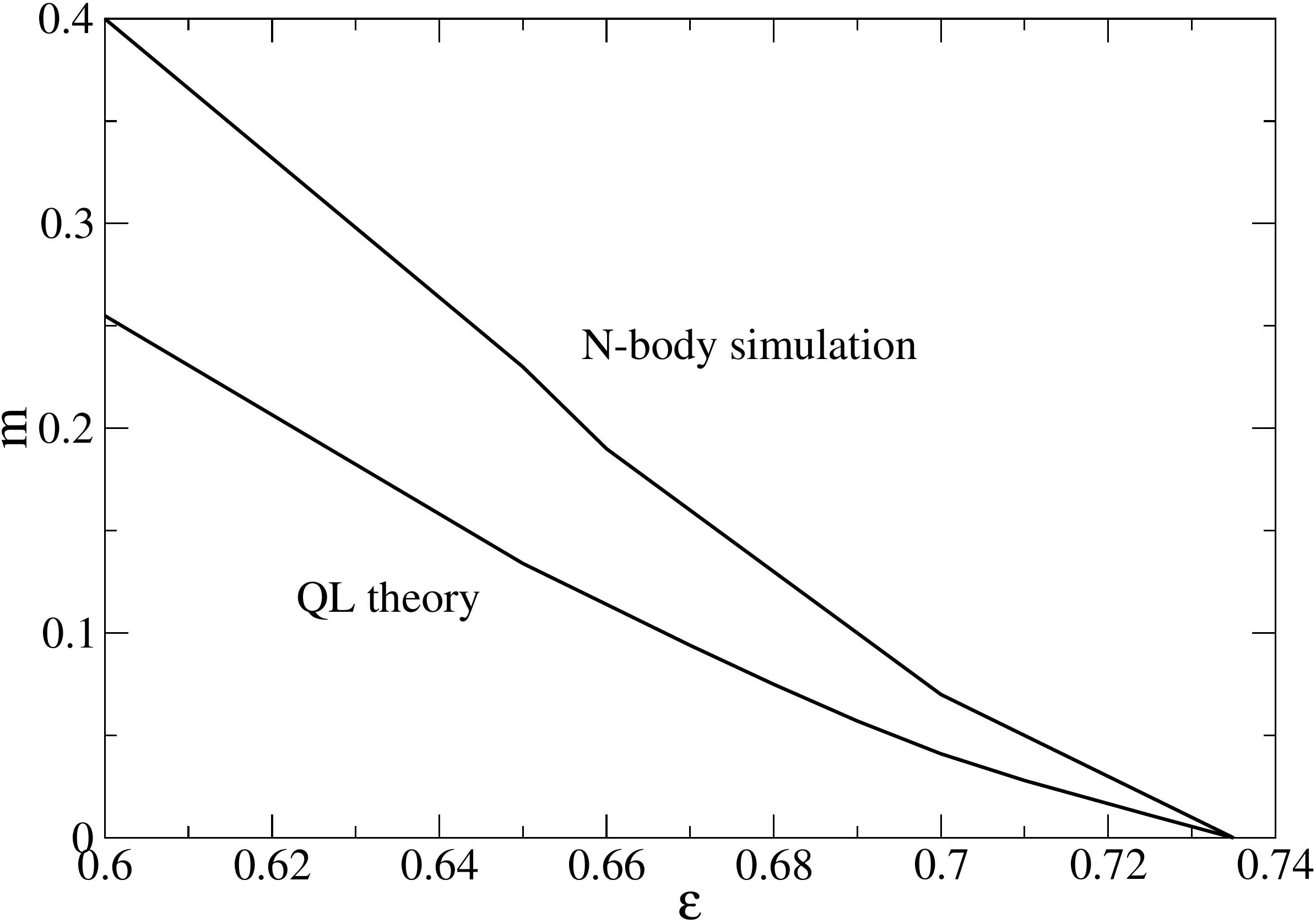}
\end{center}
\caption
{\label{energy-mag} The kinetic temperature and the magnetization of the
stationary state reached by the diffusion equation of the 
QL theory
and that of the QSS reached by the $N$-body simulation, as a function of
the energy. The homogeneous initial distribution is a Gaussian
in the velocity. }
\end{figure}

To give an easily
grasped quantitative meaning to this agreement/disagreement, we plot
in Fig. \ref{energy-mag} the kinetic temperature and the magnetization of
the final distribution of the QL theory as a function of the energy for the
Gaussian
initial distribution. The plot also shows the kinetic temperature  and the
magnetization of the QSS
reached in the $N$-body simulation at the same energy. Although
the
magnetization of the QL theory is 
different from that of the simulation, the plot shows that the energy of the
phase transition
from the unmagnetized state ($m=0$) to the magnetized state
($m\neq 0$) is well predicted by
the QL theory. We stress that this is to be interpreted
as a nonequilibrium phase transition since it refers to the magnetization of
the out-of-equilibrium QSSs. We can conclude that
the QL theory is able to localize this phase transition correctly. A more
detailed
discussion is provided in the following section.

\section{Comparison of the QSS with the prediction of the QL theory and with 
the polytropic fit}
\label{sec_comp}

\subsection{Caloric and magnetization curves}
\label{sec_cm}

In this section, we compare the caloric curve obtained from
direct $N$-body
simulations of the HMF model, starting from a homogeneous Gaussian
distribution, with the prediction of the QL theory and with the 
polytropic
fit considered in our previous papers
\cite{epjb2010,epjb2013}.

\begin{figure}[htpb]
\begin{center}
\includegraphics[width=0.45\linewidth]{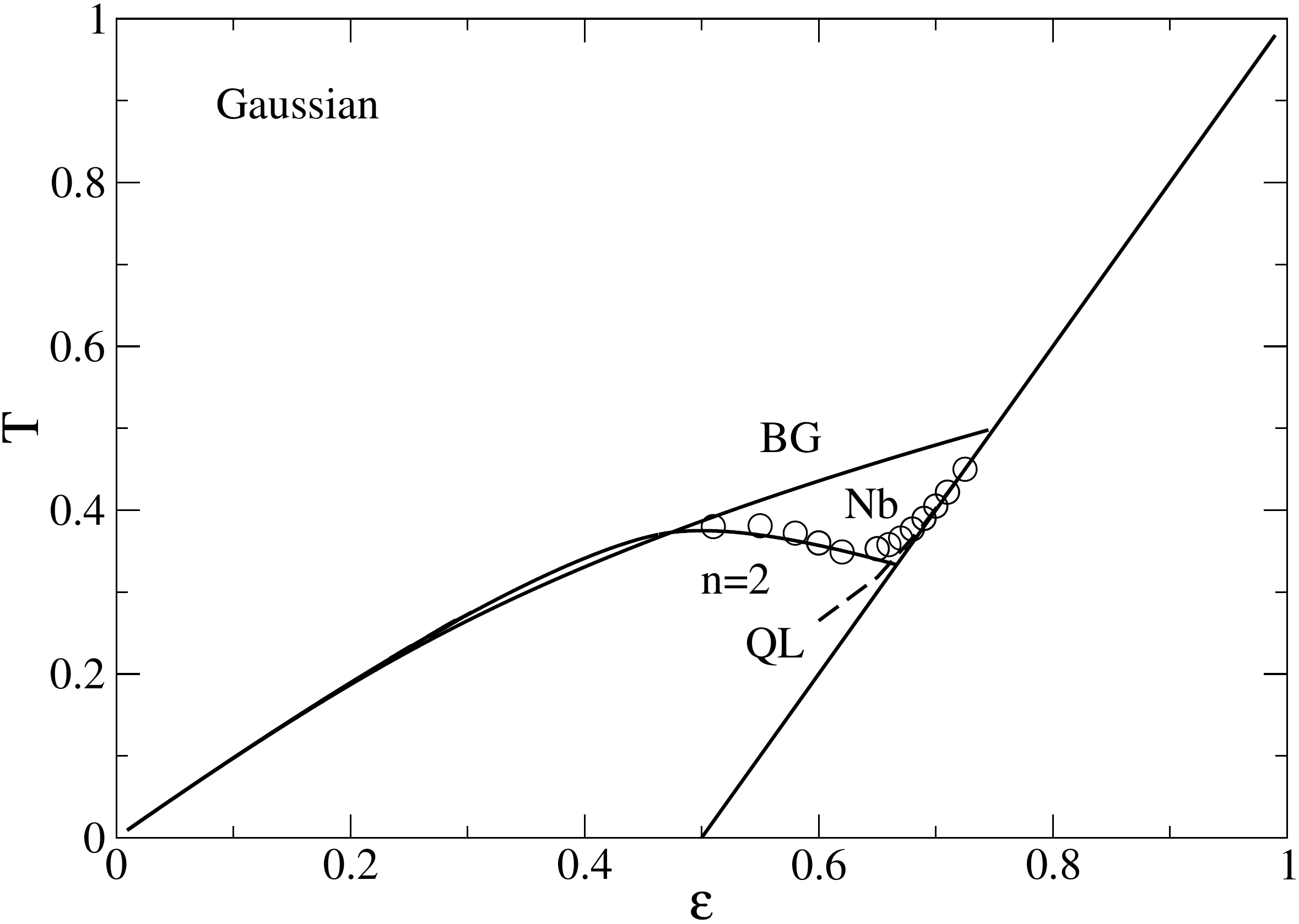}
\includegraphics[width=0.45\linewidth]{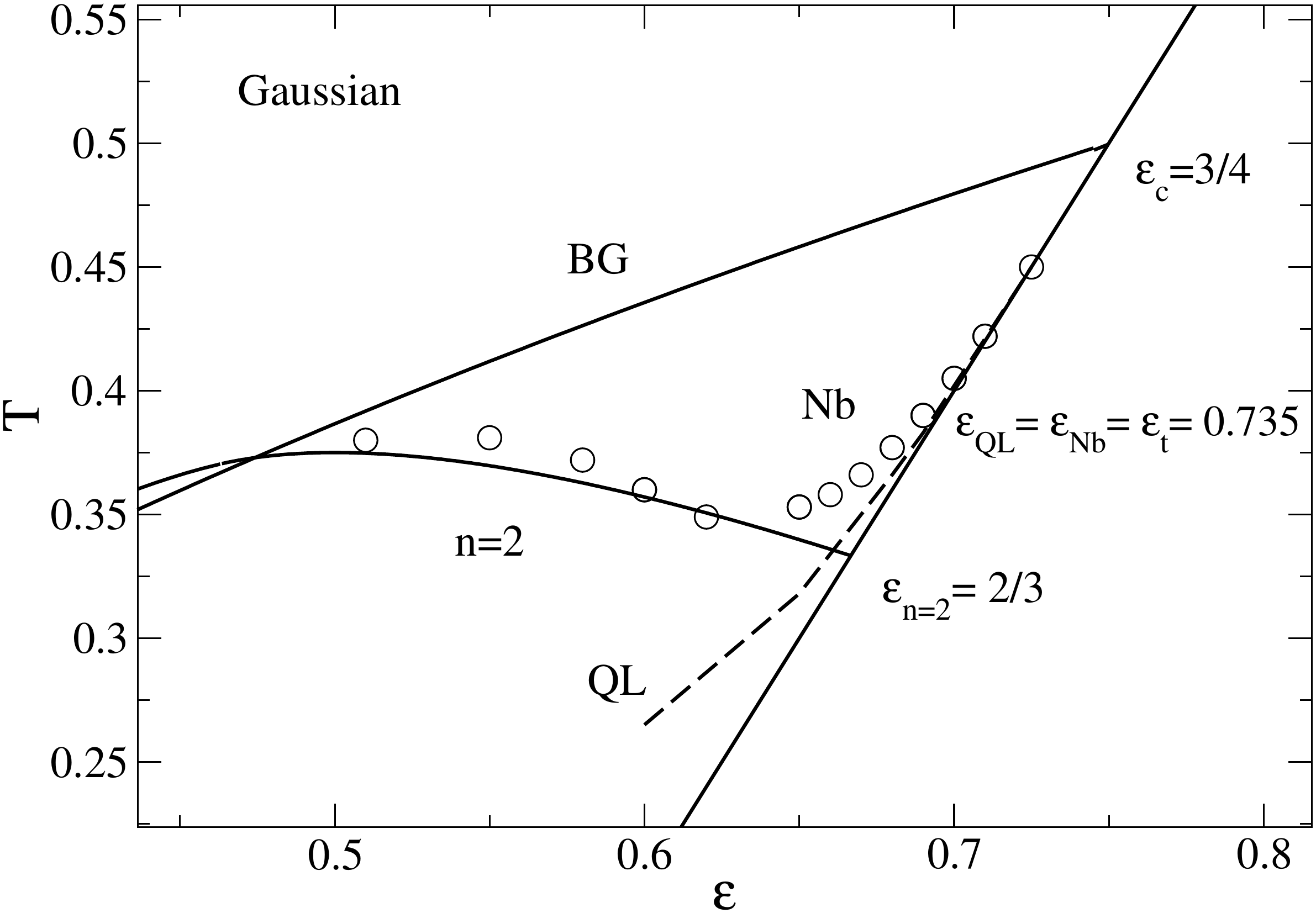}
\end{center}
\caption
{\label{gaussian} Caloric curve of the HMF model for a spatially
homogeneous Gaussian initial condition (BG: Boltzmann-Gibbs states; bullets:
results of $N$-body simulations; dashed line: prediction
of the QL theory; $n=2$: polytropic fit).}
\end{figure}

\begin{figure}[htpb]
\begin{center}
\includegraphics[width=0.45\linewidth]{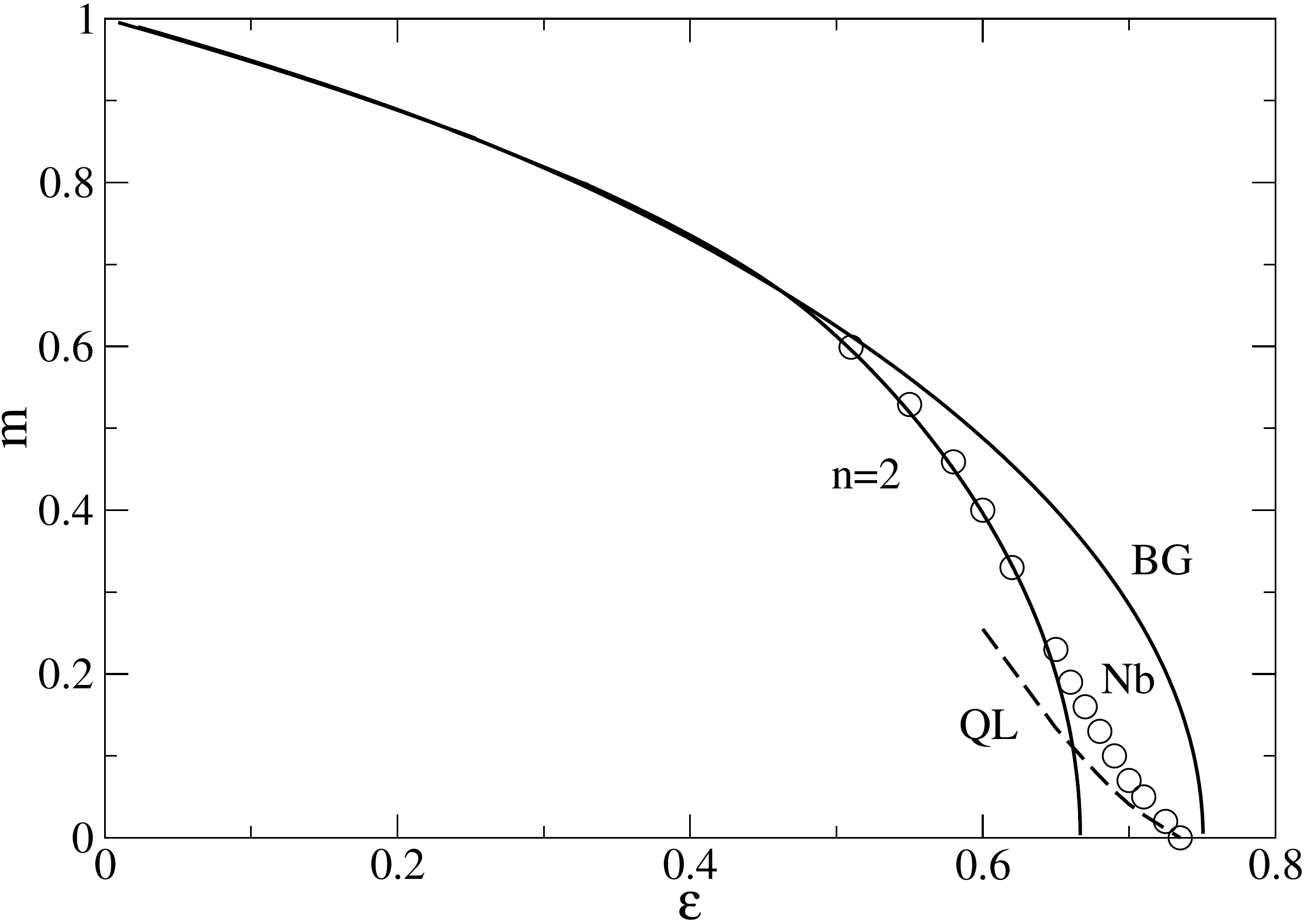}
\includegraphics[width=0.45\linewidth]{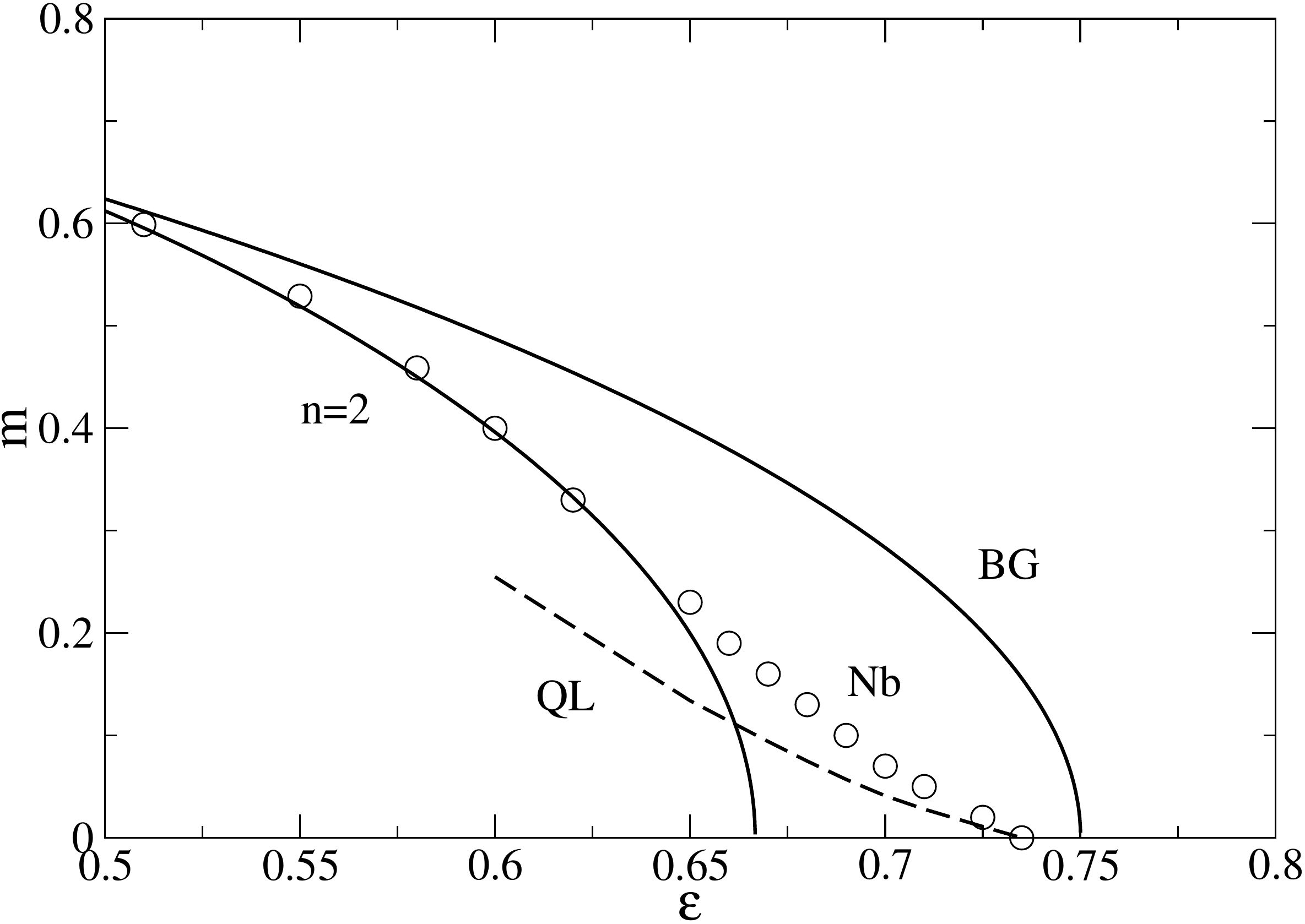}
\end{center}
\caption
{\label{magTOTAL}  Magnetization  curve of the HMF model for a spatially
homogeneous Gaussian initial condition.  }
\end{figure}

The caloric curve $T(\epsilon)$  is represented in Fig.
\ref{gaussian} and the magnetization curve $m(\epsilon)$ is represented in
Fig. \ref{magTOTAL}. We recall that, for a given energy $\epsilon$, the kinetic
temperature $T$ is related to the magnetization by Eq.
(\ref{ener_m_temp_rel}). Therefore, these two curves are equivalent. The
non-magnetized branch $m=0$ in Fig. \ref{magTOTAL} corresponds
to the line $T=2\epsilon-1$ in Fig. \ref{gaussian}.

In these curves, the solid line denoted BG corresponds to
the Boltzmann-Gibbs caloric curve. It describes the statistical
equilibrium state of the system, reached for $t\rightarrow +\infty$, as a
result of a collisional evolution (finite $N$-effects). Although this curve
is not physically relevant to interpret QSSs, that are out-of-equilibrium
structures, it is plotted for comparison. It also determines the domain of
stability/instability of the initial condition that is a
homogeneous Gaussian distribution.

The bullets are the results of direct $N$-body simulations. They form the
``experimental''  caloric curve. It is made of QSSs reached as a
result of a
violent collisionless relaxation (Vlasov dynamics). This is the curve that we
have to explain. This curve presents several striking features that were
considered as a
``surprise'' in the early works on the HMF model \cite{hmf,latora}: (i) it
differs from the BG
caloric curve for intermediate energies; (ii) it presents a region of negative
specific heats (unlike the BG caloric curve); (iii) the system remains
homogeneous (non-magnetized) slightly below the BG critical energy
$\epsilon_c=3/4$.
Although many things have been
understood on the physics of systems with long-range interactions 
\cite{houches,assisebook,campabook,physrep} since the
original works on the HMF
model (in particular the difference between QSSs and true equilibrium states),
we
believe that the HMF model has still not revealed all his secrets. In
particular, the precise nature of the QSSs, the reason for the region of
negative specific heats, and the shift of the transition point are still not
clearly understood. These are the questions that initiated the
topic more than 20 years ago and that still haven't found a definitive answer.
We stress that the  caloric
curve of Fig. \ref{gaussian} cannot be explained by Lynden-Bell's theory, as
pointed out in our previous work \cite{epjb2013}.\footnote{The
success or
failure of
the Lynden-Bell theory depends on the initial condition \cite{lbt}. There are
cases where
the Lynden-Bell
prediction works very well (see Refs. \cite{precommun,stan} and Fig. 36 of
\cite{epjb2013}) and cases,
such as the one we are
investigating here,
where it does not work \cite{epjb2013}. We can state that the
Lynden-Bell prediction does not work without making explicit calculations (which
would require a difficult numerical work) because, as found in \cite{epjb2013}
and discussed further in Sec. \ref{sec_df}, the QSSs have a compact support
(they can be fitted by polytropes of index $n=2$) while the Lynden-Bell
distribution has no cut-off in energy.} We shall
interpret this curve 
in
relation to the QL theory and to polytropic  (Tsallis) distributions.

The dashed line corresponds to the prediction of the QL theory. It is obtained
by solving the diffusion equation (\ref{vlasovaveragediff_bis}) until its
convergence towards a steady state (expected to represent the
angle-averaged QSS). From the
obtained velocity distribution $f_0(p)$, we can  compute the kinetic temperature
$T$ from Eq. (\ref{tkin}) and, using Eq. (\ref{ener_m_temp_rel}), we
can obtain the
corresponding magnetization $m$.

The solid line denoted ``$n=2$'' corresponds to the caloric curve obtained by 
assuming that the QSSs are polytropes of index $n=2$. This caloric curve
exhibits a region of negative
specific heats with $C_{\rm kin}=-5/2$. We emphasize that we do
not have any theory to predict why the QSSs should be polytropic (Tsallis)
distributions and
why their index should be $n=2$ for a Gaussian initial distribution. However, as
shown in \cite{epjb2013}, and confirmed below, polytropes of index
$n=2$ give a remarkable fit to the QSSs for a wide range of energies.

Let us now describe the curves of Figs. \ref{gaussian} and \ref{magTOTAL}. For
$\epsilon>\epsilon_{\rm c}=3/4=0.75$, the homogeneous
Gaussian distribution is stable (dynamically and thermodynamically), so the
system does not evolve (even on a collisional timescale).
For $\epsilon<\epsilon_{\rm c}$, the homogeneous
Gaussian
distribution is dynamically (Vlasov) unstable and the system rapidly evolves
towards a
QSS. The $N$-body simulations
show that the QSS is homogeneous for $\epsilon_{\rm
Nb}<\epsilon<\epsilon_{\rm c}$
and that it becomes magnetized for  $\epsilon<\epsilon_{\rm Nb}$ with
$\epsilon_{\rm Nb}\simeq 0.735$ (this is most clearly seen on the
magnetization
curve of Fig. \ref{magTOTAL}). The QL
theory predicts that the QSS is homogeneous for
$\epsilon_{\rm QL}<\epsilon<\epsilon_{\rm c}$
and that it becomes magnetized for 
$\epsilon<\epsilon_{\rm QL}$ with $\epsilon_{\rm QL}\simeq 0.735$. We
observe
that
$\epsilon_{\rm QL}\simeq \epsilon_{\rm Nb}$ in good approximation so the
QL
theory correctly predicts
the energy threshold $\epsilon_t\simeq \epsilon_{\rm QL}\simeq \epsilon_{\rm
Nb}\simeq 0.735$
below which the QSSs become magnetized (i.e.
depart from the line of homogeneous states). This is a non-trivial prediction
of the QL theory.
The polytropic caloric curve of index $n=2$ exhibits a transition energy
$\epsilon_{n=2}=2/3\simeq 0.666$ but this is not a
prediction since we have no way to say {\it a priori} why the polytropic index
should be $n=2$ without doing a $N$-body numerical simulation and making
a fit by a
polytrope \cite{epjb2013}. By contrast,
the result of the QL theory is a prediction since  it does not rely on a direct
$N$-body simulation of the HMF model; it  is directly obtained  from the
diffusion
equation (\ref{vlasovaveragediff_bis}).\footnote{We note that it is necessary
to solve a dynamical equation, which is a smoothing (or a coarse-graining) of
the
Vlasov equation, in order to predict the QSS. This confirms the claim made in
\cite{incomplete} concerning the importance of the {\it dynamics}.}
 On the other hand,
the transition energy $\epsilon_{\rm QL}\simeq 0.735$ predicted by the QL
theory
is in
much better agreement with the ``experimental'' results than the transition
energy $\epsilon_{n=2}=2/3\simeq 0.666$ corresponding to the $n=2$ polytropic
fit.
Indeed,
the points of the $N$-body simulation (QSSs) depart from
the
homogeneous branch ($m=0$) long
before the energy $\epsilon_{n=2}$ (the
$N$-body
simulation shows that the transition between un-magnetized and magnetized
states occurs at $\epsilon_{\rm Nb}=0.735$
which is sensibly larger than  $\epsilon_{n=2}=2/3\simeq 0.666$).
In addition, for
$\epsilon>\epsilon_{n=2}$ the caloric curve predicted by the QL
theory gives a much better agreement with the results of the $N$-body simulation
than
the caloric curve of a $n=2$ polytrope that predicts $m=0$. By contrast, for
$\epsilon<\epsilon_{n=2}$ the situation is reversed. In that case, the 
$n=2$ polytropes provide a much better agreement with the $N$-body simulation
than the QL theory. This is revealed not only by the values of the magnetization
and kinetic temperature that are very close to the numerical ones, but also by
the
distribution function itself as discussed in Sec. \ref{sec_df}. Therefore, we
conclude
that the QL theory works well close to the
critical energy $\epsilon_{\rm c}$ (in particular, it is able to predict the
{\it
shift} $\Delta\epsilon=\epsilon_t-\epsilon_{c}=-0.015$ of the
transition
energy from un-magnetized to magnetized QSSs) while
the
polytropic fit works well at lower energies $\epsilon<\epsilon_{n=2}$ (in
particular
it is able to account for the region of negative specific heats). These results,
which were foreseen in \cite{epjb2013}, represent an improvement in the
understanding of the caloric curve of Fig. \ref{gaussian}. However, everything
is still not understood.  First, we do not know {\it why} the QSSs are so
well-fitted by
polytropic (Tsallis) distributions 
of index $n=2$ (this is a property of incomplete mixing but it is not clear how
one can predict the efficiency of mixing \cite{incomplete}). Second, even if the
QL theory is
qualitatively correct close to $\epsilon_c$, it does not give a perfect
agreement with the results of the $N$-body simulations.

\subsection{Velocity distribution functions}
\label{sec_df}

In this section, we compare the velocity distribution functions obtained from
the $N$-body simulations, the QL theory, and the $n=2$ polytropic fit for two
different energies.

\begin{figure}[htpb]
\begin{center}
\includegraphics[width=0.45\linewidth]{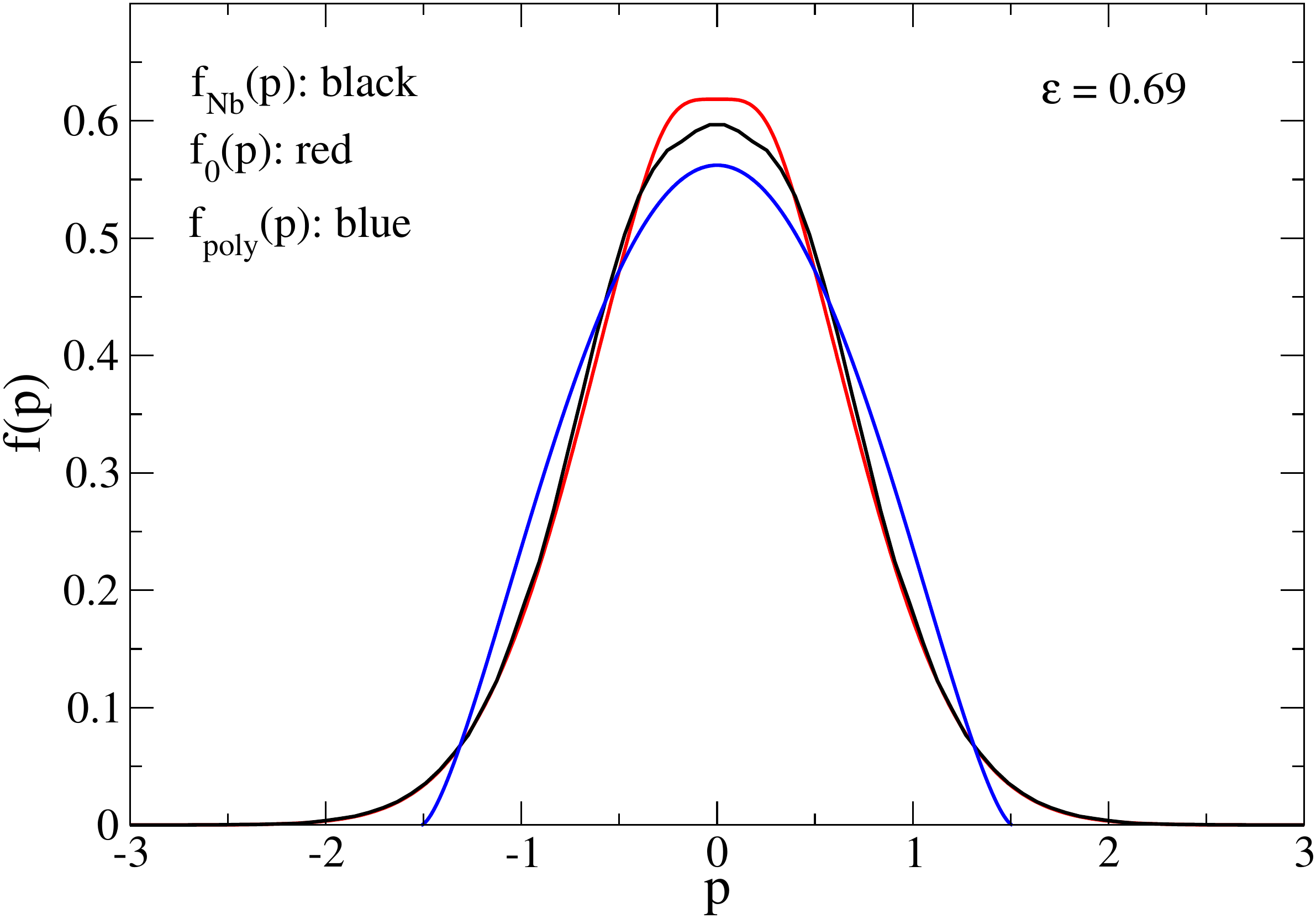}
\end{center}
\caption
{\label{nbodyPOLYTROPES69} Velocity distribution of the QSS for $\epsilon=0.69$
(black: $N$-body simulation; red: QL theory; blue: polytrope $n=2$).
}
\end{figure}

We first consider an energy $\epsilon=0.69$ close to the critical point
$\epsilon_{\rm c}=3/4=0.75$. The velocity distributions are plotted in Fig.
\ref{nbodyPOLYTROPES69}. As already discussed in Sec. \ref{sec_fc}, we find
a relatively good agreement between the QL theory and the $N$-body simulation.
This is of course consistent with the discussion of Sec. \ref{sec_cm} where we
showed that the QL theory works well for weakly inhomogeneous systems. We note
that the tail of the distribution
function has almost not evolved and coincides with the initial Gaussian
distribution. Only the core of the distribution has changed in order to make the
system dynamically stable.
The relatively good agreement between the QL theory and the $N$-body simulation
is confirmed by the values of the kinetic temperature and magnetization. We find
 $T_{\rm QL}=0.383$ and $m_{\rm QL}=0.057$  to be compared with
$T_{\rm Nb}=0.390$ and $m_{\rm Nb}=0.1$. On the other hand, the fit by a
polytrope $n=2$ is not good.\footnote{For
$\epsilon=0.69>\epsilon_{n=2}=2/3\simeq
0.666$ the polytrope $n=2$ is homogeneous so it is not necessary to average
the polytropic distribution function over the angles.} The main reason is that
the velocity
distribution of the  
polytrope $n=2$ has a compact support that is in contradiction with the
infinite extension of the Gaussian tail of the QSS.  The disagreement between
the $n=2$ polytrope  and the $N$-body simulation is
confirmed by the values of the kinetic temperature and magnetization. We find
$T_{\rm poly}=0.38$  and $m_{\rm poly}=0$ (homogeneous) to be compared
with $T_{\rm Nb}=0.390$ and $m_{\rm Nb}=0.1$ (inhomogeneous).

\begin{figure}[htpb]
\begin{center}
\includegraphics[width=0.45\linewidth]{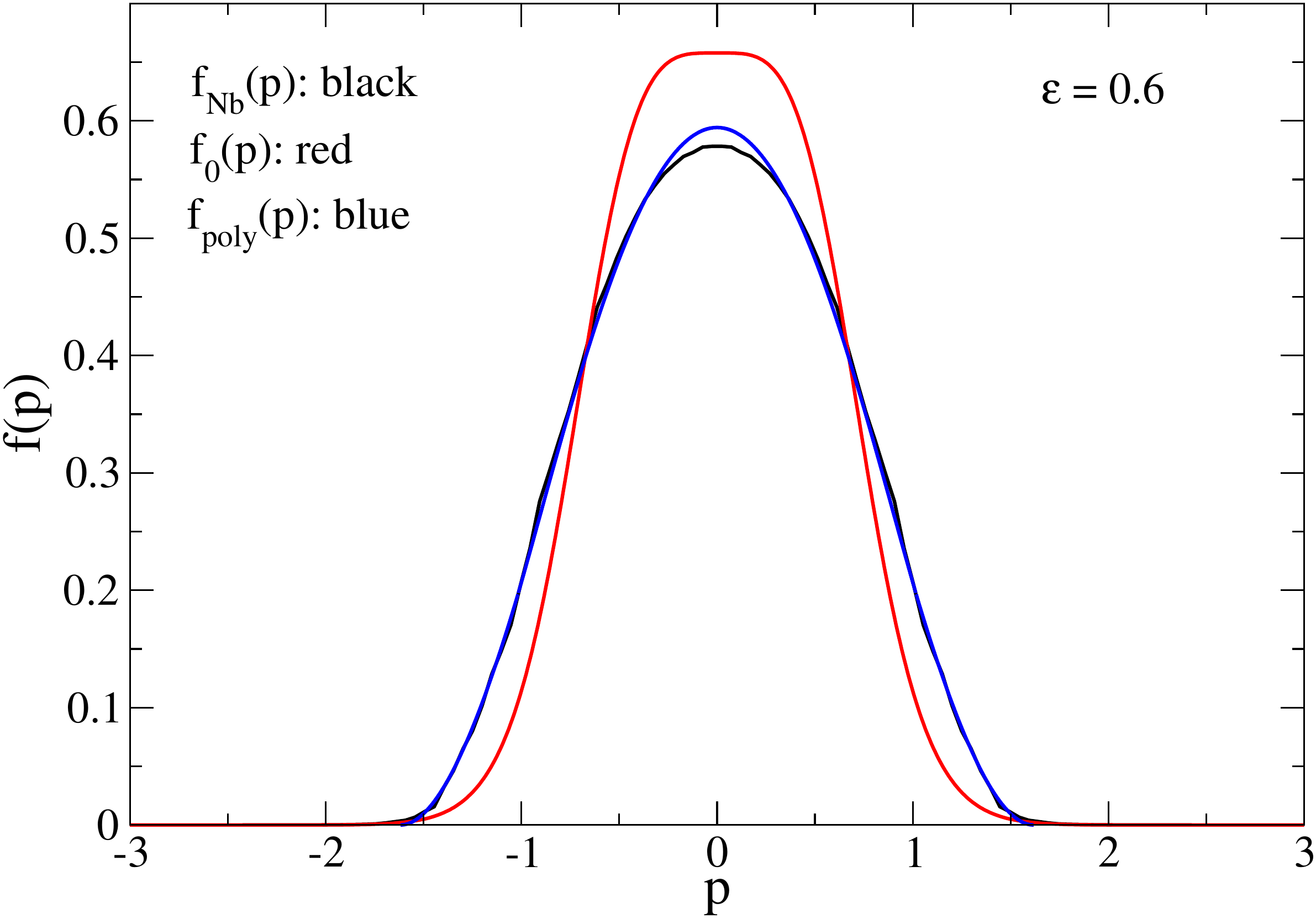}
\end{center}
\caption
{\label{nbodyPOLYTROPES60}  Velocity distribution of the QSS for $\epsilon=0.6$
(black: $N$-body simulation; red: QL theory; blue: polytrope $n=2$). }
\end{figure}

We now consider a lower energy $\epsilon=0.6$. The velocity distributions are
plotted in Fig. \ref{nbodyPOLYTROPES60}. In that case, the situation is
reversed. We observe a
remarkable agreement between the velocity distribution of the QSS reached
in
the  $N$-body simulation and the velocity distribution of a polytrope
$n=2$.\footnote{For $\epsilon=0.6<\epsilon_{n=2}=2/3\simeq
0.666$ the polytrope $n=2$ is inhomogeneous and, in order to obtain the
velocity distribution $f(p)$, we have integrated the polytropic distribution
function
$f(\theta,p)$ over the angles using the theoretical results obtained in
\cite{epjb2010,epjb2013}. As shown in \cite{epjb2013}, the excellent
agreement between the polytrope $n=2$ and the QSS is also valid for the complete
distribution function $f(\theta,p)=f(\epsilon)$.} The 
distribution function has substantially evolved from the initial Gaussian
distribution  and has acquired a compact support, which is a property of
polytropic (Tsallis) distribution with index $n\ge 1/2$
\cite{epjb2010,epjb2013}. As
discussed in \cite{epjb2013}, this ``confinement'' is a manifestation of an
incomplete relaxation (the distribution function predicted by the Lynden-Bell
theory,
which corresponds to the most mixed state, never has a compact support). The
excellent agreement between the $N$-body simulation and the $n=2$ polytropic fit
is confirmed by the values of the kinetic temperature and magnetization. We
find $T_{\rm poly}=0.357$ and $m_{\rm poly}=0.396$ to be compared with
$T_{\rm Nb}=0.360$ and $m_{\rm Nb}=0.40$. On the other hand, the prediction
of the QL theory is not good. The main reason is that this distribution has a
Gaussian tail, similarly to the initial condition, that is in contradiction with
the compact support of the QSS. The disagreement between the
QL theory  and the $N$-body simulation is
confirmed by the values of the kinetic temperature and magnetization. We find
$T_{\rm QL}=0.265$ and  $m_{\rm QL}=0.255$   to be compared
with $T_{\rm Nb}=0.360$ and $m_{\rm Nb}=0.40$.

At even lower energies (typically $\epsilon<0.55$), our study in \cite{epjb2013}
shows that the system takes a core-halo structure. The core corresponds to
the pure polytrope $n=2$ but it is now surrounded by a halo of particles. We
note that the presence of the halo does not significantly affect the value of
the kinetic temperature and magnetization of the QSS since a pure $n=2$
polytrope still gives a rather good fit to the caloric curve and magnetization
curve at low  energies (see Figs.
\ref{gaussian} and \ref{magTOTAL}).

\subsection{Summary and discussion}

In summary, for a homogeneous Gaussian initial distribution,
we have the
following results. For
$\epsilon>\epsilon_c$, the system remains in the
homogeneous Gaussian distribution which is stable. For
$\epsilon_{n=2}<\epsilon<\epsilon_{c}$, the QL theory works reasonably well. 
In particular, it is able to predict the energy $\epsilon_t\simeq
0.735$ marking the transition between unmagnetized and magnetized QSSs. More
precisely (i) for $\epsilon_{t}<\epsilon<\epsilon_c$,
the system achieves a  homogeneous modified Gaussian
distribution with a flat core; (ii) for $\epsilon_{n=2}<\epsilon<\epsilon_{t}$,
the
system achieves an  inhomogeneous modified Gaussian
distribution with a flat core. For lower energies, the QL theory does not work.
In that
case, we
observe that the QSSs are remarkably
well fitted by  polytropic (Tsallis) distributions
with index
$n=2$. They have  compact support. For
$\epsilon<\epsilon_{n=2}$, the stable polytropic distributions are
inhomogeneous.
These polytropic
distributions are able to account for the
region of negative specific heats in the out-of-equilibrium caloric curve,
unlike the Boltzmann distribution and (presumably) the Lynden-Bell
distribution, and unlike the QL theory. At even lower energies, the
system takes a core-halo structure with a polytropic core and a halo of
particles.

The QL theory and the polytropic fit have different, but well-defined, domains
of validity. This may be qualitatively 
understood as follows. The QL theory is based on the assumption that the system
remains slightly inhomogeneous during its evolution from the initial condition
to the QSS (weak mixing). Therefore, the QL theory is expected to be valid close
to the
instability threshold $\epsilon_c$. In that case, the distribution function
does not substantially change from the initial condition. It just changes a
little in a manner to become dynamically (Vlasov)
stable.\footnote{This is the
stabilization mechanism of the QL theory. The initial condition $f_0(p,t=0)$ is
dynamically unstable meaning that the complex pulsation $\omega_I(t=0)$ is
strictly positive. Qualitatively, this is because the velocity dispersion
(kinetic temperature) is too low [see Eq. (\ref{stabeff})]. The initially
positive complex
pulsation $\omega_I(t=0)>0$ causes the distribution function to diffuse [see
Eqs. (\ref{vlasovaveragediff_bis}) and (\ref{defdiffhmf_b_bis})].  By
diffusing, the distribution function spreads
and its velocity dispersion increases. As a result, the complex pulsation
$\omega_I(t)$ decreases [see Eq. (\ref{disperhmfsmallsol})]. The
diffusion stops when $\omega_I(t)$, hence $D(p,t)$, vanishes leading the
system to a Vlasov stable, or marginally stable, QSS.} As
we have seen, this change
only concerns the core of the distribution which becomes flatter and wider
(``hotter''). The tail of the distribution does not change and remains
Gaussian, as the initial state. When we depart significantly from the
instability threshold, the evolution of the system is
expected to be more violent and the system is expected to be strongly
inhomogeneous at least during the early stage of its evolution (strong mixing).
In that case, the distribution function is expected to change
substantially from the initial condition. This is the regime where the
Lynden-Bell \cite{lb} theory should apply in principle. However, it turns out
that, in many cases, the relaxation to the Lynden-Bell statistical equilibrium
state is incomplete \cite{incomplete}. In case of incomplete relaxation,
polytropic distributions may provide a good
fit of the QSS because they have a compact support that is typical of an
incomplete relaxation (in case of incomplete relaxation high energy states
are less populated than what is predicted from the statistical theory of
Lynden-Bell) \cite{lbt,epjb2010,epjb2013}. This is probably why the polytropic
fit works well for low values of the energy. In that case, the distribution
function changes from an initial Gaussian to a polytrope with a compact support.

\section{The case of a homogeneous  semi-elliptical initial condition}
\label{sec_comp_bis}

In this section, we show how the previous results are modified when we consider
a homogeneous semi-elliptical initial condition  (homogeneous  polytrope of
index $n=1$) instead of a homogeneous 
Gaussian initial condition.

\subsection{The $N$-body simulations and the comparison with the results of the
diffusion equation}
\label{sec_fcbis}

\begin{figure}[htpb]
\begin{center}
\includegraphics[width=0.45\linewidth]{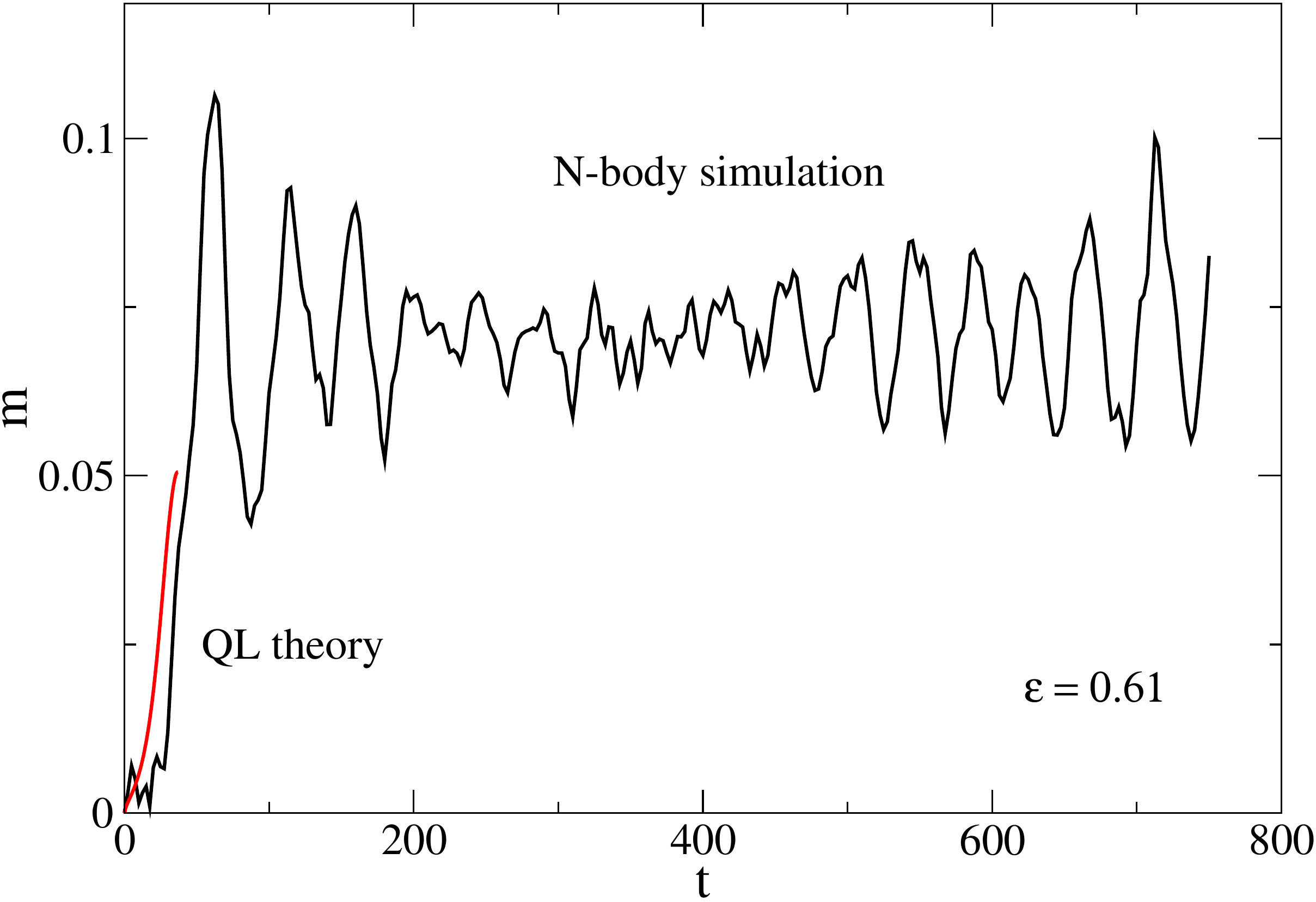}
\includegraphics[width=0.45\linewidth]{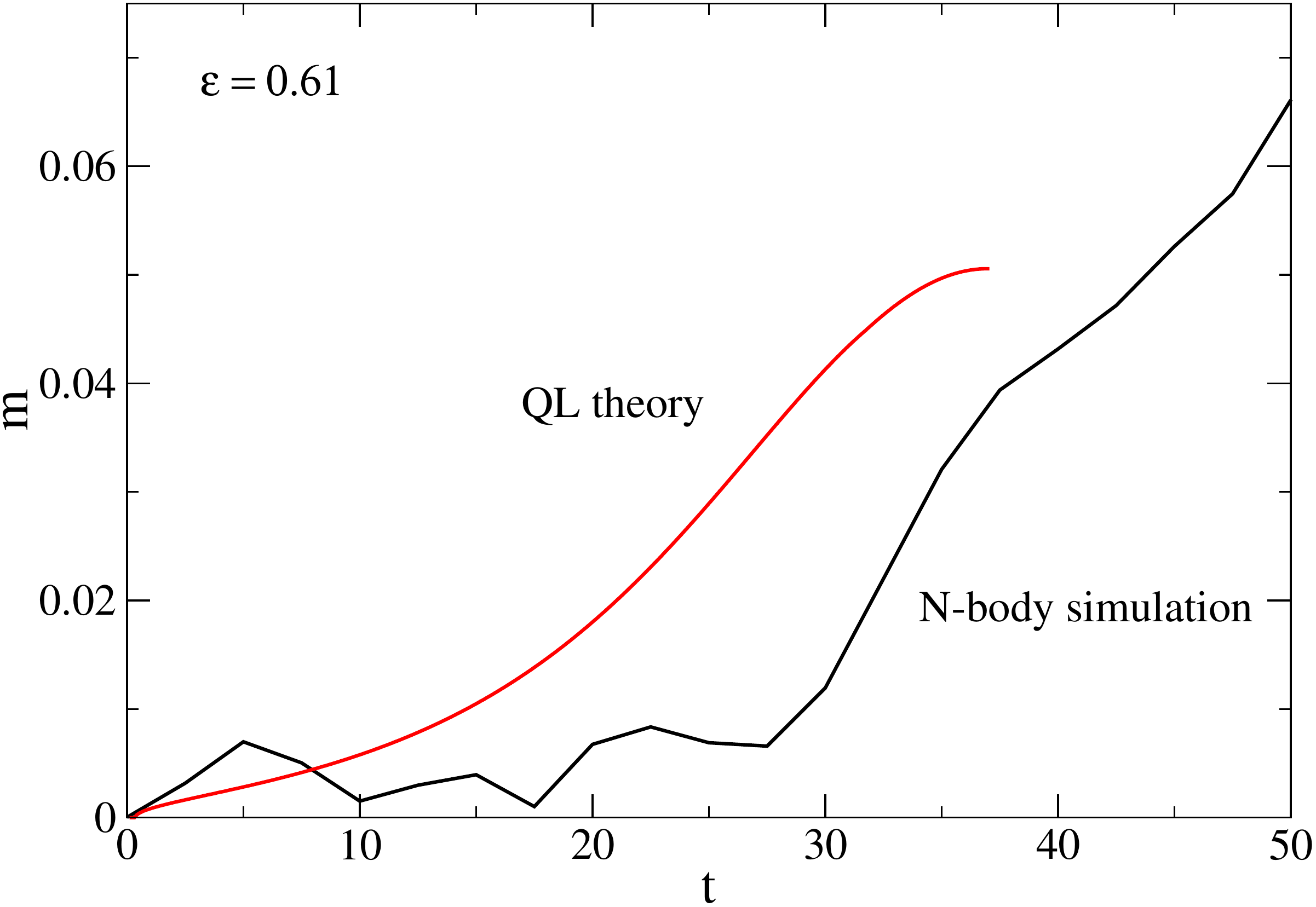}
\end{center}
\caption
{\label{mnbody_061SEMIELLIPSE} Magnetization as a function of time at energy
$\epsilon=0.61$ for a homogeneous semi-elliptical
velocity initial distribution. In
order to optimize the agreement between the  $N$-body simulation (black curve)
and the QL theory (red curve) for the chosen value of $N=2^{18}$, we have solved
the diffusion equation with  $\chi(0)=10^{-6}$. As in the
Gaussian case,
a different choice of $\chi(0)$ affects the time scale, but not the final
distribution or the final value of
the magnetization.
}
\end{figure}

In Fig. \ref{mnbody_061SEMIELLIPSE}, we plot the temporal evolution of the
magnetization obtained from the $N$-body simulations and by solving the
diffusion
equation of the QL theory for $\epsilon=0.61$. The interpretation is similar to
that given previously for the initial Gaussian distribution. For
$\epsilon=0.61$, the final magnetization predicted by the QL theory ($m_{\rm
QL}=0.05$) is relatively close to the value obtained from the $N$-body
simulation ($m_{\rm Nb}=0.07$) by
averaging over the oscillations.

\begin{figure}[htpb]
\begin{center}
\includegraphics[width=0.45\linewidth]{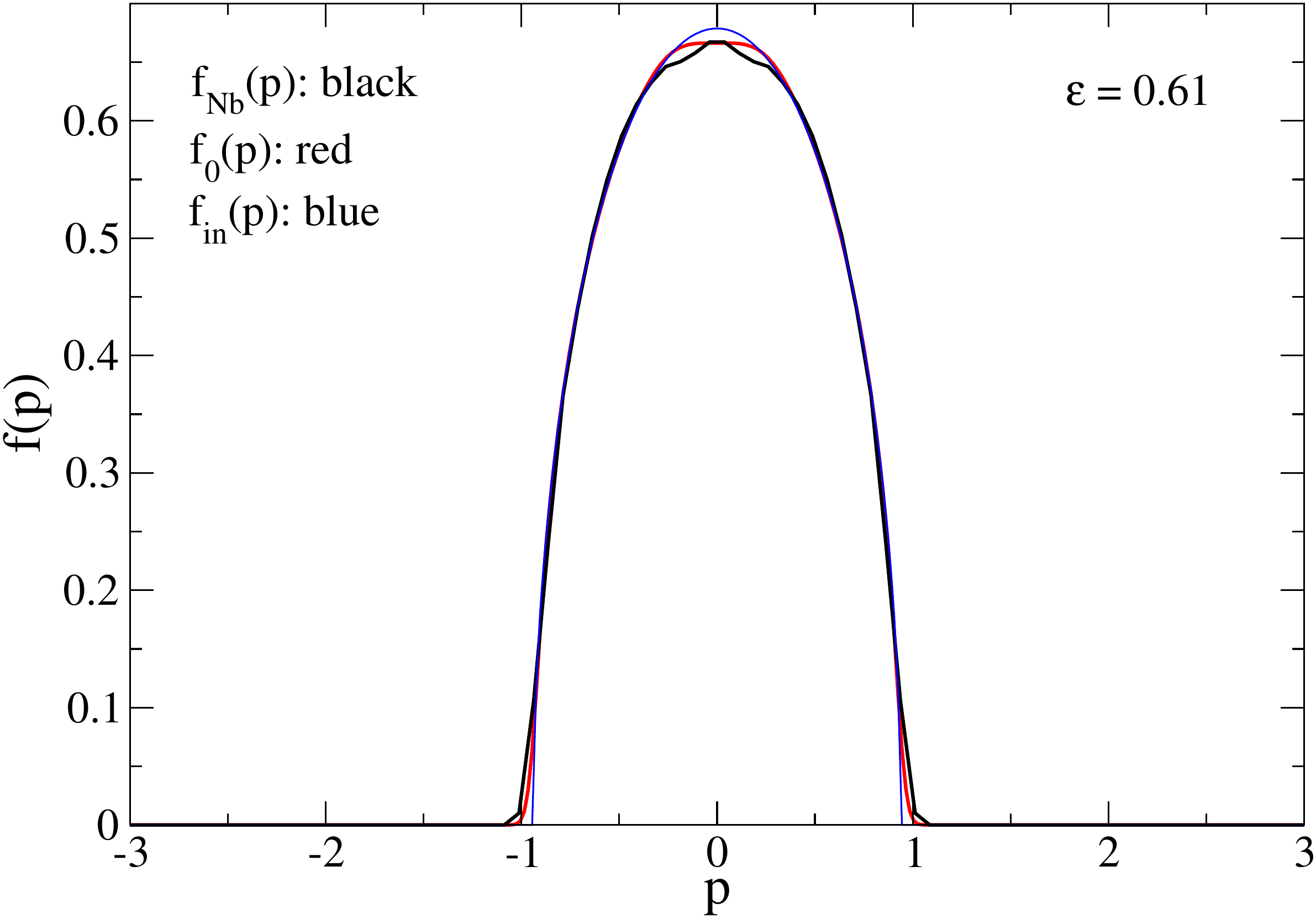}
\includegraphics[width=0.45\linewidth]{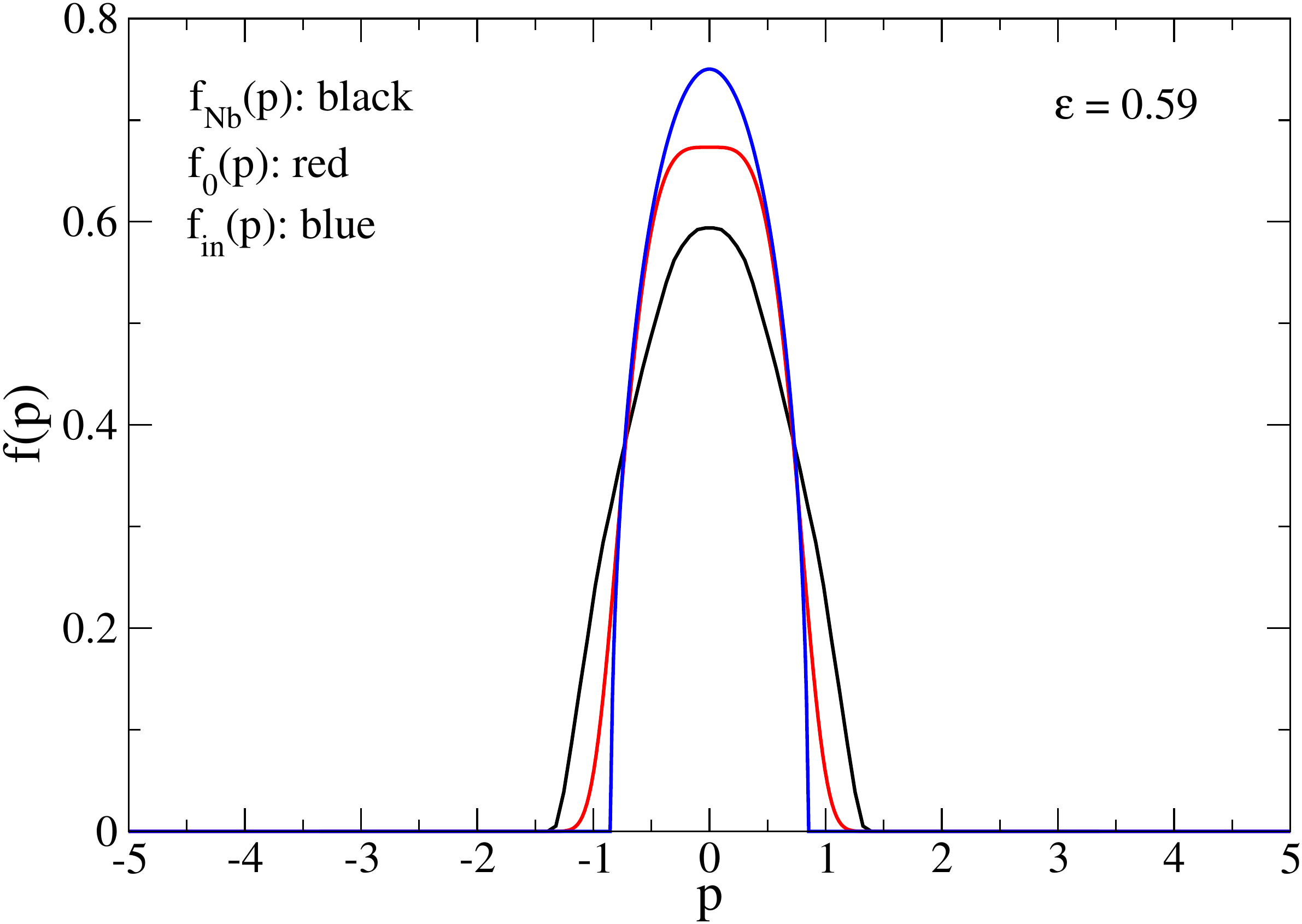}
\end{center}
\caption
{\label{f_ell_061_ql_nbody} The final distribution function $f_0(p)$ of the
diffusion equation and the velocity distribution
$f_{\rm Nb}(p)$ of the QSS of the $N$-body simulation, starting from a
homogeneous semi-elliptical initial distribution with energy $\epsilon = 0.61$
(left pannel) and  $\epsilon = 0.59$ (right pannel).
We have also plotted the initial condition for comparison. We
emphasize that the initial
distribution is Vlasov unstable while the final distribution is Vlasov stable
even when the profile does not seem to have changed a lot (left panel). }
\end{figure}

In Fig. \ref{f_ell_061_ql_nbody}, we compare the velocity distribution function
of the QSS reached in the $N$-body simulation and the final distribution
function of the diffusion equation. Like the initial distribution function, the
final distribution function has a compact support (actually, the distribution
function keeps a compact support at all times). We consider two energies
$\epsilon =0.61$ and $\epsilon =0.59$. For $\epsilon=0.61$,
which is close to
the instability threshold $\epsilon^*_c =\epsilon_{n=1}=5/8=0.625$ of the
homogeneous semi-elliptical distribution ($n=1$
polytrope), the agreement between the
QL theory and the $N$-body simulation is good except in the central region (and
to a lesser extent close to the maximum velocity). We note
that the distribution function has not changed much as compared to the initial
semi-elliptical distribution.  Only the core of the distribution
and the region close to the maximum velocity have changed  in order to make the
distribution  dynamically stable (without this small change, the distribution
would be unstable). In particular, the core of the distribution
becomes flatter and wider (``hotter''). For
the lower energy $\epsilon=0.59$, the disagreement is more pronounced
everywhere. The distribution function in the $N$-body simulation
has substantially evolved from the initial
semi-elliptical distribution  and has spread (while keeping a compact support)
contrary to the distribution of the QL theory. These results are similar to
those obtained in the case of a Gaussian initial distribution.
We note, however, that in the semi-elliptical case, close to the instability
threshold $\epsilon^*_c$, the distribution function changes from the
initial
distribution function not only in the core but also in the tail, close to the
maximum velocity. By contrast, in the Gaussian case, close to the instability
threshold $\epsilon_c$, the tail of the distribution almost does not change from
the initial condition.

\subsection{Caloric and magnetization curves}
\label{sec_cmbis}

\begin{figure}[htpb]
\begin{center}
\includegraphics[width=0.45\linewidth]{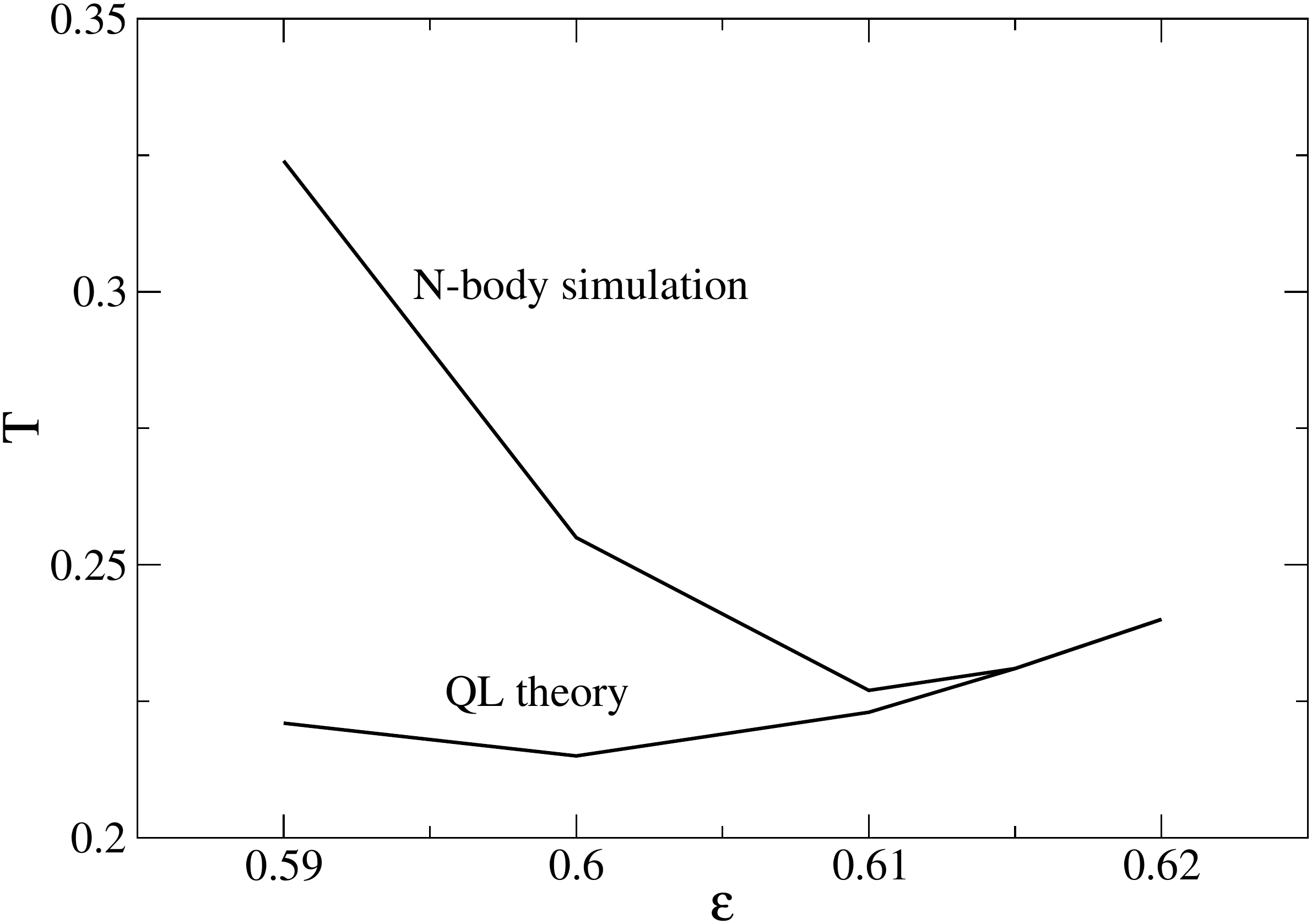}
\includegraphics[width=0.45\linewidth]{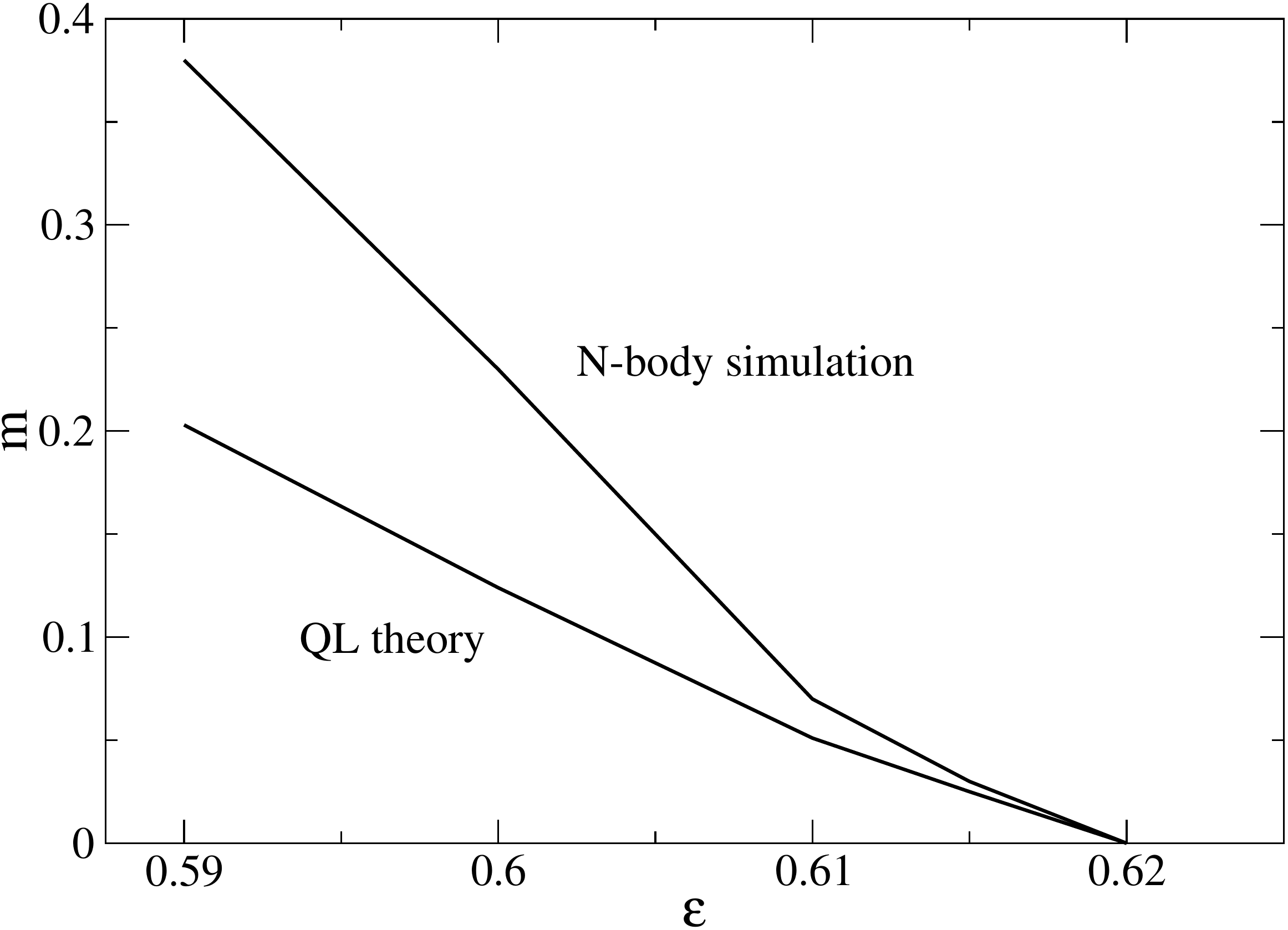}
\end{center}
\caption
{\label{etkinSEMIELLIPSE} The kinetic temperature and the magnetization of the
stationary state reached by the diffusion equation of the 
QL theory
and that of the QSS reached by the $N$-body simulation, as a function of
the energy.
}
\end{figure}

In Fig. \ref{etkinSEMIELLIPSE},  we plot the kinetic temperature and the
magnetization of the final distribution of the QL theory and of the $N$-body
simulation as a function of the energy. The two curves approach each other
close to the instability threshold. In particular, the QL theory is able to
predict the energy of the out-of-equilibrium phase transition from the
unmagnetized state ($m=0$) to the magnetized state ($m\neq 0$) as discussed
below.

\begin{figure}[htpb]
\begin{center}
\includegraphics[width=0.45\linewidth]{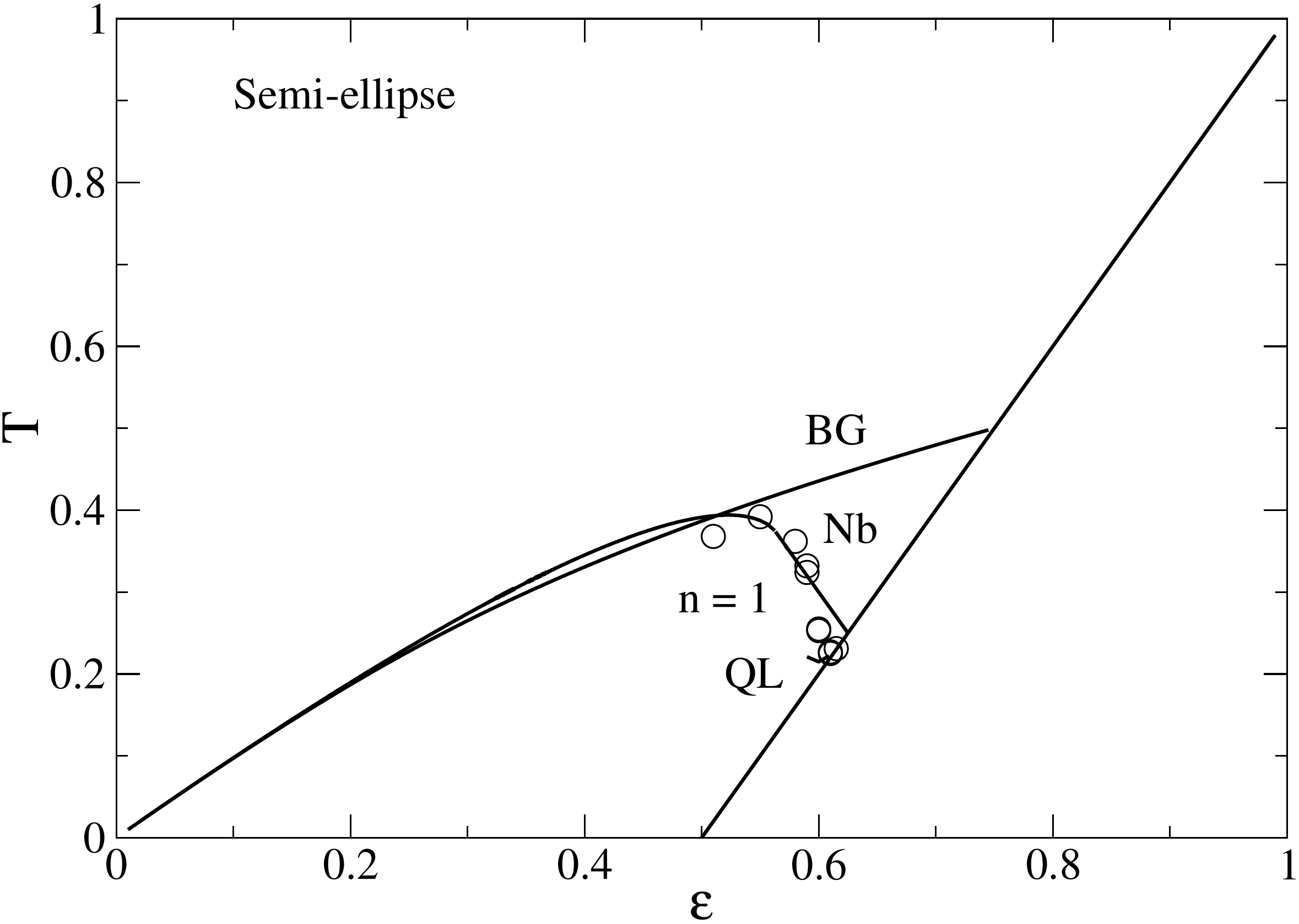}
\includegraphics[width=0.45\linewidth]{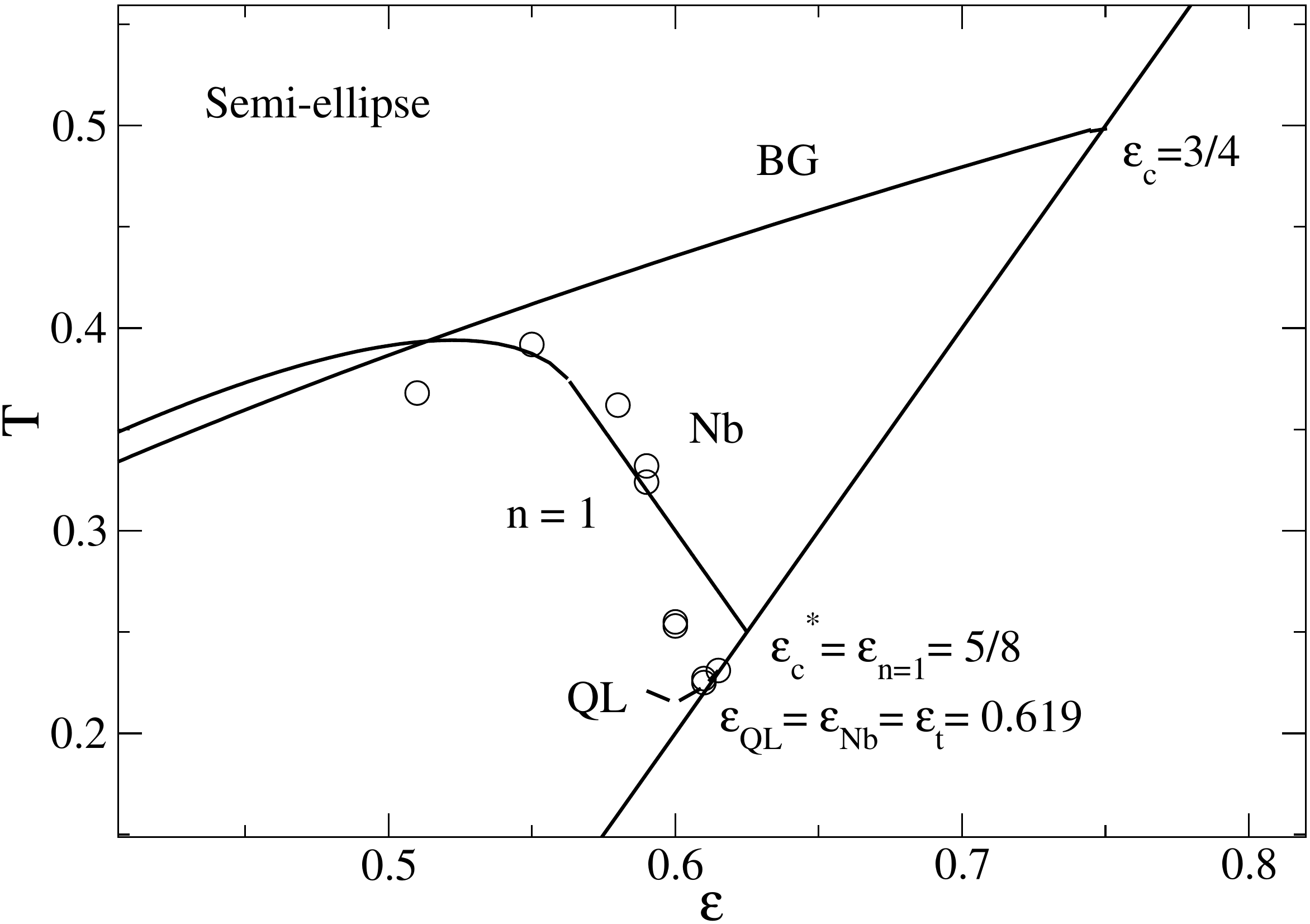}
\end{center}
\caption
{\label{ellipse} Caloric curve of the HMF model for a spatially
homogeneous semi-elliptical initial condition (BG: Boltzmann-Gibbs states;
bullets:
results of $N$-body simulations; dashed line: prediction
of the QL theory; $n=1$: polytropic fit). }
\end{figure}

\begin{figure}[htpb]
\begin{center}
\includegraphics[width=0.45\linewidth]{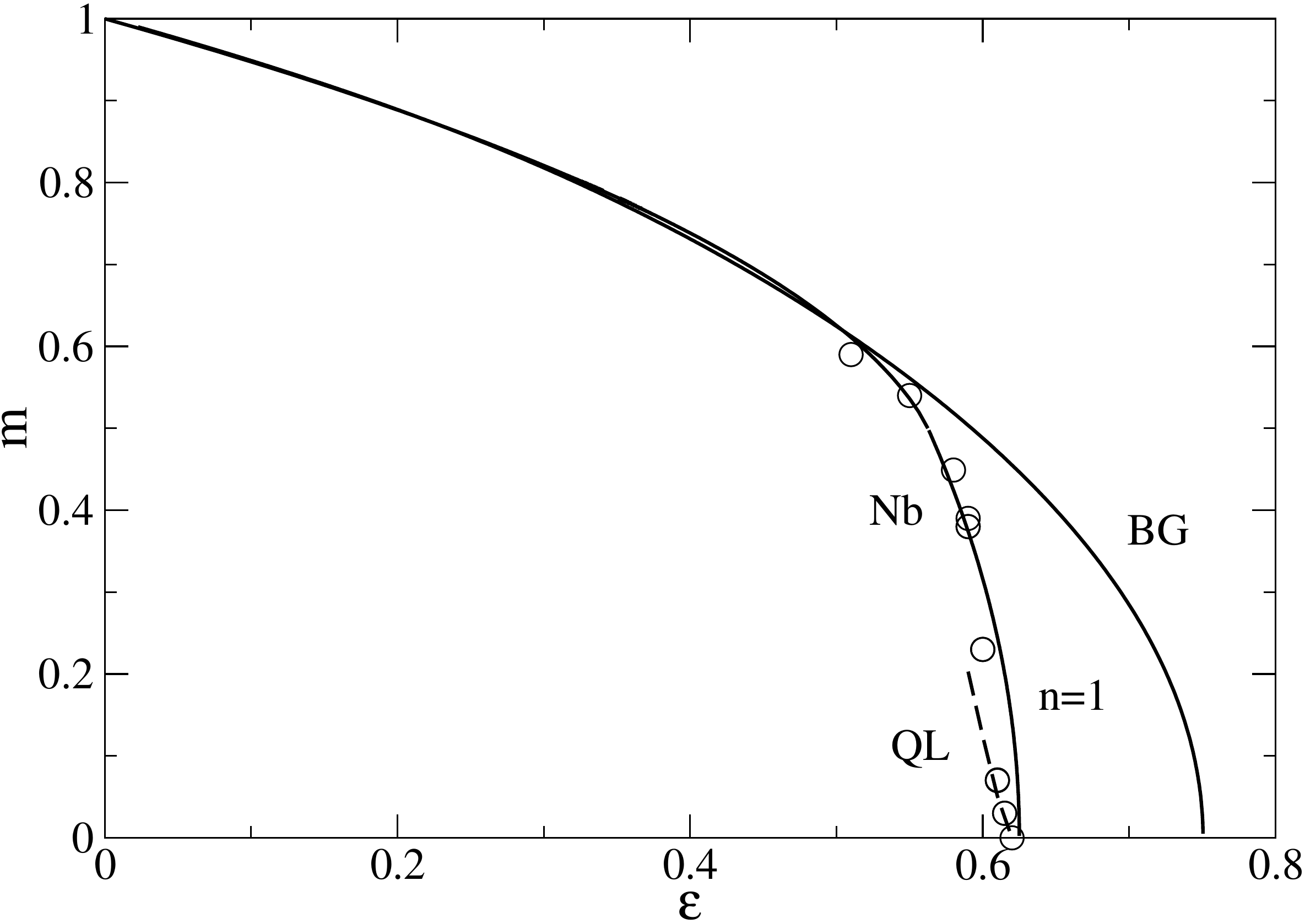}
\includegraphics[width=0.45\linewidth]{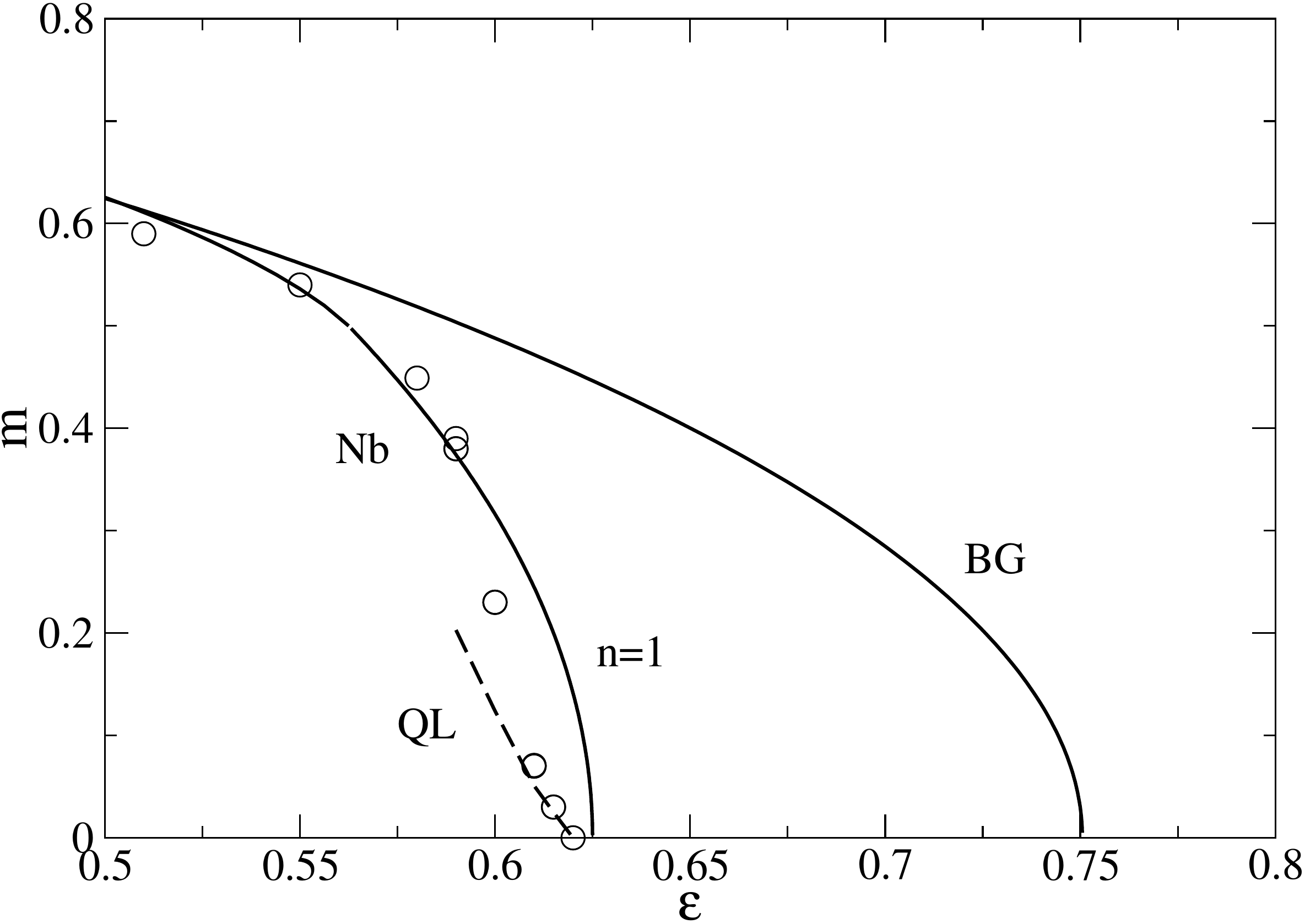}
\end{center}
\caption
{\label{magTOTALellipse}  Magnetization  curve of the HMF model for a spatially
homogeneous semi-elliptical initial condition.}
\end{figure}

In Figs. \ref{ellipse} and \ref{magTOTALellipse}, we compare the
caloric curve $T(\epsilon)$  and the magnetization curve  $m(\epsilon)$ 
obtained from direct $N$-body simulations of the HMF model, starting from a
homogeneous semi-elliptical distribution, with the
prediction
of the QL theory and with the fit corresponding to a polytrope $n=1$ considered
in our previous papers \cite{epjb2010,epjb2013}. The bullets are the results of
direct $N$-body simulations. The dashed line
corresponds to the prediction of the QL theory. The solid line denoted ``$n=1$''
corresponds to the caloric curve obtained by assuming that the QSSs are
polytropes of index $n=1$. This caloric curve exhibits a region of negative
specific heats with $C_{\rm kin}=-1/2$. As shown in \cite{epjb2013}, and
confirmed below,
polytropes of index $n=1$ give a remarkable fit to the QSSs for a wide range of
energies.

The initial condition,
which is a spatially homogeneous
semi-elliptical distribution, corresponds to a polytrope of index $n=1$.
Therefore, for
$\epsilon>\epsilon_{\rm c}^*=\epsilon_{n=1}=5/8=0.625$, the homogeneous
semi-elliptical 
distribution is dynamically stable and the
system does not evolve on a collisionless timescale.
For $\epsilon<\epsilon_{\rm c}^*$, the homogeneous semi-elliptical 
distribution is dynamically (Vlasov) unstable and the system rapidly evolves
towards a QSS.

Slightly below $\epsilon_{\rm c}^*$, the QL theory is in good agreement with
$N$-body simulations. It predicts that the QSS remains homogeneous (non
magnetized) for $\epsilon_{QL}\simeq 0.619<\epsilon<\epsilon_{\rm c}^*$ and that
it
becomes inhomogeneous for $\epsilon<\epsilon_{QL}$ (the value 
$\epsilon_{QL}\simeq 0.619$ has been extrapolated from the magnetization curve
of Fig. \ref{etkinSEMIELLIPSE}). The transition
energy $\epsilon_{QL}\simeq 0.619$ predicted by the QL theory is in good
agreement
with the value $\epsilon_{\rm Nb}\simeq 0.619$ obtained from the $N$-body
simulations (also extrapolated from the magnetization curve
of Fig. \ref{etkinSEMIELLIPSE}).
Therefore, for a homogeneous semi-elliptical initial condition, the
out-of-equilibrium transition energy is
$\epsilon_t\simeq \epsilon_{\rm QL}\simeq \epsilon_{\rm Nb}\simeq 0.619$ (it can
be compared to the value $\epsilon_t\simeq 0.735$ obtained for a homogeneous
Gaussian initial condition). For
$\epsilon_{t}<\epsilon<\epsilon_{\rm c}^*$,  the distribution is not a
homogeneous  polytrope of index $n=1$ since this distribution is unstable.
The observed distribution of the QSS is close to the distribution predicted by
the QL theory as analyzed in Sec. \ref{sec_dfbis}.

For smaller energies, the QL theory does not give a good agreement with
the $N$-body simulations anymore. In that case, we observe (in agreement with
our earlier paper \cite{epjb2013}) that the QSS is well-fitted by an
inhomogenous $n=1$ polytrope. This is revealed not only by
the values of the magnetization and kinetic temperature that are very close to
the numerical ones, but also by the
distribution function itself as discussed in Sec. \ref{sec_dfbis}. We have,
however, no theory to justify why the $n=1$ polytropic fit works so well.

In conclusion, the QL theory works well close to the 
critical energy $\epsilon_{\rm c}^*$ (in particular it is able to predict the
{\it shift} $\Delta\epsilon=\epsilon_t-\epsilon_{c}^*=-0.006$ of the
transition energy from un-magnetized to magnetized QSSs) while
the polytropic fit works well at lower energies (in particular
it is able to account for the region of negative specific heats). 
These conclusions are similar to those obtained for a spatially
homogeneous Gaussian initial condition. We note, however, that the
shift $\Delta\epsilon=\epsilon_t-\epsilon_{c}^*=-0.006$ obtained in the
semi-elliptical case is smaller than the
shift $\Delta\epsilon=\epsilon_t-\epsilon_{c}=-0.015$ obtained in the Gaussian
case. We also note that, in the semi-elliptical case, the transition energy
$\epsilon_t\simeq 0.619$ predicted by the QL theory is below the 
transition energy $\epsilon_{\rm c}^*=\epsilon_{n=1}=5/8=0.625$ corresponding to
$n=1$
polytropes while in the Gaussian case, the transition energy 
$\epsilon_t\simeq 0.735$ predicted by the QL theory is above the 
transition energy $\epsilon_{n=2}=2/3\simeq 0.666$ corresponding to $n=2$
polytropes. Therefore, the fact
that we reach similar conclusions in these two different situations is
encouraging.

\subsection{Velocity distribution functions}
\label{sec_dfbis}

In this section, we compare the velocity distribution functions obtained from
the $N$-body simulations, the QL theory, and the $n=1$ polytropic fit for two
different energies.

\begin{figure}[htpb]
\begin{center}
\includegraphics[width=0.45\linewidth]{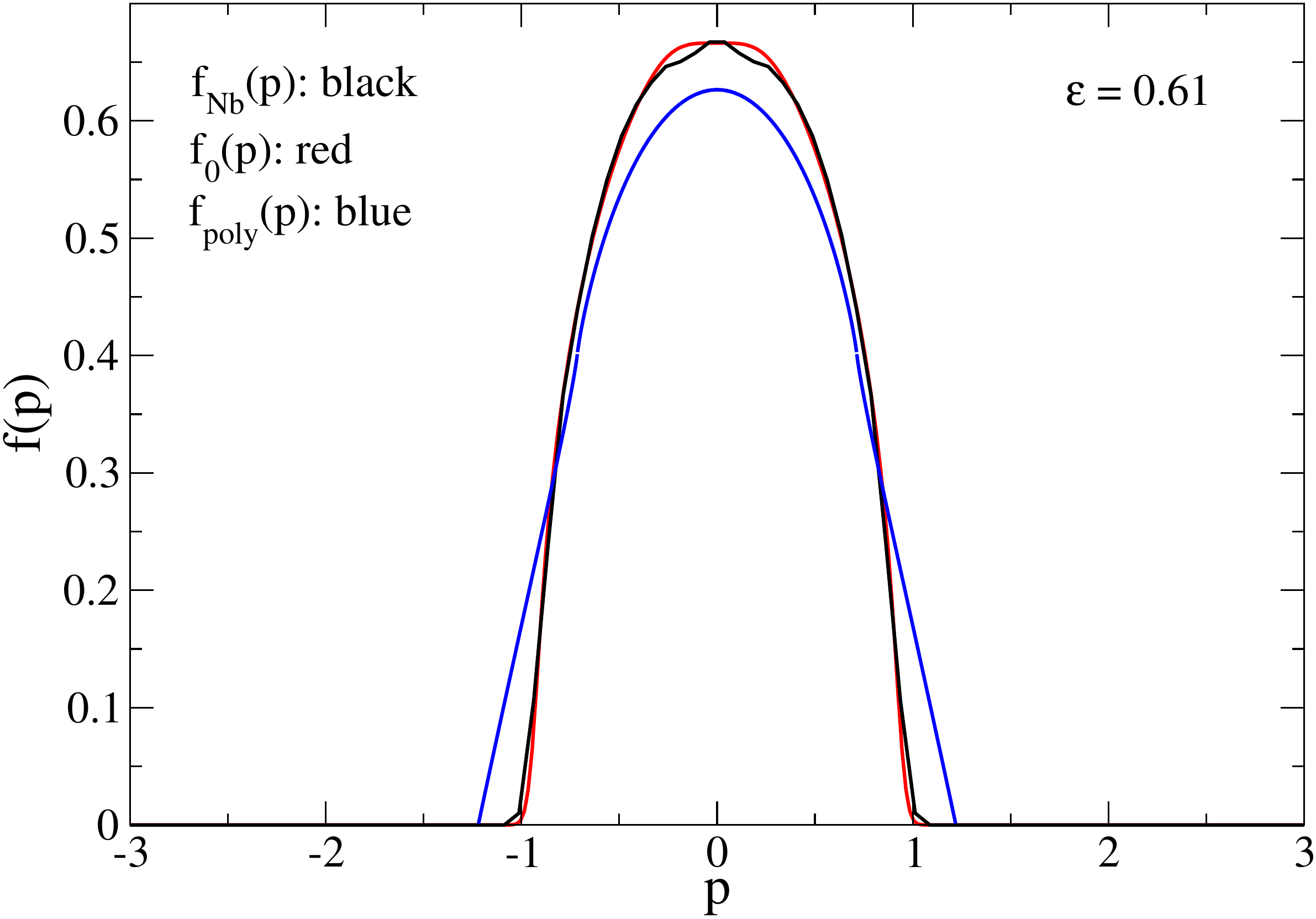}
\end{center}
\caption
{\label{compNb0polyE0p61ellispe}  Velocity distribution of the QSS for
$\epsilon=0.61$
(black: $N$-body simulation; red: QL theory; blue: polytrope $n=1$). }
\end{figure}

We first consider an energy $\epsilon=0.61$ close to the critical point
$\epsilon_{\rm c}^*=5/8=0.625$. The velocity distributions are plotted in Fig.
\ref{compNb0polyE0p61ellispe}. As already discussed in Sec. \ref{sec_fcbis}, we
find
a relatively good agreement between the QL theory and the $N$-body simulation.
This relatively good agreement
is confirmed by the values of the kinetic temperature and magnetization. We find
 $T_{\rm QL}=0.223$ and $m_{\rm QL}=0.05$  to be compared with
$T_{\rm Nb}=0.227$ and $m_{\rm Nb}=0.07$.
On the other hand, the fit by an inhomogeneous
polytropic distribution of index $n=1$ is not good. Although it has a compact
support, it is more
spread than the QSS obtained from the $N$-body simulation.   The disagreement
between the $n=1$ polytrope  and the $N$-body simulation is
confirmed by the values of the kinetic temperature and magnetization. We find
$T_{\rm poly}=0.28$ and $m_{\rm poly}=0.245$ to be compared
with $T_{\rm Nb}=0.227$ and $m_{\rm Nb}=0.07$.

\begin{figure}[htpb]
\begin{center}
\includegraphics[width=0.45\linewidth]{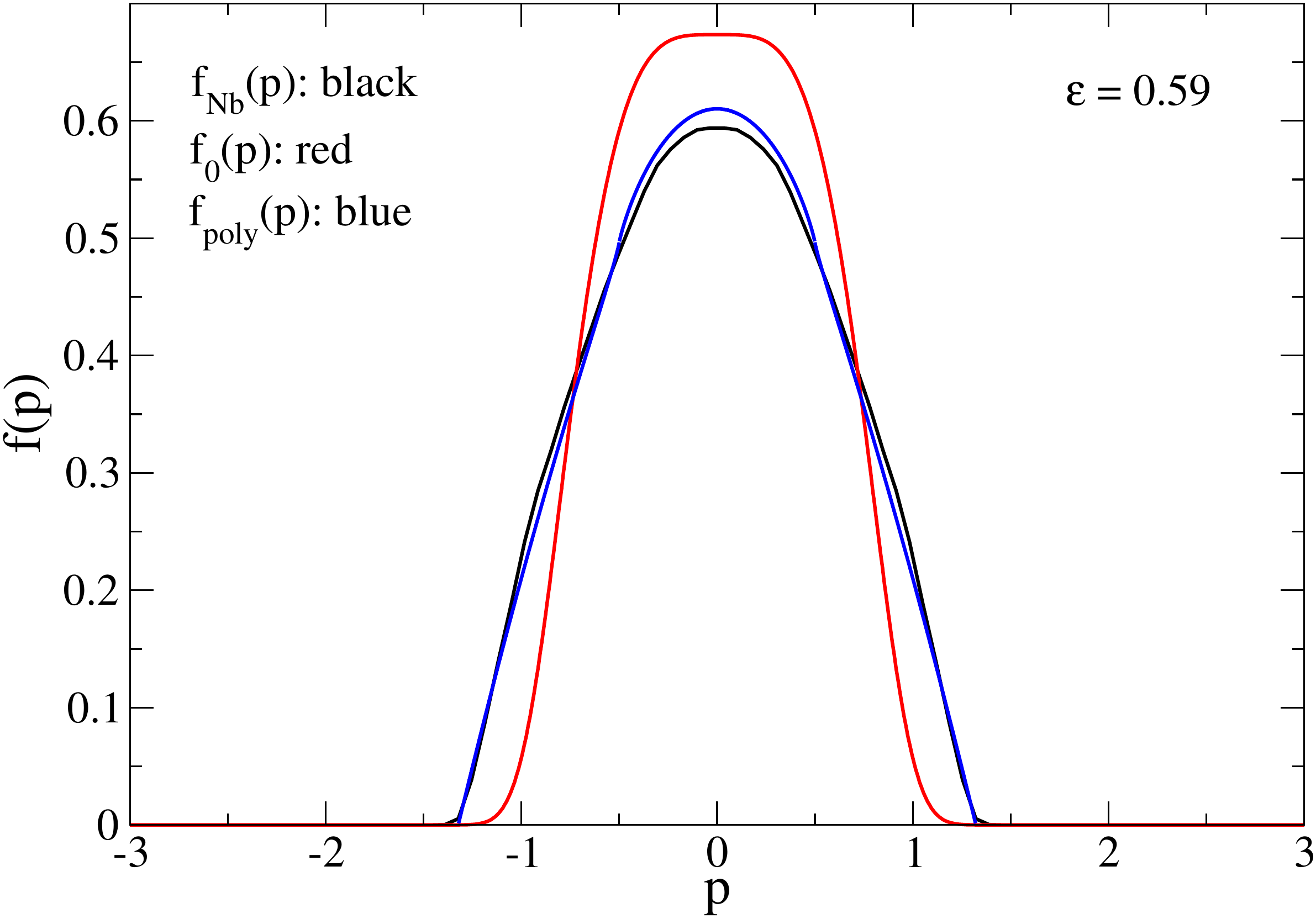}
\end{center}
\caption
{\label{compNb0polyE0p59ellispe}  Velocity distribution of the QSS for
$\epsilon=0.59$
(black: $N$-body simulation; red: QL theory; blue: polytrope $n=1$). }
\end{figure}

We now consider a lower energy $\epsilon=0.59$.  The velocity distributions are
plotted in Fig.
\ref{compNb0polyE0p59ellispe}. In that case, the situation is
reversed. We observe a
remarkable agreement between the velocity distribution of the QSS obtained from
the  $N$-body simulation and the velocity distribution of a polytrope
$n=1$.\footnote{As shown in \cite{epjb2013}, the excellent
agreement between the polytrope $n=1$ and the QSS is also valid
for the complete
distribution function $f(\theta,p)=f(\epsilon)$.} This is
confirmed by the values of the kinetic temperature and magnetization. We
find $T_{\rm poly}=0.32$ and $m_{\rm poly}=0.374$ to be compared with
$T_{\rm Nb}=0.324$ and $m_{\rm Nb}=0.38$. On the other hand, the prediction
of the QL theory is not good. This
distribution has not spread enough from the initial condition (see Fig.
\ref{f_ell_061_ql_nbody}). The disagreement between the
QL theory  and the $N$-body simulation is
confirmed by the values of the kinetic temperature and magnetization. We find
$T_{\rm QL}=0.22$ and  $m_{\rm QL}=0.203$   to be compared
with $T_{\rm Nb}=0.324$ and $m_{\rm Nb}=0.38$.

At even lower energies (e.g. $\epsilon=0.55$), our study in \cite{epjb2013}
shows that the system takes a core-halo structure. The core corresponds to
the pure polytrope $n=1$ but it is now surrounded by a halo of particles.

\subsection{Summary and discussion}

In summary, for a homogeneous semi-elliptical initial distribution,
we have the following results. For
$\epsilon>\epsilon_c^*=\epsilon_{n=1}=0.625$, the system remains in the
homogeneous  semi-elliptical  distribution which is stable. Close to
$\epsilon_{c}^*$, the QL theory works reasonably well. 
In particular, it is able to predict the energy $\epsilon_t\simeq
0.619$ marking the transition between unmagnetized and magnetized QSSs. 
For lower energies, the QL theory does not work.
In that
case, we
observe that the QSSs are remarkably
well fitted by inhomogeneous polytropic distributions (Tsallis distributions)
with index
$n=1$. These polytropic
distributions are able to account for the
region of negative specific heats in the out-of-equilibrium caloric curve,
unlike the Boltzmann and Lynden-Bell distributions. At even lower energies, the
system takes a core-halo structure with a polytropic core and a halo of
particles.

Since the initial condition is a spatially homogeneous polytrope
of index $n=1$,
and since it appears that the QSSs are often well-fitted by inhomogeneous
polytropes of index $n=1$, we could expect that the QSS could be fitted by
an inhomogeneous polytrope of index $n=1$ {\it as soon as}
$\epsilon<\epsilon_c^*=\epsilon_{n=1}$. Actually, this is not the case. There is
a small range
of energies $[\epsilon_t,\epsilon_c^*]$ where the QSS remains homogeneous.
Within this small range of energies, the QL theory works
relatively well, unlike the polytropic fit. Therefore, even if the caloric
curves differ in the detail, the
results obtained for a semi-elliptical initial distribution are relatively
similar to those obtained for a Gaussian initial distribution. In consequence,
the phenomenology that we have described seems to be relatively general despite
the fact that the nature of the QSS strongly depends on the initial condition.

A striking and mysterious outcome of our study (see also
\cite{epjb2010,epjb2013}) is the finding that, in many cases, the QSSs are well
fitted by polytropic (Tsallis) distributions. The value of the polytropic index
depends on the class of initial condition. For the semi-elliptical initial
distribution, which is a polytrope of index $n=1$, the polytrope that fits the
QSS has the same index as the initial state, i.e., $n=1$. Although not
considered in the present paper, we found in \cite{epjb2013} that for the
waterbag initial distribution, which is a polytrope of index $n=1/2$, the
polytrope that fits the QSS also has the same index, i.e., $n=1/2$. By contrast,
for the Gaussian initial distribution, which can be viewed as a polytrope of
index $n=\infty$, the polytrope that fits the QSS has a  different index, $n=2$.
We have no theory to explain these striking results. However, they confirm that
polytropic (Tsallis) distributions  can be useful in systems with
long-range interaction as claimed in \cite{lbt}.

\section{Conclusion}

In this work, we have considered the QSSs of long-range
systems, using the paradigmatic HMF model,
that are reached when the system is initially in a spatially homogeneous state
which is Vlasov
unstable. The Vlasov equation admits an infinity of
stable stationary states, each of them representing a possible QSS of the
$N$-body system. The prediction of
the QSS reached by any given initial unstable state is still practically
an unsolved problem although progress has been done lately
\cite{houches,assisebook,campabook,physrep}.
The Lynden-Bell theory \cite{lb} is in a sense conceptually satisfying since it
is based, as the equilibrium Boltzmann-Gibbs theory,
on the assumption of complete mixing of the dynamics. The differences between
the Boltzmann theory and the Lynden-Bell theory come from the difference in the
number of integrals of motion between the full $N$-body dynamics
(collisional) and the Vlasov dynamics (collisionless). In particular, the
Vlasov equation conserves an infinite number of Casimir integrals that must
be accounted for in the  Lynden-Bell theory. This leads
to a great difficulty in
the concrete application of the Lynden-Bell theory to generic initial conditions
but, at least in principle, one has the tool to determine
the QSS as a function of the initial state. Unfortunately, the basic assumption
of the Lynden-Bell theory, i.e. complete mixing, seems very far from
being satisfied in general \cite{incomplete}. Without this starting point, the
construction of alternative theories are conceptually more difficult since
it is not yet possible to quantify in a sensible way the departure from complete mixing, and even if it were possible, it would still
be uncertain how to translate it in a theory of the dynamical behavior of the
system. Thus, for the moment one can only resort to
semi-empirical approaches. In Ref. \cite{epjb2013} we have proposed, in
the absence
of complete mixing, to substitute the Lynden-Bell entropy
with other Casimirs of the Vlasov equation, e.g. the Tsallis entropy leading to
polytropic QSSs. This is, however, essentially empirical because we have
no way to predict {\it a priori} if the QSS will be a polytrope and, if
so, what the value of the polytropic index will be. However, the polytropic fit
works remarkably well for a wide range of energies, sufficiently far from the
instability threshold, and the polytropic index appears to be  relatively {\it
universal} \cite{epjb2013} for a given class of initial conditions.\footnote{The
quality of the polytropic fit is impressive and the universality of the
polytropic index for some classes of equivalence of the initial condition
($n=2$ for Gaussian, $n=1$ for semi-elliptical and $n=1/2$
for waterbag)
remains relatively mysterious. Also mysterious is the fact that the collisional
evolution of the system can be fitted by a sequence of polytropic
distributions with a time-dependent index $n(t)$ both in
the homogeneous \cite{ccgm} and in the inhomogeneous \cite{epjb2013} phase.
These
exciting results represent some of the last unsolved mysteries of
long-range systems.} Another proposal has been given
in Ref. \cite{levin2011} where a core-halo structure of the QSS was
hypothesized from the start. We showed in
\cite{epjb2013} that our approach based on polytropic distributions is
consistent with this core-halo approach and generalizes it.
At low energies, we find that the QSS has a core-halo structure with a
polytropic core surrounded by a halo of particles. In the
case of waterbag initial conditions (that correspond to polytropes of index
$n=1/2$) we recover the ``uniform'' core-halo structure found
in \cite{levin2011}.\footnote{To some extent, this core-halo structure can be
understood from energetic considerations (see Fig. 34 in  \cite{epjb2013}).}
For other initial conditions, we obtain more
general ``polytropic'' core-halo structures.
In the present work, we have developed  a QL theory
that has proven to work with good approximation close to the instability
threshold, i.e., in the energy range where the
polytropic fit proposed in Ref. \cite{epjb2013} is not good. Contrary
to time-independent approaches that try to predict the
form of the
distribution of the QSS either by the optimization of a
functional or by an assumption about the form of its analytical
expression,  the QL theory requires to
solve a dynamical (diffusion) equation that smoothes out the Vlasov
equation.\footnote{Even in the context of the Lynden-Bell theory, it may be
necessary to solve a dynamical (relaxation) equation in order to take into
account the problem of incomplete relaxation as discussed in \cite{incomplete}.}
The first
approaches are implicitly based on the assumption that the evolution of the
system is violent and strongly inhomogeneous so that a change of structure of
the initial distribution function occurs. The QL theory, on the other hand, is
based on the assumption that throughout the
transient from the initial distribution to the QSS the distribution remains
always almost homogeneous. The first approaches are expected to work if the
instability of the
initial distribution is strong, i.e., if the initial condition is far from
the instability threshold. The QL theory is expected to
work if the instability of the
initial distribution is weak, i.e., if the initial condition is close to
the instability threshold. These general features have been verified in
this work.

Summarizing, the dynamical process leading to a QSS seems to
be best described by different approaches depending on the
intial state of the system. The Lynden-Bell theory is based on an 
assumption of efficient mixing (ergodicity) whose validity is hard to 
establish {\it a priori}. The polytropic and core-halo fits seem to work well
when the  Lynden-Bell theory fails. The quasilinear theory is a fully predictive
theory that is valid in a perturbative regime close to the instability
threshold. These approaches have
complementary domains of validity. Hopefully, in a near future, these methods
will be encompassed into a single more general theory.

\appendix

\section{The decrease of $\omega_{{\rm I}}(t)$ with time}
\label{omegadecrease}
We have seen in the numerical integration of the diffusion equation that the function $f_0(p,t)$ keeps approximately
its functional form. This can be exploited to give an argument
showing that $\omega_{{\rm I}}(t)$ decreases monotonically with
time.
We start from the dispersion relation (\ref{disperhmfeven_b}), that for
convenience we recall here:
\begin{equation}\label{disperhmfeven_b_app}
1 +\pi \int \dd p \, \frac{pf_0'(p,t)}
{p^2 + \omega_{{\rm I}}^2(t)} = 0 \,\, .
\end{equation}
Let us write the function $f_0$ at time $t + \dd t$ as
\begin{equation}\label{funcmodified}
f_0(p,t+\dd t) = \alpha f_0(\alpha p,t) + \delta f (p,t),
\end{equation}
where $\alpha = 1 - \xi$ with $\xi \ll 1$, and where we assume that
\begin{equation}\label{deltafsmall}
|\delta f (p,t)| \ll |f_0(p,t) - \alpha f_0(\alpha p,t)|
\end{equation}
By neglecting the term $\delta f$ in Eq. (\ref{funcmodified}) we obtain 
$\omega_{{\rm I}}(t+ \dd t) < \omega_{{\rm I}}(t)$ as we show below. The
fact that $\alpha < 1$ is obtained from
Eq. (\ref{derkinener_b}). Now, writing the dispersion relation (\ref{disperhmfeven_b_app})
for $f_0(p,t + \dd t)$ by using Eq. (\ref{funcmodified}) and neglecting $\delta f$, we have
\begin{eqnarray}\label{disperhmfeven_b_app_mod}
&&1 +\pi \int \dd p \, \frac{pf_0'(p,t + \dd t)}
{p^2 + \omega_{{\rm I}}^2(t+\dd t)} =
1 +\pi \int \dd p \, \frac{\alpha^2 pf_0'(\alpha p,t)}
{p^2 + \omega_{{\rm I}}^2(t+\dd t)} \nonumber \\ &=&
1 +\pi \int \dd p \, \frac{pf_0'(p,t)}
{(p^2/\alpha^2) + \omega_{{\rm I}}^2(t+\dd t)} = 0 \,\, ,
\end{eqnarray}
where the last passage has been obtained by a trivial change of variable in the integral.
If $\omega_{{\rm I}}(t+\dd t)$ is not smaller than $\omega_{{\rm I}}(t)$, then
$(p^2/\alpha^2) + \omega_{{\rm I}}^2(t+\dd t)> p^2 + \omega_{{\rm I}}^2(t)$ for any $p$. Considering that the integrand
in the dispersion relation is negative definite, we infer that the right-hand side of Eq. (\ref{disperhmfeven_b_app_mod})
would be necessarily positive. The only way to make it vanish is to have $\omega_{{\rm I}}(t+ \dd t) < \omega_{{\rm I}}(t)$.

\section{Proof of Eq. (\ref{disperhmfsmall})}
\label{proofomegasmall}

We write $p/(p^2+\omega_{\rm I}^2(t)) = 1/p - \omega_{\rm
I}^2(t)/(p^2+\omega_{\rm I}^2(t))$. Substituting this decomposition
in Eq. (\ref{disperhmfeven_b}), the first term gives rise to the integral in Eq.
(\ref{disperhmfsmall}). For the second
term we have
\begin{eqnarray}
&&-\pi \omega_{\rm I}^2(t) \int \dd p \, \frac{f_0'(p,t)}{p(p^2+\omega_{\rm
I}^2(t))}
= -\pi {\rm sign}[\omega_{\rm I}(t)] \int \dd x \, \frac{f_0'(\omega_{\rm
I}(t)x,t)}{x(x^2+1)}
\nonumber \\
&=& -\pi |\omega_{\rm I}(t)| \int \dd x \, \frac{f_0''(0,t)}{x^2+1} +
O(\omega_{\rm I}^2(t))
= -\pi^2 |\omega_{\rm I}(t)| f_0''(0,t) + O(\omega_{\rm I}^2(t)).
\label{proofomsmall}
\end{eqnarray}

\section{The modified Gaussian}
\label{appmodgau}

Here, we show how to solve the system of equations
(\ref{eqforparameters_1})-(\ref{eqforparameters_3}) and to obtain
the values of $A$, $\beta$ and $p_1$ as a function of $\epsilon_{\rm kin}$.
Using the complementary error function
${\rm erfc}(x)$ defined by
\begin{equation}\label{erfcdef}
{\rm erfc}(x) = 1 - {\rm erf}(x) = \frac{2}{\sqrt{\pi}} \int_x^\infty \dd y \, {\rm e}^{-y^2}
\,\, ,
\end{equation}
the three equations become
\begin{eqnarray}\label{eqforparametersb_1}
2Ap_1 {\rm e}^{-\beta \frac{p_1^2}{2}} + \sqrt{\frac{2\pi}{\beta}}A \, {\rm erfc}\left(\sqrt{\frac{\beta}{2}}p_1\right)
&=& \frac{1}{2\pi}, \\
\label{eqforparametersb_2}
1 - \pi \beta \sqrt{\frac{2\pi}{\beta}}A \, {\rm
erfc}\left(\sqrt{\frac{\beta}{2}}p_1\right) &=& 0, \\
\label{eqforparametersb_3}
Ap_1 {\rm e}^{-\beta \frac{p_1^2}{2}} \left( \frac{p_1^2}{3} + \frac{1}{\beta}\right) + \sqrt{\frac{\pi}{2\beta^3}}A \,
{\rm erfc}\left(\sqrt{\frac{\beta}{2}}p_1\right) &=&
\frac{\epsilon_{\rm kin}}{2\pi}.
\end{eqnarray}
We know that for the energy $\epsilon_{\rm kin} = {1}/{4}$,
corresponding to the
critical point of the model,
for which the Gaussian velocity distribution function is marginally stable, the
solution of these three equations
is $\beta=2$, $p_1=0$ and $A = {1}/{2\pi^{{3}/{2}}}$. Starting from this
solution, it is not difficult to find numerically
the solution for other values of $\epsilon_{\rm kin}$. In fact, the system of
three equations is of the form
\begin{eqnarray}\label{systemeq}
F_1(A,\beta,p_1) &=& 0, \\
F_2(A,\beta,p_1) &=& 0, \\
F_3(A,\beta,p_1;\epsilon_{\rm kin}) &=& 0 \, .
\end{eqnarray}
It is then possible to obtain closed expressions for ${\partial A}/{\partial
\epsilon_{\rm kin}}$,
${\partial \beta}/{\partial \epsilon_{\rm kin}}$ and ${\partial
p_1}/{\partial \epsilon_{\rm kin}}$. Then, starting
from the known solution for $\epsilon_{\rm kin} = {1}/{4}$, one can compute the
solutions for other values by numerical
integration.


\begin{thebibliography}{}


\bibitem{houches} {\small {\it Dynamics and thermodynamics of systems
with long range interactions}, edited by T. Dauxois {\it et al.},
Lecture Notes in Physics {\bf 602}, (Springer, 2002).}
\bibitem{assisebook} {\small {\it Dynamics and thermodynamics of systems
with long range interactions: Theory and experiments}, edited by
A. Campa {\it et al.}, AIP Conf. Proc. {\bf 970} (AIP, 2008). }
\bibitem{campabook}   
A. Campa, T. Dauxois, D. Fanelli, and S. Ruffo, {\it
Physics of Long-Range Interacting Systems} (Oxford University Press, 2014).
\bibitem{physrep}
A. Campa, T. Dauxois, and S. Ruffo, Phys Rep. {\bf 480}, 57 (2009).
\bibitem{hmf}  
M. Antoni and S. Ruffo, Phys. Rev. E {\bf 52}, 2361
(1995).
\bibitem{ybbdr} 
Y. Yamaguchi, J. Barr\'e, F. Bouchet, T. Dauxois, S. Ruffo, Physica A {\bf
337}, 36 (2004).
\bibitem{incomplete} 
P.H. Chavanis, Physica A {\bf 365}, 102 (2006).
\bibitem{lb} 
D. Lynden-Bell, MNRAS {\bf 136}, 101 (1967).
\bibitem{precommun}  
A. Antoniazzi, D. Fanelli, J. Barr\'e, P.H.
Chavanis, T. Dauxois, and S. Ruffo, Phys. Rev. E {\bf 75}, 011112 (2007).
\bibitem{stan} 
F. Staniscia, P.H. Chavanis, G. de Ninno, and D. Fanelli,
Phys. Rev. E {\bf 80}, 021138 (2009).
\bibitem{epjb2013}
A. Campa and P.H. Chavanis, Eur. Phys. J. B {\bf 86}, 170 (2013).
\bibitem{tsallisbook}
C. Tsallis, {\it Introduction to Nonextensive
Statistical Mechanics} (Springer, 2009).
\bibitem{btnew}  
J. Binney and S. Tremaine, {\it Galactic Dynamics} (Princeton Series in
Astrophysics, 2008).
\bibitem{thlb}
S. Tremaine, M. H\'enon and D. Lynden-Bell, Mon. Not. R. astr. Soc.  {\bf 219},
285 (1986).
\bibitem{jsm2010}
A. Campa and P.H. Chavanis, J. Stat. Mech. {\bf 6}, 06001 (2010).
\bibitem{epjb2010}
P.H. Chavanis and A. Campa, Eur. Phys. J. B {\bf 76}, 581 (2010).
\bibitem{assise}
P.H. Chavanis, AIP Conf. Proc. {\bf 970}, 39 (2008).
\bibitem{nicholson}  
D.R. Nicholson, {\it Introduction to Plasma Theory}
(Krieger Publishing Company, Florida, 1992).
\bibitem{epjp}
P.H. Chavanis, Eur. Phys. J. Plus {\bf 127},
19 (2012).
\bibitem{csr}
P.H. Chavanis, J. Sommeria and R. Robert, Astrophys. J.  {\bf 471},
385 (1996).
\bibitem{cg}
P.H. Chavanis, Mon. Not. R. astr. Soc.  {\bf 300},
981 (1998).
\bibitem{case}
K. M. Case, Ann. Phys. {\bf 7}, 349 (1959).
\bibitem{vankampen}
N.G. Van Kampen, Physica  {\bf 21}, 949 (1955).
\bibitem{landau}
L. Landau, J. Phys. U.S.S.R.  {\bf 10}, 25 (1946).
\bibitem{nyquisthmf}
P.H. Chavanis and L. Delfini, Eur. Phys. J. B {\bf 69}, 389 (2009).
\bibitem{nyquistaa}
P.H. Chavanis, Eur. Phys. J. B {\bf 85}, 229 (2012).
\bibitem{ccgm} 
A. Campa,  P.H. Chavanis, A. Giansanti, and G. Morelli,
Phys. Rev. E {\bf 78}, 040102 (2008).
\bibitem{mk} 
H. Morita and K. Kaneko, Phys. Rev. Lett. {\bf 96}, 050602 (2006).
\bibitem{latora} 
V. Latora, A. Rapisarda, and C. Tsallis, Phys. Rev. E 
{\bf 64}, 056134 (2001).
\bibitem{lbt} 
P.H. Chavanis, Eur. Phys. J. B {\bf 53}, 487 (2006).
\bibitem{levin2011}
R. Pakter and Y. Levin, Phys. Rev. Lett. {\bf 106}, 200603 (2011).


\end{thebibliography}
\end{document}